\documentclass[prd,nofootinbib,onecolumn,preprint,11pt]{revtex4}

\usepackage{amssymb,amsmath}
\usepackage{latexsym}

\usepackage{graphicx}
\usepackage{bm}
\usepackage{float}
\usepackage{subfigure}
\usepackage{amsmath}
\usepackage{amsfonts}
\usepackage{color}
\usepackage{slashed}
\usepackage{fancyhdr}

\newcommand{\mpl}{M_{\text{pl}}}

\newcommand{\dd}{\text{d}}

\newcommand{\Pb}{\overline{\Pi}}

\begin{document}

\title{Retarded Green's Function Of A Vainshtein System And Galileon Waves}

\author{Yi-Zen Chu and Mark Trodden}
\affiliation{
Center for Particle Cosmology, Department of Physics and Astronomy,\\
University of Pennsylvania, Philadelphia, PA 19104
}

\begin{abstract}
\noindent Motivated by the desire to test modified gravity theories exhibiting the Vainshtein mechanism, we solve in various physically relevant limits, the retarded Galileon Green's function (for the cubic theory) about a background sourced by a massive spherically symmetric static body. The static limit of our result will aid us, in a forthcoming paper, in understanding the impact of Galileon fields on the problem of motion in the solar system. In this paper, we employ this retarded Green's function to investigate the emission of Galileon radiation generated by the motion of matter lying deep within the Vainshtein radius $r_v$ of the central object: acoustic waves vibrating on its surface, and the motion of compact bodies gravitationally bound to it. If $\lambda$ is the typical wavelength of the emitted radiation, and $r_0$ is the typical distance of the source from the central mass, with $r_0 \ll r_v$, then, compared to its non-interacting massless scalar counterpart, we find that the Galileon radiation rate is suppressed by the ratio $(r_v/\lambda)^{-3/2}$ at the monopole and dipole orders at high frequencies $r_v/\lambda \gg 1$. However, at high enough multipole order, the radiation rate is enhanced by powers of $r_v/r_0$. At low frequencies $r_v/\lambda \ll 1$, and when the motion is non-relativistic, Galileon waves yield a comparable rate for the monopole and dipole terms, and are amplified by powers of the ratio $r_v/r_0$ for the higher multipoles.
\end{abstract}

\maketitle

\section{Introduction and Motivation}

The study of the problem of motion in General Relativity (GR) has been a central theme in testing its validity. For instance, the post-Newtonian program, computing GR corrections to the Newtonian gravitational potential between massive bodies, is crucial to understanding gravity in our solar system. In the past decade or more, the post-Newtonian analysis of weak field gravity has also been developed to very high order in perturbation theory because of the need to model gravitational waves (GWs) from inspiraling compact binaries, which is expected to be a major source for detectors like Advanced LIGO. At the same time, the discovery of cosmic acceleration and its associated cosmological constant problem, has prompted many attempts to modify how gravity operates at large (astrophysical) length scales. One such example is the family of scalar field theories known as Galileons~\cite{Nicolis:2008in}, building on interesting properties of a limit of the Dvali-Gabadadze-Porrati model~\cite{Dvali:2000hr}. These scalar fields couple to the stress-energy of ordinary matter, and therefore alter large scale dynamics, but due to their self interactions they exhibit the Vainshtein screening effect~\cite{Vainshtein:1972sx,Deffayet:2001uk}, such that masses close to the central source of gravity (lying within a so-called Vainshtein radius $r_v$) do not feel their presence, and thus allowing Galileons to evade solar system tests of GR to date. Furthermore, models of this type have become even more interesting since it has been discovered that they may be extended in multiple ways to yield other new field theories~\cite{Luty:2003vm,Nicolis:2004qq,Deffayet:2009wt,Deffayet:2009mn,deRham:2010eu,Deffayet:2010zh,Padilla:2010de,Deffayet:2010qz,Hinterbichler:2010xn,Goon:2010xh,Khoury:2011da,Goon:2011qf,Trodden:2011xh,Goon:2011uw,Burrage:2011bt,Goon:2011xf,Zhou:2011ix,Goon:2012mu,Goon:2012dy,Gabadadze:2012tr} with related attractive properties, and that they arise as a limit of ghost-free massive gravity~\cite{deRham:2010ik,deRham:2010kj,Hinterbichler:2011tt}

In this paper, we wish to lay the groundwork for understanding analytically, and in some detail, the impact of such Vainshtein screened scalar fields on the problem of motion. To achieve concrete results we will consider the cubic Galileon theory about flat spacetime and couple the Galileon field to the trace of the stress-energy of matter. We will place a large mass $M$ at the origin of our coordinate system. The central goal of this paper is to solve the retarded Green's function of the linearized equations of motion of the Galileon fluctuating around the background field sourced by $M$. In a paper in preparation~\cite{ACHT}, we use the static limit of our Green's function here to investigate the conservative portion of the dynamics; to compute the effective potential between well separated test masses orbiting around $M$.

In the current paper we address the dissipative aspect of the dynamics: does motion of matter lying well within the Vainshtein radius of $M$ produce Galileon radiation that can carry energy-momentum away to infinity? This question arises as an issue of principle because the Galileon is a massless scalar field, and one would expect the motion of sources of massless fields to create radiation. Yet, it is not clear how much radiation would actually be produced, because it could perhaps be suppressed by the Vainshtein mechanism.

Beyond issues of principle, we are also motivated by the possibility that one can constrain modified gravity theories by demanding that the power loss from binary pulsar systems, such as the Hulse-Taylor binary PSR B1913+16, had better not deviate too far from the predictions of GR, since observations have confirmed the latter to high precision\footnote{For recent work on this topic, see~\cite{deRham:2012fw}}. Furthermore, we look forward to the prospect that, as already alluded to, within the next decade or so GW detectors may be able to directly listen in on the spacetime ripples generated by such compact binary systems. Once this is possible, we may hope to use these GW signals to search for or put further constraints on the existence of the Vainshtein mechanism. This requires that we develop a quantitative prediction of the Galileon waves themselves, beyond just an estimate of the power loss through scalar emission. Even though the two comparable point mass Galileon problem possibly requires different techniques to solve due to the importance of the nonlinearities of the field equations, the results of our current paper, which assume the existence of a very large central mass, may perhaps be seen as an approximation to the situation where the inspiraling binary consists of rather unequal masses, say $M_1 \ll M_2$. Yet another possible source of GWs is the oscillations of neutron stars themselves; as such, we will also consider a toy problem of surface waves on a spherical body stimulating Galileon waves.

The outline of the paper is as follows. In section \eqref{Setup}, we set up our problem in a quantitative manner. In section \eqref{GreensFunctionResults} we summarize the results for the Galileon retarded Green's function about the background field of the central mass $M$; and following that in section \eqref{GreensFunctionDerivation} we step through its derivation. In section \eqref{CurvedSpacetime} we describe how the linearized equations of motion of the Galileon field propagating about the background sourced by $M$ is equivalent to a minimally coupled massless scalar in some curved spacetime. Then in section \eqref{RadiationSection} we move on to use the retarded Galileon Green's function to study two examples of radiative processes: Galileon radiation produced by surface waves on an otherwise spherical central body, and that generated by $n$ point masses gravitationally bound to $M$. In appendix \eqref{GradientsOfMinimallyCoupledMasslessScalarField}, we work out both the curved and Minkowski spacetime minimally coupled massless scalar analog of the Li\'{e}nard-Wiechert potentials in electromagnetism; this spacetime calculation complements the frequency space one in section \eqref{NBodyProblem_Radiation}.

{\it Notation} \qquad A few words on notation. The speed of light is set to unity. The Galileon field is $\Pi$. The background spacetime is (3+1)-dimensional Minkowski, with metric in Cartesian coordinates given by 
\begin{align}
\dd s^2 = \eta_{\mu\nu} \dd x^\mu \dd x^\nu, \quad \eta_{\mu\nu} = \text{diag}[1,-1,-1,-1]
\end{align}
so that,
\begin{align}
\partial^2 \Pi \equiv \eta^{\mu\nu} \partial_\mu \partial_\nu \Pi, \qquad (\partial \Pi)^2 \equiv \eta^{\mu\nu} \partial_\mu \Pi \partial_\nu \Pi.
\end{align}
In spherical coordinates $(t,r,\theta,\phi)$, with $\theta \in [0,\pi]$ and $\phi\in[0,2\pi)$, the metric reads instead
\begin{align}
\dd s^2 = \dd t^2 - \dd r^2 - r^2 \Omega_\text{AB} \dd x^\text{A} \dd x^\text{B}, \\
\Omega_\text{AB} = \text{diag}[1,\sin^2\theta], \qquad x^\text{A} = (\theta,\phi) \ ,
\end{align}
where $\dd \Omega \equiv \dd\theta\dd\phi\sqrt{\Omega}$ (with $\sqrt{\Omega} \equiv \sqrt{\det \Omega_\text{AB}} = \sin\theta$) is the infinitesimal solid angle.

A hat on a variable representing spatial location, e.g. $\widehat{x}$, denotes the unit vector, $\widehat{x} \equiv \vec{x}/|\vec{x}|$. In particular, $\widehat{x}$ only depends on the spherical coordinate angles $\widehat{x} = \widehat{x}[\theta,\phi]$. Finally, the Planck 
mass is defined in terms of Newton's gravitational constant $G_\text{N}$ as $\mpl \equiv 1/\sqrt{32\pi G_\text{N}}$.

\section{Setup}
\label{Setup}

We would like to understand Galileon radiative processes taking place in the background Galileon field $\Pb[r]$ generated by a massive central body of mass $M$, which we would take to be static (time independent) and spherically symmetric. To model this mass $M$ we shall treat it as a point particle at rest, located at the spatial origin $\vec{0}$ of the coordinate system. The Galileon radiation we are investigating is generated by matter (described by stress-energy tensor $\delta T^{\mu\nu}$) lying well within the Vainshtein radius $r_v$ of the central mass. (The Vainshtein radius $r_v$, as we will see very shortly, is the radius below which the Galileon field $\Pb$ generated by the mass $M$ is increasingly governed by non-linear self interactions; well outside Vainshtein, the theory is linear and there $M$ generates a $1/r$ Coulomb potential.) This matter distribution $\delta T^{\mu\nu}$ is meant to be viewed as a perturbation relative to the mass $M$, but can otherwise be arbitrary. For instance $\delta T^{\mu\nu}$ may describe slight deviation of the mass $M$ from an exact spherical configuration (its multipole moments), and/or a deviation from time independence; it could also be the energy-momentum of $n$ light compact bodies, with masses $\{m_a \ll M|a=1,\dots,n\}$, orbiting around $M$.

We will take for the Galileon theory the simplest cubic $\Pi$ Lagrangian; in actuality, several other terms are allowed in 4 spacetime dimensions. The Galileon, being a scalar, couples to the trace of the stress-energy tensor of matter. The total action for our setup is therefore
\begin{align}
\label{FullAction}
S_\Pi + S_M + \delta S
\end{align}
where, in Cartesian coordinates,
{\allowdisplaybreaks\begin{align}
\label{Action_Pi}
S_\Pi &\equiv \int \dd^4 x \left( \frac{1}{2} (\partial\Pi)^2 + \frac{1}{\Lambda^3} \partial^2 \Pi (\partial\Pi)^2 \right), \ \Lambda > 0, \\
\label{Action_CentralMass}
S_M &\equiv \frac{M}{\mpl} \int \dd t' \Pi[ t',\vec{0} ], \\
\label{Action_Perturbations}
\delta S &\equiv \int \dd^4 x \frac{\Pi}{\mpl} \delta T, \qquad \delta T \equiv \delta T^\mu_{\phantom{\mu}\mu}.
\end{align}}

Because we are interested in radiation that can propagate to infinity, we note that far away from the source $M$, contributions to the Galileon stress-energy tensor $\mathfrak{T}^{\mu\nu}$ due to the cubic-in-$\Pi$ terms in $S_\Pi$ fall away more rapidly than their quadratic counterparts (provided, of course, that $\Pi$ and its gradients falls off with increasing $r$). Expressed in Cartesian coordinates, the asymptotic Galileon energy-momentum tensor is thus that of the non-interacting massless scalar in flat spacetime,
\begin{align}
\label{GalileonAsymptoticStressTensor}
\mathfrak{T}_{\mu\nu}[r\to\infty] = \partial_\mu \Pi \partial_\nu \Pi - \frac{1}{2} \eta_{\mu\nu} (\partial\Pi)^2 \ .
\end{align}
The background Galileon field sourced by the mass $M$ is the exact solution to $\delta (S_\Pi + S_M)/\delta \Pb = 0$. Since we have a static spherically symmetric source, $\Pb$ must only depend on $r$, and so the Euler-Lagrange equation from varying $S_\Pi + S_M$ becomes
\begin{align}
\label{BackgroundPiEquation_TotalDivergence}
\partial_r \left( r^2 \left\{ \partial_r \Pb - \frac{4}{\Lambda^3 r} \left( \partial_r \Pb \right)^2 \right\} \right) \
= -\frac{r^2 M}{\mpl} \delta^{(3)}[\vec{x}]
\end{align}
Integrating both sides of \eqref{BackgroundPiEquation_TotalDivergence} over a sphere of radius $r$ centered at $\vec{0}$ then yields
\begin{align}
\label{BackgroundPiEquation}
\partial_r \Pb - \frac{4}{\Lambda^3 r} \left( \partial_r \Pb \right)^2 = -\frac{M}{4\pi \mpl r^2}.
\end{align}
(Even though we have modeled the central body as a point mass, \eqref{BackgroundPiEquation} is valid outside any isolated static spherically symmetric matter distribution as long as we replace $M$ with the integral $4\pi\int_0^\infty \tau^\mu_{\phantom{\mu}\mu}[r'] r'^2 \dd r'$, where $\tau^\mu_{\phantom{\mu}\nu}$ is the stress-energy tensor of the body.) Eq. \eqref{BackgroundPiEquation} is a quadratic equation in $\Pb'[r]$. Defining the Vainshtein radius 
\begin{align}
\label{VainshteinRadius}
r_v \equiv \frac{1}{\Lambda}\left(\frac{4}{\pi} \frac{M}{\mpl}\right)^{1/3} \ ,
\end{align}
we obtain
\begin{align}
\label{BackgroundPiSolution}
\frac{\Pb'[r]}{\Lambda^3} = \frac{r}{8} \left( 1 - \sqrt{ 1 + \left(\frac{r_v}{r}\right)^3 } \right).
\end{align}
There is a second solution for $\Pb'[r]$ in which the negative sign in front of the square root is replaced with a plus sign. This solution is proportional to $r$ and hence blows up as $r \to \infty$. Since the stress-tensor depends on $\Pb'[r]$ -- see eq. \eqref{GalileonAsymptoticStressTensor} and note that including the contributions to $\mathfrak{T}^{\mu\nu}$ from the cubic-in-$\Pi$ terms in $S_\Pi$ would only exacerbate the problem -- we may discard this second solution on the grounds that the energy-momentum of $\Pb$ measured by an asymptotic observer cannot be infinite. 

It is possible to integrate eq. \eqref{BackgroundPiSolution} exactly in terms of hypergeometric functions,
\begin{align}
\label{BackgroundPiSolution_2F1}
\frac{\Pb[r]}{\Lambda^3}
= \frac{r^2}{16} \left( 1 - 4 \left(\frac{r_v}{r}\right)^{\frac{3}{2}} \,_2 F_1\left[ -\frac{1}{2}, \frac{1}{6}; \frac{7}{6}; -\frac{r^3}{r_v^3} \right] \right) 
- \frac{\Gamma\left[ -\frac{2}{3} \right] \Gamma\left[ \frac{7}{6} \right]}{8 \sqrt{\pi}} r_v^2 \ ,
\end{align}
where we have chosen the asymptotic boundary condition to be $\Pb[r \to \infty] = 0$.

For later use, note the following limits of equations \eqref{BackgroundPiSolution} and \eqref{BackgroundPiSolution_2F1}. When $r \ll r_v$,
\begin{align}
\label{BackgroundPiSolution_2F1_SmallRadius}
\Pb[r] &= -\frac{M}{\mpl \pi r_v} \left( \frac{\Gamma[-\frac{2}{3}] \Gamma[\frac{7}{6}]}{2\sqrt{\pi}} + \sqrt{\frac{r}{r_v}} + \mathcal{O}\left[ \left(\frac{r}{r_v}\right)^2 \right] \right),
\end{align}
and
\begin{align}
\label{BackgroundPiSolution_SmallRadius_IofII}
\Pb'[r] 
&= - \frac{\Lambda^3 r}{8} \left( \left(\frac{r_v}{r}\right)^{3/2} - 1 + \mathcal{O}\left[ \left(\frac{r}{r_v}\right)^{\frac{3}{2}} \right] \right), \\
\label{BackgroundPiSolution_SmallRadius_IIofII}
\Pb''[r] 
&= \frac{\Lambda^3}{16} \left( \left(\frac{r_v}{r}\right)^{3/2} + 2 + \mathcal{O}\left[ \left(\frac{r}{r_v}\right)^{\frac{3}{2}} \right] \right).
\end{align}
On the other hand, when $r \gg r_v$,
\begin{align}
\label{BackgroundPiSolution_2F1_LargeRadius}
\Pb[r] &= \frac{M}{4 \pi \mpl r} \left( 1 + \mathcal{O}\left[ \left(\frac{r_v}{r}\right)^3 \right] \right),
\end{align}
and
\begin{align}
\label{BackgroundPiSolution_LargeRadius_IofII}
\Pb'[r]
&= - \frac{\Lambda^3 r}{16} \left( \left(\frac{r_v}{r}\right)^3 + \mathcal{O}\left[ \left(\frac{r_v}{r}\right)^6 \right]  \right), \\
\label{BackgroundPiSolution_LargeRadius_IIofII}
\Pb''[r] 
&= \frac{\Lambda^3}{8} \left( \left(\frac{r_v}{r}\right)^3 + \mathcal{O}\left[ \left(\frac{r_v}{r}\right)^6 \right]  \right).
\end{align}
Notice that the first term of the small radius limit $\Pb'[r] \sim 1/\sqrt{r}$ in \eqref{BackgroundPiSolution_SmallRadius_IofII} can be obtained by dropping the linear term $\partial_r\Pb[r]$ in eq. \eqref{BackgroundPiEquation}, whereas the first term in \eqref{BackgroundPiSolution_LargeRadius_IofII} of the large radius limit $\Pb'[r] \sim 1/r^2$, which is the usual Coulomb force law, comes about from dropping the non-linear $(\partial_r \Pb)^2$ piece in eq. \eqref{BackgroundPiEquation}. This is the quantitative statement that, close to the matter source ($r \ll r_v$), the dynamics of Galileons are primarily governed by their nonlinear self interactions. The theory is linear when the observer is well outside the Vainshtein radius. 

More physically, to see the Vainshtein effect at work, imagine a test point mass $m \ll M$, at spatial location $\vec{Z}$, orbiting the central body. Its action takes the same form as $S_M$ in eq. \eqref{Action_CentralMass}, except we evaluate the Galileon field about the background $\Pb$ generated by $M$,
\begin{align}
S_m \equiv \frac{m}{\mpl} \int \dd t \Pb\left[\vec{Z}[t]\right].
\end{align}
(Strictly speaking, Lorentz invariance says $\dd t$ needs to be replaced with $\dd t \sqrt{1-\dot{\vec{Z}}[t]^2}$; however, we are working in the non-relativistic regime, where the square root is very close to unity.) By employing the definition of the Vainshtein radius in eq. \eqref{VainshteinRadius}, followed by integrating the first term on the right-hand-sides of equations \eqref{BackgroundPiSolution_SmallRadius_IofII} and \eqref{BackgroundPiSolution_LargeRadius_IofII} without worrying too much about the overall numerical factors,
\begin{align}
S_m &\sim \int \dd t \frac{G_\text{N} M m}{|\vec{Z}[t]|} \left(\frac{|\vec{Z}[t]|}{r_v}\right)^{\frac{3}{2}}, 	&|\vec{Z}[t]| \ll r_v \\
&\sim \int \dd t \frac{G_\text{N} M m}{|\vec{Z}[t]|} \ , 														&|\vec{Z}[t]| \gg r_v 
\end{align}
Thus, the Galileon potential experienced by a test point mass orbiting close to the central mass $M$ is the Newtonian gravitational potential $G_\text{N} M/|\vec{Z}|$ multiplied by a suppression factor of $(|\vec{Z}|/r_v)^{3/2} \ll 1$. Only when the test mass travels well outside Vainshtein does the suppression factor drop out and the Galileon potential become comparable in strength to that of regular gravity.\footnote{The reader concerned about the stability of the Galileon model is referred to~\cite{Nicolis:2004qq}.}

With the exact background solution $\Pb'[r]$ in hand, we now substitute
\begin{align}
\Pi[x] = \Pb[r] + \varphi[x]
\end{align}
in \eqref{FullAction}, and drop terms cubic-in-$\varphi$. The resulting linearized equation of motion of $\varphi$ about the background $\Pb$ reads 
\begin{align}
\label{LinearizedEOM}
\mathcal{W}_x \varphi[x] = \frac{\delta T[x]}{\mpl} \ ,
\end{align}
where the differential operator $\mathcal{W}_x$ is 
\begin{align}
\mathcal{W}_x \varphi[x] \equiv \left( e_1 \partial_t^2 - e_2 \partial_r^2 - \frac{2}{r} e_3 \partial_r - \frac{1}{r^2} e_3 \vec{L}^2 \right) \varphi[x]
\end{align}
and $\vec{L}^2$ is the angular part of the Laplacian in Euclidean 3-space, usually called the negative of the ``angular momentum squared'' operator, given by
\begin{align}
\label{NegativeLSquared}
\vec{L}^2 \varphi \equiv \frac{1}{\sqrt{\Omega}} \partial_\text{A} \left( \sqrt{\Omega} \Omega^\text{AB} \partial_\text{B} \varphi \right).
\end{align}
Here
{\allowdisplaybreaks\begin{align}
\label{e1}
e_1[r] &\equiv 1 - \frac{8 \overline{\Pi}'}{\Lambda^3 r} - \frac{4 \overline{\Pi}''}{\Lambda^3}
			= \frac{1}{4} \left( 3\sqrt{\frac{(2 r^3 + r_v^3)^2}{r^3(r^3+r_v^3)}} -2 \right), \\
\label{e2}
e_2[r] &\equiv 1 - \frac{8 \overline{\Pi}'}{\Lambda^3 r}
			= \sqrt{1 + \left(\frac{r_v}{r}\right)^3}, \\
\label{e3}
e_3[r] &\equiv 1 - \frac{4 \overline{\Pi}'}{\Lambda^3 r} - \frac{4 \overline{\Pi}''}{\Lambda^3} 
			= \frac{1}{4}\sqrt{\frac{(4 r^3 + r_v^3)^2}{r^3(r^3+r_v^3)}}.
\end{align}}
If we assume $\delta T$ does not implicitly depend on $\Pi$, then $\varphi$ is entirely sourced by $\delta T$, the (for now arbitrary) matter perturbations. The key to solving $\varphi$ in terms of $\delta T$ is the retarded Green's function defined by the equation, 
\begin{align}
\label{GreensFunctionEquation}
\mathcal{W}_x G[x,x'] = \mathcal{W}_{x'} G[x,x'] 
= \frac{1}{rr'} \delta[t-t'] \delta[r-r']\delta[\phi-\phi']\delta[\cos\theta-\cos\theta'], 
\end{align}
where the $\delta$s are the Dirac delta functions. The solution to the linearized Galileon equation \eqref{LinearizedEOM} about the static spherically symmetric background $\Pb$ is now (in Cartesian coordinates)
\begin{align}
\label{PiGeneralSolution}
\varphi[x] = \int \dd^4 x' G[x,x'] \frac{\delta T[x']}{\mpl}.
\end{align}
We write $x$ to represent a collective label for $(t,r,\theta,\phi)$ and $x'$ for $(t',r',\theta',\phi')$, so that the requirement that the signal does not precede the turning on of the source requires $G[x,x'] = 0$ for $t < t'$. 

Note that it is not obvious that the appropriate spacetime dependence multiplying the $\delta$-functions in the Green's function equation \eqref{GreensFunctionEquation} is $(rr')^{-1}$, and therefore we will justify this in appendix \eqref{ODEGreensFunction_Appendix} below.

When solving eq. \eqref{GreensFunctionEquation} it is important to remember the following boundary condition. Since the Green's function is the field generated by a unit point mass, if we let $\delta T$ describe a static point mass sitting at the origin,
\begin{align}
\delta T[x'] \equiv \delta M \delta^{(3)}[\vec{x}'], \qquad \delta M/M \ll 1,
\end{align}
this merely amounts to shifting the mass of the central body by $M \to M + \delta M$. Then we already know what to expect from the linear solution represented by the integral in eq. \eqref{PiGeneralSolution}. It should be the linear-in-$\delta M$ piece of the full $\Pb[r]$ solution in eq. \eqref{BackgroundPiSolution_2F1} upon the replacement $M \to M + \delta M$. Remember that the mass dependence in the full solution of eq. \eqref{BackgroundPiSolution_2F1} is contained entirely in $r_v$ via eq. \eqref{VainshteinRadius}. Perturbing $M \to M + \delta M$ in eq. \eqref{BackgroundPiSolution_2F1} up to linear order in $\delta M$ yields,
\begin{align}
\Pb[r;M+\delta M] = \Pb[r;M] + \delta \Pb[r] ,
\end{align}
with
\begin{align}
\label{deltaPiBackground}
\delta \Pb[r] =
\frac{\delta M}{\mpl} \frac{1}{2\pi r_v} \left(  
\frac{\Gamma\left[\frac{1}{3}\right] \Gamma\left[\frac{1}{6}\right]}{6\sqrt{\pi}} - \sqrt{\frac{r}{r_v}} \,_2F_1\left[\frac{1}{6},\frac{1}{2};\frac{7}{6};-\frac{r^3}{r_v^3}\right]
\right) .
\end{align}
Hence, when solving $G[x,x']$ below, we must obtain from eq. \eqref{PiGeneralSolution},
\begin{align}
\label{ell=0_BoundaryCondition}
\frac{\delta M}{\mpl} \int_{-\infty}^\infty \dd t' G[x,x'] 
= \frac{\delta M}{\mpl} \int_{-\infty}^\infty \dd t G[x,x'] 
= \delta \Pb[r], \qquad x \equiv (t,\vec{x}); \ x' \equiv (t',\vec{0}).
\end{align}
(The second equality follows from the time translation symmetry of the problem at hand.) For later use let us note that the small and large radius limits are, respectively,
\begin{align}
\label{StaticGreensFunction_rp=0_SmallRadius}
\int_{-\infty}^\infty \dd t' G[x,x'] \to \frac{1}{2\pi r_v} \left(  
\frac{\Gamma\left[\frac{1}{3}\right] \Gamma\left[\frac{1}{6}\right]}{6\sqrt{\pi}} - \sqrt{\frac{r}{r_v}} 
\right), \qquad r \to 0
\end{align}
and (using eq. \eqref{2F1_zTo1/z} below)
\begin{align}
\label{StaticGreensFunction_rp=0_LargeRadius}
\int_{-\infty}^\infty \dd t' G[x,x'] \to \frac{1}{4\pi r}, \qquad r\to\infty .
\end{align}

\section{Retarded Galileon Green's Function In Background Sourced By A Massive Spherically Symmetric Static Body}

In this section we solve, in different limits of physical interest, the Galileon retarded Green's function about the spherically symmetric background $\Pb[r]$ in equation \eqref{BackgroundPiSolution}. The first subsection summarizes all the results in a coherent manner, and in subsequent subsections we step through the derivation systematically.

\subsection{Overview of results}
\label{GreensFunctionResults}

Our solution of the retarded Galileon Green's function $G[x,x']$ obeying equation \eqref{GreensFunctionEquation} is composed of an integral over all angular frequencies $\omega$ and an infinite mode sum over all harmonics $(\ell,m)$:\footnote{Strictly speaking, we need to specify a contour for the Fourier integral below, but since we do not need it in this paper, we shall leave this question for the future.}
\begin{align}
\label{GreensFunctionResult_ModeExpansion}
G[x,x'] &= \int_{-\infty}^{+\infty} \frac{\dd \omega}{2\pi} e^{-i\omega(t-t')} 
\omega \sum_{\ell = 0}^\infty \widetilde{g}_\ell\big[\omega r, \omega r'\big] \sum_{m=-\ell}^{+\ell} Y_\ell^m[\theta,\phi] \overline{Y_\ell^m}[\theta',\phi'].
\end{align}
We will make frequent use of the dimensionless variables
\begin{align}
\label{xiDef}
\xi \equiv \omega r, \ \xi' \equiv \omega r', \ \xi_v \equiv \omega r_v.
\end{align}
because we will shortly show that the radial Green's function $\widetilde{g}_\ell$ depends on $r$, $r'$ and $r_v$ solely through $\xi$, $\xi'$ and $\xi_v$ respectively. In eq. \eqref{GreensFunctionResult_ModeExpansion}, the $Y_\ell^m$s are the usual spherical harmonics spanning a complete set of functions defined on a sphere of unit radius embedded in 3 spatial dimensions (the over bar means complex conjugation); they obey the eigenvalue equation
\begin{align}
\label{SphericalHarmonics_EigenEquation}
\vec{L}^2 Y_\ell^m = -\ell(\ell+1) Y_\ell^m.
\end{align}

Because the background $\Pb$ is static, the Green's function reflects the time translation symmetry of the setup at hand. Moreover, spherical symmetry tells us the radial Green's functions $\widetilde{g}_\ell$ do not depend on the azimuthal number $m$.

The separation of variables method of mode expansion employed in eq. \eqref{GreensFunctionResult_ModeExpansion} reduces the problem of solving the linear partial differential equation for $G[x,x']$ in eq. \eqref{GreensFunctionEquation} to a linear second order ordinary differential equation (ODE) for the radial Green's function $\widetilde{g}_\ell$. Inserting the ansatz in eq. \eqref{GreensFunctionResult_ModeExpansion} into eq. \eqref{GreensFunctionEquation}, and using a Fourier representation of $\delta[t-t']$ and the completeness relation for the spherical harmonics, one may read off the ODE for $\widetilde{g}_\ell$ in frequency space by equating the coefficient of $\exp[-i\omega(t-t')] Y_\ell^m[\theta,\phi] \overline{Y_\ell^m}[\theta',\phi']$ on both sides of the Green's function equation. We then arrive at
\begin{align}
\label{GreensFunctionEquation_RadialModeEquation}
\left( - e_2 \partial_\xi^2 - \frac{2}{\xi} e_3 \partial_\xi - e_1 + \frac{\ell(\ell+1)}{\xi^2} e_3 \right) \widetilde{g}_\ell[\xi,\xi'] 
= \frac{\delta[\xi-\xi']}{\xi\xi'} .
\end{align}
We have carried out a change of variables according to the rules in eq. \eqref{xiDef}; for the $e_{1,2,3}$ this amounts to simply replacing every $r$ variable with its corresponding $\xi$ variable. That eq. \eqref{GreensFunctionEquation_RadialModeEquation} no longer depends explicitly on the radii nor on the angular frequency $\omega$ means that the solution for the radial Green's function $\widetilde{g}_\ell$, cannot depend on the lengths $r,r',r_v$ or frequency $\omega$ explicitly. 

The radial Green's function $\widetilde{g}_\ell$ has a discontinuous first derivative at $\xi = \xi'$ because its second derivatives at $\xi=\xi'$ needs to yield $\delta[\xi-\xi']/(\xi\xi')$. Therefore we need to distinguish between the two regions $|\xi| > |\xi'|$ or $|\xi| < |\xi'|$. We therefore let $r_>$ and $r_<$ represent the larger and smaller of the two radii $r$ and $r'$, and define
\begin{align}
\xi_> \equiv \omega r_>, \qquad \xi_< \equiv \omega r_<.
\end{align}

We now proceed to summarize the results for $\widetilde{g}_\ell$.

{\it Radiative Limit} \qquad Of central importance in this paper, is the situation where the emitter lies deep inside the Vainshtein radius while the observer sits far outside ($r \gg r_v \gg r'$). Taking the limit where one of the radii is much smaller than $r_v$ and the other much larger, more specifically $r_< \ll r_v$ and $|\xi_>| \gg |\xi_v|^{3/2}$, we obtain
\begin{eqnarray}
\label{GreensFunctionResult_RadiativeLimit_IofIII}
\widetilde{g}_\ell[\xi,\xi'] &=&
\left\{
\begin{array}{lr}
h_\ell^{(1)}[\xi_>] C_0^\text{(rad)}[\xi_v] \sqrt[4]{\xi_<} J_{-\frac{1}{4}} \left[ \sqrt{3}\xi_</2 \right] & \ell=0 \\
& \\
h_\ell^{(1)}[\xi_>] C_\ell^\text{(rad)}[\xi_v] \sqrt[4]{\xi_<} J_{\frac{1}{4}(2\ell+1)}\left[ \sqrt{3}\xi_</2 \right] & \ell>0 \\
\end{array}
\right. \ ,
\end{eqnarray}
where $J_\nu$ is the Bessel function and the $h_\ell^{(1)}$ is the spherical Hankel function of the first kind. 

In the high frequency regime $|\xi_v| \gg \ell$, the coefficients $C_\ell^\text{(rad)}$ are
\begin{eqnarray}
\label{GreensFunctionResult_RadiativeLimit_IIofIII}
C_\ell^\text{(rad)} &=&
\left\{
\begin{array}{lr}
\frac{\sqrt{\pi/2}}{\xi_v^{3/4}} e^{i \pi\frac{7}{8} - i\xi_v \mathcal{I}_\infty} & \ell=0 \\
& \\
\frac{\sqrt{\pi/2}}{\xi_v^{3/4}} e^{i \pi\left(\frac{\ell}{4}+\frac{5}{8}\right) - i\xi_v \mathcal{I}_\infty} & \ell>0 \\
\end{array}
\right. \ ,
\end{eqnarray}
where
\begin{align}
\label{UInfinity}
\mathcal{I}_\infty &\equiv \int_0^\infty \left( 1-\sqrt{-U[\vartheta]} \right) \dd \vartheta \approx 0.253 \ ,
\end{align}
and
\begin{equation}
\label{WKBMomenta}
U[\vartheta] \equiv -\frac{1}{4} \frac{3(1+2\vartheta^3) - 2\sqrt{\vartheta^3(1+\vartheta^3)}}{1+\vartheta^3} \ .
\end{equation}
and is plotted in Fig.\eqref{WKBPotential} below. The suppression factor of $1/\xi_v^{3/4}$ in  \eqref{GreensFunctionResult_RadiativeLimit_IIofIII}  indicates that high frequency Galileon signals are indeed Vainshtein screened, at least for small $\ell$s.

\begin{figure}
\begin{center}
\includegraphics[width=3.2in]{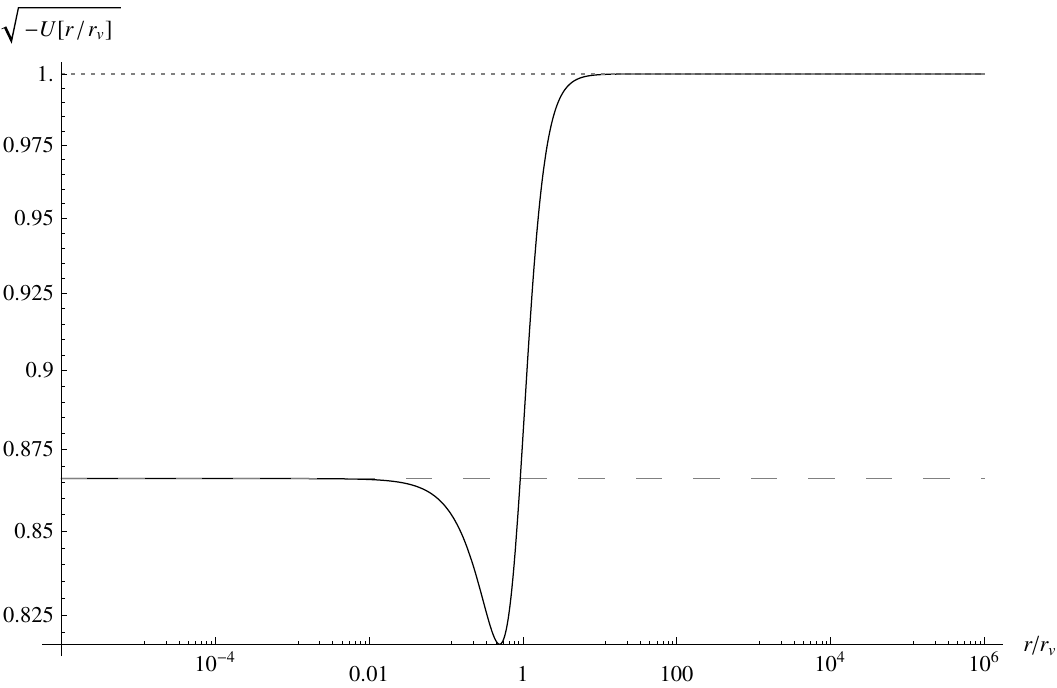}
\caption{A Log-Log plot of the WKB ``momenta'' $\sqrt{-U[r/r_v]}$ (solid line) -- see eq. \eqref{WKBMomenta} -- as a function of the ratio $r/r_v$. The asymptotics are: $\sqrt{-U[0]} = \sqrt{3}/2$ (long-dashed line) and $\sqrt{-U[\infty]} = 1$ (short-dashed line). The turning point, which is a global minimum, is at $\sqrt{-U[r/r_v = 1/2]} = \sqrt{2/3}$.}
\label{WKBPotential}
\end{center}
\end{figure}

In the low frequency regime, $|\xi_v| \ll 1$, the coefficients $C_\ell^\text{(rad)}$ are
\begin{eqnarray}
\label{GreensFunctionResult_RadiativeLimit_IIIofIII}
C_\ell^\text{(rad)} &=&
\left\{
\begin{array}{lr}
i \frac{\sqrt[8]{3} \pi}{\Gamma[\frac{1}{4}]} & \ell=0 \\
& \\
i \sqrt{\pi} \xi_v^{\frac{\ell-1}{2}} \frac{\Gamma\left[-\frac{2 \ell }{3}-\frac{1}{3}\right] \Gamma\left[\frac{5}{6}-\frac{\ell }{3}\right]}{2^{2\ell+1} 3^{\frac{\ell}{4}+\frac{1}{8}} \Gamma \left[\frac{\ell}{2}+\frac{3}{4}\right] \Gamma [-\ell]} \left(1 + \frac{\cos\left[\frac{1}{6} (2\ell+1)\pi\right]}{\sin[\pi\ell]} \right) & \ell>0 \\
\end{array}
\right. \ ,
\end{eqnarray}

It is worth pointing out that although $\sin[\pi\ell]$ appears in the denominator, and this expression contains $\Gamma$-functions whose arguments appear they could be negative integers; these $\ell\geq 1$ terms are all in fact non-singular. (This remark also applies to the related equation eq. \eqref{GreensFunctionResult_Static_rp<<rv<<r} below.) We list the first ten $C_\ell^\text{(rad)}$s here, to 3 significant figures:
\begin{center}
  \begin{tabular}{c|c|c|c|c|c|c|c|c|c|c}
    \hline
    $\ell$ & 1 & 2 & 3 & 4 & 5 & 6 & 7 & 8 & 9 & 10 \\ \hline
    $C_\ell^\text{(rad)}/(i \sqrt{\pi} \xi_v^{\frac{\ell-1}{2}})$ & 0.324 & 0.127 & 0.0386 & 0.0100 & 0.00230 & 0.000482 & 0.0000932 & 0.0000168 & 2.86 $\times 10^{-6}$ & 4.61 $\times 10^{-7}$ \\ 
    \hline
  \end{tabular}
\end{center}
Notice from equation \eqref{GreensFunctionResult_RadiativeLimit_IIIofIII} that the $\ell=0,1$ modes do not contain $r_v$; this is a direct consequence of the fact that the leading order terms of the static Green's function in this same $r_> \gg r_v \gg r_<$ limit (eq. \eqref{GreensFunctionResult_Static_rp<<rv<<r}) do not contain any $r_v$ for $\ell=0,1$. In fact, we may further compare the cubic Galileon radial Green's function $\widetilde{g}_\ell$ with its non-interacting massless cousin $\widetilde{g}^\text{(Flat)}_\ell = i h_\ell^{(1)}[\xi_>] j_\ell[\xi_<]$ (see eq. \eqref{MinimallyCoupledMasslessScalarG_Fourier} and \eqref{AdditionTheorem} below). Let us consider the non-relativistic limit $|\xi'| \ll 1$, where the reciprocal of the characteristic frequency of the motion is much smaller than the characteristic distance of the source to the central mass $M$. In this limit, we may replace the Bessel functions with their small argument limits, and find
\begin{align}
\label{VainshteinAmplification_IofII}
\widetilde{g}_0 = \widetilde{g}^\text{(Flat)}_0 = i h_\ell^{(1)}[\xi_>], \qquad \qquad
\widetilde{g}_1 = \widetilde{g}^\text{(Flat)}_1 = \frac{i}{3} \xi_< h_\ell^{(1)}[\xi_>]
\end{align}
and for $\ell \geq 2$,
\begin{align}
\label{VainshteinAmplification_IIofII}
\widetilde{g}_\ell[r_> \gg r_v \gg r_<] &\to \kappa_\ell \xi_v^\ell \left(\frac{r_<}{r_v} \right)^{\frac{\ell+1}{2}} h_\ell^{(1)}[\xi_>] \\
\widetilde{g}^\text{(Flat)}_\ell[r_> \gg r_v \gg r_<] &\to \kappa'_\ell \xi_v^\ell \left(\frac{r_<}{r_v} \right)^\ell h_\ell^{(1)}[\xi_>] . \nonumber
\end{align}
Here, $\kappa_\ell$ and $\kappa'_\ell$ are constants that depend solely on $\ell$. This teaches us that, while high frequency Galileon power loss is Vainshtein screened, low frequency signals generated from deep within the Vainshtein radius of $M$ are comparable to or even Vainshtein enhanced relative to the non-interacting massless scalar. The results of the radiative processes described in section \eqref{RadiationSection} will reflect these observations.

{\it The WKB High Frequency Limit} \qquad When $|\xi|, |\xi'|, |\xi_v| \gg \max[1,\ell]$, we may apply the WKB approximation. Let us first define
\begin{equation}
\label{PhiDef}
\Phi_{\lessgtr} \equiv \int_0^{r_{\lessgtr}/r_v} \sqrt{-U[\vartheta]} \dd \vartheta.
\end{equation}
where it is worth noting that $\sqrt{-U[0]} = \sqrt{3}/2$ and $\sqrt{-U[\infty]} = 1$. The leading order WKB solution is
\begin{align}
\label{GreensFunctionResult_WKB}
\widetilde{g}_\ell[\xi,\xi'] 
&= \left( U[\xi/\xi_v] U[\xi'/\xi_v] \xi(\xi^3+\xi_v^3) \xi'(\xi'^3+\xi_v^3) \right)^{-\frac{1}{4}} \\
&\qquad \qquad 
\times \left( \frac{i}{2} \exp\left[ i \xi_v \left( \Phi_> - \Phi_< \right) \right] 
+ \frac{C_\ell^{++}}{2} \exp\left[ i \xi_v \left( \Phi_> + \Phi_< \right) \right] \right) \ ,
\nonumber
\end{align}
where
\begin{eqnarray}
\label{GreensFunctionResult_WKB_C++}
C^{++}_\ell  &=&
\left\{
\begin{array}{lr}
e^{i\frac{\pi}{4}} & \ell=0 \\
& \\
-e^{i\pi \left(\frac{3}{4}-\frac{\ell}{2}\right)} & \ell>0 \\
\end{array}
\right. \ .
\end{eqnarray}
When $\max[1,\ell] \ll |\xi_v| \ll |\xi|, |\xi'|$, the radial Green's functions become
\begin{align}
\label{GreensFunctionResult_WKB_LargeRadii}
\widetilde{g}_\ell[\xi,\xi'] 
&= \frac{1}{\xi \xi'} \left( \frac{i}{2} \exp\left[ i \left( \xi_> - \xi_< \right) \right] 
+ \frac{C_\ell^{++}}{2} \exp\left[ i \left( \xi_> + \xi_< - 2 \xi_v \mathcal{I}_\infty \right) \right] \right) \ ,
\end{align}
whereas when  $\ell \ll |\xi|, |\xi'| \ll |\xi_v|$ they become
\begin{align}
\label{GreensFunctionResult_WKB_SmallRadii}
\widetilde{g}_0[\xi,\xi'] 
&= \frac{2}{\sqrt{3} \xi_v^{3/2}\sqrt[4]{\xi\xi'}} \left( \frac{i}{2} \exp\left[ i\frac{\sqrt{3}}{2} \left( \xi_> - \xi_< \right) \right] 
+ \frac{C_\ell^{++}}{2} \exp\left[ i \frac{\sqrt{3}}{2} \left( \xi_> + \xi_< \right) \right] \right) .
\end{align}
In this limit, observe that the $r_>$ and $t-t'$ dependent portion of the combination $e^{-i\omega (t-t')} \widetilde{g}_\ell$ in the mode expansion eq. \eqref{GreensFunctionResult_ModeExpansion} is
\begin{align}
\frac{1}{\sqrt[4]{r_>}} \exp\left[i \frac{\sqrt{3}}{2} \omega \left( r_> - \frac{2}{\sqrt{3}} (t-t') \right)\right],
\end{align}
and if we imagine a source located much closer to $M$ than the observer is ($r' \ll r \ll r_v$) the Green's function tells us the observer will receive purely outgoing radial waves. It is not entirely clear from the outset that this would be the case, particularly viewed from the curved spacetime picture (which we describe in section \eqref{CurvedSpacetime} below), because one may think that the radiation from the source could backscatter off the spacetime geometry and return to the observer, thereby mimicking an ingoing radial wave. (This scenario may in fact occur in the low frequency limit, where the longer wavelength of the Galileon waves may grow more sensitive to the curvature of the background effective geometry.) The phase $r - (2/\sqrt{3})(t-t')$ also indicates these outgoing waves, if they are propagating only in the radial direction, are superluminal because $2/\sqrt{3} > 1$. 

Of particular importance is the WKB radiative limit, $r_> \gg r_v \gg r_<$. Here the radial Green's function is
\begin{eqnarray}
\label{GreensFunctionResult_WKB_rp<<rv<<r}
\widetilde{g}_\ell[\xi,\xi'] &=&
\left\{
\begin{array}{lr}
\frac{\sqrt{2}i}{\xi_> \sqrt[4]{3 \xi_<} \xi_v^{3/4}} e^{i\left( \xi_> - \xi_v \mathcal{I}_\infty - \frac{\pi}{8}\right)} \cos\left[ \frac{\sqrt{3}}{2} \xi_< - \frac{\pi}{8} \right] & \ell=0 \\
& \\
\frac{\sqrt{2}i}{\xi_> \sqrt[4]{3 \xi_<} \xi_v^{3/4}} e^{i\left( \xi_> - \xi_v \mathcal{I}_\infty + \pi\frac{5-2\ell}{8}\right)} \cos\left[ \frac{\sqrt{3}}{2} \xi_< + \pi\frac{5-2\ell}{8} \right] & \ell>0 \\
\end{array}
\right. \ .
\end{eqnarray}

{\it Large Mode Number} \qquad It is useful to note that the asymptotic expansion of the Bessel function $J_\nu[z]$ exhibits an exponential suppression at large order $\nu$, for $z < \nu$. Therefore, the presence of $J_{(1/4)(2\ell+1)}[\sqrt{3}\xi_</2]$ in eq. \eqref{GreensFunctionResult_RadiativeLimit_IofIII} means, at least for the radiation problem ($r_> \gg r_v \gg r_<$) --  the main object of this paper -- we may neglect mode numbers much larger than $|\xi_<|$.

{\it The Static Limit} \qquad We may also obtain the zero frequency (static) limit of the Green's function, defined as
\begin{align}
\label{GreensFunctionEquation_StaticDef}
G^\text{(static)}[\vec{x},\vec{x}'] = \int_{-\infty}^{+\infty} G\left[x \equiv (t,\vec{x}), x' \equiv (t',\vec{x}')\right] \dd t
\end{align}
(It does not actually matter, because of time translation symmetry of the situation, whether we integrate with respect to $t$ or $t'$ in eq. \eqref{GreensFunctionEquation_StaticDef}.) As the name suggests, the static Green's function does not depend on time. We may interpret $G^\text{(static)}[\vec{x},\vec{x}']$ as the Galileon potential between two static point sources, both of unit mass, in the background $\Pb[r]$. Putting the mode expansion in eq. \eqref{GreensFunctionResult_ModeExpansion} into the integral over all time in eq. \eqref{GreensFunctionEquation_StaticDef} yields the mode expansion for the static Green's function
\begin{align}
\label{GreensFunctionResult_ModeExpansion_Static}
G^\text{(static)}[\vec{x},\vec{x}'] 
= \lim_{\omega\to 0} \omega \sum_{\ell = 0}^\infty \widetilde{g}_\ell[\xi,\xi'] \sum_{m=-\ell}^{+\ell} Y_\ell^m[\theta,\phi] \overline{Y_\ell^m}[\theta',\phi'] .
\end{align}
The exact result can be expressed in terms of hypergeometric functions as
\begin{align}
\label{GreensFunctionResult_Static_Exact}
G^\text{(static)}[\vec{x},\vec{x}']
&= \frac{1}{2\pi r_v} \left(  
\frac{\Gamma\left[\frac{1}{3}\right] \Gamma\left[\frac{1}{6}\right]}{6\sqrt{\pi}}
- \sqrt{\frac{r_>}{r_v}} \,_2F_1\left[\frac{1}{6},\frac{1}{2};\frac{7}{6};-\frac{r_>^3}{r_v^3}\right] 
\right) \nonumber\\
&+ \frac{1}{r_v} \sum_{\ell = 1}^\infty \sum_{m = -\ell}^\ell \frac{ Y_\ell^m[\theta,\phi] \overline{Y_\ell^m}[\theta',\phi'] }{ 2\ell+1 }
\left( \frac{r_<}{r_v} \right)^{\frac{\ell+1}{2}} \ _2F_1\left[ \frac{1}{6}-\frac{\ell}{6}, \frac{1}{2}+\frac{\ell}{2}; \frac{7}{6}+\frac{\ell}{3}; -\frac{r_<^3}{r_v^3} \right] \nonumber\\
&\times \Bigg( 
2 \left( \frac{r_v}{r_>} \right)^{\frac{\ell}{2}} \ _2F_1\left[ \frac{\ell}{6}+\frac{1}{3}, -\frac{\ell}{2}; \frac{5}{6}-\frac{\ell}{3}; -\frac{r_>^3}{r_v^3} \right] \\
&\qquad \qquad \qquad \qquad
+ \frac{\ell! \Gamma\left[ -\frac{1}{6}(2\ell+1) \right]}{\sqrt{\pi} \Gamma\left[\frac{1}{3}(2\ell+1)\right]} \left( \frac{r_>}{r_v} \right)^{\frac{\ell+1}{2}} \ _2F_1\left[ \frac{1}{6}-\frac{\ell}{6}, \frac{1}{2}+\frac{\ell}{2}; \frac{7}{6}+\frac{\ell}{3}; -\frac{r_>^3}{r_v^3} \right] 
\Bigg) \ . \nonumber
\end{align}
Via the identity in eq. \eqref{2F1_zTo1/z} below, the $\ell=0$ mode (the first line on the right hand side of eq. \eqref{GreensFunctionResult_Static_Exact}) is equivalent to
\begin{align}
\label{GreensFunctionResult_Static_Exact_ell=0}
\frac{1}{4\pi r_>} \,_2F_1\left[ \frac{1}{3}, \frac{1}{2}; \frac{4}{3}; -\frac{r_v^3}{r_>^3} \right] .
\end{align}
Also, since
\begin{align}
\,_2 F_1\left[ \alpha,\beta;\gamma; z \right] &=
\frac{\Gamma[\alpha+\beta-\gamma]\Gamma[\gamma]}{\Gamma[\alpha]\Gamma[\beta]} (1-z)^{\gamma-\alpha-\beta} \,_2 F_1[\gamma-\alpha,\gamma-\beta;\gamma-\alpha-\beta+1;1-z] \nonumber\\
&\qquad \qquad + \frac{\Gamma[\gamma-\alpha-\beta]\Gamma[\gamma]}{\Gamma[\gamma-\alpha]\Gamma[\gamma-\beta]} \,_2 F_1[\alpha,\beta;\alpha+\beta-\gamma+1;1-z] ,
\end{align}
and using that $1/\Gamma[-m] = 0$ if $m$ is a positive integer or zero, one of the static mode functions can be written as, for $\ell$ even,
\begin{align}
\left( \frac{r_v}{r} \right)^{\frac{\ell}{2}} \ _2F_1\left[ \frac{\ell}{6}+\frac{1}{3}, -\frac{\ell}{2}; \frac{5}{6}-\frac{\ell}{3}; -\frac{r^3}{r_v^3} \right]
&= \sqrt{\pi} \left( \frac{r_v}{r} \right)^{\frac{\ell}{2}} \frac{\Gamma\left[ \frac{5-2\ell}{6} \right]}{\Gamma\left[ -\frac{\ell-1}{2} \right] \Gamma\left[ \frac{\ell+5}{6} \right]} \,_2F_1\left[ -\frac{\ell}{2}, \frac{\ell+2}{6}; \frac{1}{2}; 1 + \frac{r^3}{r_v^3} \right],
\end{align}
and, for $\ell$ odd,
\begin{align}
\left( \frac{r_v}{r} \right)^{\frac{\ell}{2}} \ _2F_1\left[ \frac{\ell}{6}+\frac{1}{3}, -\frac{\ell}{2}; \frac{5}{6}-\frac{\ell}{3}; -\frac{r^3}{r_v^3} \right]
&= -2 \sqrt{1+\frac{r^3}{r_v^3}} \sqrt{\pi} \left( \frac{r_v}{r} \right)^{\frac{\ell}{2}} \frac{\Gamma\left[ \frac{5-2\ell}{6} \right]}{\Gamma\left[ -\frac{\ell}{2} \right] \Gamma\left[ \frac{\ell+2}{6} \right]} \,_2F_1\left[ -\frac{\ell-1}{2}, \frac{\ell+5}{6}; \frac{3}{2}; 1 + \frac{r^3}{r_v^3} \right] .
\end{align}
(Note that $\,_2F_1[-\ell/2,\dots; 1+(r/r_v)^3]$ and $\,_2F_1[-(\ell-1)/2,\dots; 1+(r/r_v)^3]$ are, respectively, $(\ell/2)$th (even $\ell$) and $(1/2)(\ell-1)$th (odd $\ell$) order polynomials in the variable $1+(r/r_v)^3$.)

Taking the limit $r,r' \gg r_v$ hands us the Green's function to the Laplacian in Euclidean 3-space
\begin{align}
\label{GreensFunctionResult_Static_rrp>>rv}
G^\text{(static)}[\vec{x},\vec{x}'] &= \frac{1}{4\pi |\vec{x} - \vec{x}'|},
\end{align}
plus corrections that begin at relative order $(r_v/r)^3$ and $(r_v/r')^3$. This is to be expected, since far outside the Vainshtein radius, the central mass becomes irrelevant and we ought to recover the theory of a massless scalar in flat spacetime.

Next, taking the limit $r,r' \ll r_v$ leads us to
\begin{align}
\label{GreensFunctionResult_Static_rrp<<rv}
G^\text{(static)}[\vec{x},\vec{x}'] &= \frac{1}{2\pi r_v} \left(
\frac{\sqrt{rr'/r_v^2}}{\left\vert \sqrt{r/r_v}\widehat{x} - \sqrt{r'/r_v}\widehat{x}' \right\vert} 
- \sqrt{\frac{r}{r_v}} - \sqrt{\frac{r'}{r_v}} + \frac{\Gamma\left[\frac{1}{3}\right] \Gamma\left[\frac{1}{6}\right]}{6\sqrt{\pi}} 
\right) ,
\end{align}
a result obtained independently in~\cite{ACHT}, and which will play a key role in our field theory based analysis of the conservative portion of the Galileon two body problem taking place in the background field $\Pb$ of $M$. Here $\widehat{x} = \widehat{x}[\theta,\phi]$ and $\widehat{x}' = \widehat{x}'[\theta',\phi']$ are the unit radial vectors of the observer and source, respectively, and vertical bars denote the Euclidean length. We have expressed every occurrence of the two radii in eq. \eqref{GreensFunctionResult_Static_rrp<<rv} as a small ratio, $r/r_v$ or $r'/r_v$, to highlight the Vainshtein mechanism at work.

In section \eqref{CurvedSpacetime} below, where we shall view the Galileon propagating on the background $\Pb$ as a minimally coupled massless scalar propagating in a particular curved spacetime, we shall re-derive eq. \eqref{GreensFunctionResult_Static_rrp<<rv} in an alternate manner.

For $r_> \gg r_v \gg r_<$, we obtain
\begin{align}
\label{GreensFunctionResult_Static_rp<<rv<<r}
G^\text{(static)}[\vec{x},\vec{x}'] 
= \frac{1}{4\pi r_>} + \frac{1}{r_>} \sqrt{\frac{r_<}{r_v}} &\sum_{\ell = 1}^\infty \sum_{m=-\ell}^\ell \frac{Y_\ell^m\left[ \theta,\phi \right] \overline{Y_\ell^m} \left[ \theta',\phi' \right]}{2\ell+1} \left(\frac{\sqrt{ r_v r_< }}{2 r_>}\right)^\ell \\
&\times \frac{\Gamma \left[-\frac{2 \ell}{3}-\frac{1}{3}\right] \Gamma \left[\frac{5}{6}-\frac{\ell}{3}\right]}{\sqrt{\pi } \Gamma[-\ell]} 
\left(\frac{\cos \left[\frac{1}{6} (2\ell+1)\pi\right]}{\sin[\pi\ell]} + 1 \right) \ , \nonumber
\end{align}
plus corrections that begin at order $(r_</r_v)^3$ and $(r_v/r_>)^3$ relative to these displayed terms. Notice for $\ell=0,1$, $r_v$ drops out of these leading order terms. This tells us for the monopole and dipole terms, when the wavelength of Galileon signals are much longer than the Vainshtein radius $r_v$, the Vainshtein mechanism becomes less effective.

{\it Well Inside Vainshtein} \qquad Let both the observer and emitter lie well inside the Vainshtein radius, $r,r' \ll r_v$. In the low frequency limit, $|\xi_v|\ll 1$, the radial Green's function becomes,
\begin{align}
\label{GreensFunctionResult_rrp<<rv_Static_ell=0}
\widetilde{g}_0[\xi,\xi'] &= \frac{\pi (\xi \xi')^{\frac{1}{4}}}{2 \xi_v^{3/2}} J_{-\frac{1}{4}} \left[\sqrt{3}\xi_</2\right] 
\left( \frac{4 \sqrt[4]{3}\sqrt{\pi} \Gamma[\frac{1}{3}] \Gamma[\frac{7}{6}]}{\Gamma[\frac{1}{4}]^2} \sqrt{\xi_v} J_{-\frac{1}{4}} \left[\sqrt{3}\xi_>/2\right] 
- \sqrt{2} J_{\frac{1}{4}} \left[\sqrt{3}\xi_>/2\right] \right) \ ,
\end{align}
for $\ell=0$, and
\begin{align}
\label{GreensFunctionResult_rrp<<rv_Static_ell>0}
\widetilde{g}_\ell[\xi,\xi'] &= i \frac{\pi (\xi \xi')^{\frac{1}{4}}}{2 \xi_v^{3/2}}  J_{\frac{1}{4}(2\ell+1)} \left[\sqrt{3}\xi_</2\right]
\Bigg( H^{(1)}_{\frac{1}{4}(2\ell+1)} \left[\sqrt{3}\xi_>/2\right] \nonumber\\
&\qquad \qquad
+\left( \frac{2}{i(-)^\ell-1} - \frac{i}{\xi_v^{\ell +\frac{1}{2}}} \frac{4^{\ell +1} \Gamma\left[-\frac{\ell}{3}-\frac{1}{6}\right] \Gamma\left[\frac{\ell}{2}+\frac{5}{4}\right]^2 \ell!}{3^{\frac{\ell}{2}+\frac{5}{4}} \pi^{3/2} \Gamma\left[\frac{2 (\ell +2)}{3}\right]} \Bigg) J_{\frac{1}{4}(2\ell+1)} \left[\sqrt{3}\xi_>/2\right] \right) ,
\end{align}
for $\ell\geq 1$. In the WKB limit, i.e. $|\xi_v| \gg \max[1,\ell]$, the radial Green's function instead reads
\begin{eqnarray}
\label{GreensFunctionResult_rrp<<rv_WKB}
\widetilde{g}_\ell[\xi,\xi'] &=&
\left\{
\begin{array}{lr}
i \frac{\pi (\xi \xi')^{\frac{1}{4}}}{2 \xi_v^{3/2}} e^{i\frac{\pi}{4}} H^{(1)}_{\frac{1}{4}} \left[\sqrt{3}\xi_>/2\right]
J_{-\frac{1}{4}} \left[\sqrt{3}\xi_</2\right] & \ell=0 \\
& \\
 i \frac{\pi (\xi \xi')^{\frac{1}{4}}}{2 \xi_v^{3/2}} H^{(1)}_{\frac{1}{4}(2\ell+1)} \left[\sqrt{3}\xi_>/2\right]
J_{\frac{1}{4}(2\ell+1)} \left[\sqrt{3}\xi_</2\right] & \ell>0 \\
\end{array}
\right. \ .
\end{eqnarray}

{\it Well Outside Vainshtein} \qquad When both observer and emitter lie well outside the Vainshtein radius, $r,r' \gg r_v$, we recover the theory of a minimally coupled massless scalar in flat spacetime, with the radial Green's function
\begin{align}
\label{GreensFunctionResult_rrp>>rv}
\widetilde{g}_\ell[\xi,\xi'] = i h_\ell^{(1)}[\xi_>] \left( j_\ell[\xi_<] + C^{(hh)}_\ell  h_\ell^{(1)}[\xi_<] \right),
\end{align}
where in the low frequency limit $|\xi_v| \ll 1$,
\begin{eqnarray}
C^{(hh)}_\ell &=&
\left\{
\begin{array}{lr}
< \mathcal{O}\left[ \xi_v \right] & \ell=0 \\
& \\
i \xi_v^{2 \ell +1} \frac{\Gamma \left[-\frac{2
   \ell }{3}-\frac{1}{3}\right] \Gamma \left[-\frac{\ell
   }{3}-\frac{1}{6}\right] \left(\csc \left[\frac{1}{6} (2 \pi  \ell
   +\pi )\right]+\cot \left[\frac{1}{6} (2 \pi  \ell +\pi )\right] \csc
   [\pi  \ell ]\right)}{2^{2 \ell +3} 3 \sqrt{\pi} ((2 \ell -1)!!)^2 \Gamma
   [-\ell ]}& \ell>0 \\
\end{array}
\right. \ ,
\end{eqnarray}
and in the high frequency limit $|\xi_v| \gg \max[1,\ell]$,
\begin{eqnarray}
C^{(hh)}_\ell &=&
\left\{
\begin{array}{lr}
-\frac{1}{2}\left( 1 + e^{-i\frac{\pi}{4} - 2i \mathcal{I}_\infty \xi_v} \right) & \ell=0 \\
& \\
-\frac{1}{2}\left( 1 - e^{i\pi\left(\frac{1}{4}+\frac{\ell}{2}\right) - 2i \mathcal{I}_\infty \xi_v} \right) & \ell>0 \\
\end{array}
\right. \ .
\end{eqnarray}

\subsection{Solving The Radial Green's Function}
\label{GreensFunctionDerivation}
In this section we derive the results presented in the preceding section. In appendix \eqref{ODEGreensFunction_Appendix}, we review the relevant facts about solving Green's functions for linear second order ODEs, and also justify the $(rr')^{-1}$ measure on the right hand side of eq. \eqref{GreensFunctionEquation}. The algorithm for obtaining $\widetilde{g}_\ell$ is as follows. 

{\it General Solution of Radial Green's Function} \qquad We need to first solve for the two linearly independent homogeneous solutions $R_\ell^{(1)}[\xi]$ and $R_\ell^{(2)}[\xi]$ to the ODE in eq. \eqref{GreensFunctionEquation_RadialModeEquation}, namely
\begin{align}
\label{GreensFunctionEquation_HomogeneousRadialModeEquation}
\left( - e_2 \partial_\xi^2 - \frac{2}{\xi} e_3 \partial_\xi - e_1 + \frac{\ell(\ell+1)}{\xi^2} e_3 \right) R_\ell^{(1,2)}[\xi] = 0 .
\end{align}
In the notation of eq. \eqref{ODEDifferentialOperator}, $p_2 = -e_2$ and $p_1 = -2 e_3/\xi$. Next we normalize the solutions $R_\ell^{(1,2)}$ such that they satisfy
\begin{align}
\label{WronskianCondition_Dynamic}
e_2[\xi,\xi_v] \left( R_\ell^{(1)}[\xi] (R_\ell^{(2)})'[\xi] - (R_\ell^{(1)})'[\xi] R_\ell^{(2)}[\xi] \right) = \frac{1}{\xi^2} .
\end{align}
Then the general solution to our radial Green's function is
\begin{align}
\label{GreensFunction_RadialModeAnsatz}
\widetilde{g}_\ell[\xi,\xi'] 
= \mathcal{C}_\ell R_\ell^{(1)}[\xi_>] R_\ell^{(2)}[\xi_<] - (1-\mathcal{C}_\ell) R_\ell^{(1)}[\xi_<] R_\ell^{(2)}[\xi_>] 
+ C_\ell^{11} R_\ell^{(1)}[\xi] R_\ell^{(1)}[\xi'] + C_\ell^{22} R_\ell^{(2)}[\xi] R_\ell^{(2)}[\xi'],
\end{align}
where the constants $\mathcal{C}_\ell$, $C_\ell^{11}$ and $C_\ell^{22}$ do not depend on $\xi$ nor $\xi'$, but depend on $\xi_v$. The $C_\ell$ and $1-C_\ell$ terms have discontinuous first derivatives and hence contribute to the coefficient of the $\delta$-functions on the right hand side of the Green's function equation in eq. \eqref{GreensFunctionEquation_RadialModeEquation}.

For $\ell \geq 1$, retarded boundary conditions and the demand for non-singular solutions will fix these constants uniquely. That is, we shall require that, whenever the observer is very far away from $M$, $r \gg r_v$, and the source is closer to the central mass than the observer, $r > r'$, then the observer ought to receive purely outgoing Galileon waves. Furthermore, on physical grounds, we will admit only solutions that are nonsingular when either the observer or the source is situated close to the central body.\footnote{The observer placed close to the central mass will experience a Galileon force ($\propto 1/\sqrt{r}$) due to $M$ that blows up as $r \to 0$; but here we are requiring that, as long as the observer is not sitting on top of the secondary source, i.e. the $\delta T$, she should not measure Galileon forces due to $\delta T$ that grow without bound.} For $\ell=0$, in addition to regularity and the retarded condition, we shall also need to invoke Gauss' law applied to the curved spacetime Helmholtz equation (see eq. \eqref{GreensFunctionEquation_CurvedSpacePicture_NormalizeAtZero}) to fix these constants uniquely.

In the following subsections, we will first solve for $\widetilde{g}_\ell$ in the zero frequency (static) and high frequency (WKB) limits. We will shortly also derive the $R_\ell^{(1,2)}$s in terms of Bessel and Hankel functions in the limits $r,r' \gg r_v$ and $r,r' \ll r_v$. This means we can fix the form of $\widetilde{g}_\ell$ in the limits $r,r' \ll r_v$, $r,r' \gg r_v$ and $r_> \gg r_v \gg r_<$ up the $\xi_v$-dependent constants $\mathcal{C}_\ell$, $C_\ell^{11}$ and $C_\ell^{22}$. We will then proceed to fix these constants -- at least within the low and high frequency limits, $|\xi_v| \ll 1$ and $|\xi_v| \gg \max[1,\ell]$, respectively -- by ensuring they agree with the static and WKB $\widetilde{g}_\ell$ results in the same limits.

{\it Well Outside Vainshtein} \qquad That we have just shown that the radial wave equation in eq. \eqref{GreensFunctionEquation_HomogeneousRadialModeEquation} reduces to that in Minkowski spacetime when $r \gg r_v$ implies that we may use the known solution there to read off the $R_\ell^{(1,2)}$s. The solution in flat (and, importantly, empty) Minkowski spacetime is textbook material, and we may represent it as
\begin{align}
\label{MinimallyCoupledMasslessScalarG_Fourier}
\frac{\delta[t-t'-|\vec{x}-\vec{x}'|]}{4\pi |\vec{x}-\vec{x}'|} 
= \int \frac{\dd \omega}{2\pi} e^{-i\omega(t-t')} \frac{e^{i\omega |\vec{x}-\vec{x}'|}}{4\pi |\vec{x}-\vec{x}'|} .
\end{align}
Then, using
\begin{align}
\label{AdditionTheorem}
\frac{e^{i\omega |\vec{x}-\vec{x}'|}}{4\pi |\vec{x}-\vec{x}'|} 
= i \omega \sum_{\ell = 0}^\infty \sum_{m = -\ell}^\ell Y_\ell^m[\theta,\phi] \overline{Y_\ell^m}[\theta',\phi'] j_\ell[\xi_<] h^{(1)}_\ell[\xi_>],
\end{align}
we can deduce that the $R_\ell^{(1,2)}[\xi]$ in the limit $r \gg r_v$ must be a linear combination of $j_\ell$ and $h_\ell^{(1)}$. 

The $h^{(1)}_\ell$ implement retarded boundary conditions, since the spherical Hankel function of the first kind may be understood as
\begin{equation}
h^{(1)}_\ell[z] = - i (-z)^\ell \left( \frac{1}{z} \frac{\dd}{\dd z} \right)^\ell \frac{e^{i z}}{z} \ ,
\end{equation}
Note that $h_\ell^{(1)}[\xi]$ only contains a factor of $\exp[+i\xi]$ and does not contain $\exp[-i \xi]$. Thus, using the asymptotic expansion of the Hankel function for large argument, we see that
\begin{align}
e^{-i\omega(t-t')} h_\ell^{(1)}[\xi] \to (-i)^{\ell+1}\frac{\exp\left[-i\omega(t-t'-r)\right]}{\omega r} \left( 1 + \mathcal{O}\left[ \left( \omega r \right)^{-1} \right] \right),
\end{align}
describes radially purely outgoing waves at unit speed propagating to infinity. A similar discussion shows that $h^{(2)}_\ell = (h^{(1)}_\ell)^*$ implements advanced boundary conditions, and because $j_\ell$ can be expressed as a linear combination of $h^{(1,2)}_\ell$, it describes a superposition of ingoing and outgoing waves. In the ansatz of eq. \eqref{GreensFunction_RadialModeAnsatz}, we see that we have to choose $R_\ell^{(1)}[\xi_>] = h_\ell^{(1)}[\xi_>]$ and set $\mathcal{C}_\ell = 1$ and $C^{22}_\ell = 0$ to ensure retarded boundary conditions. (We are able to deduce from eq. \eqref{AdditionTheorem} that $h_\ell^{(1)}$ and $j_\ell$ are already appropriately normalized to obey the Wronskian condition in eq. \eqref{WronskianCondition_Dynamic} for $r,r' \gg r_v$.) This means we have determined $\widetilde{g}_\ell[r,r' \gg r_v]$ to take the form in eq. \eqref{GreensFunctionResult_rrp>>rv}.

{\it Well Inside Vainshtein} \qquad Let us now understand the forms of $R_\ell^{(1,2)}[\xi]$ evaluated close to the central source ($r \ll r_v$). We exploit equations \eqref{BackgroundPiSolution_SmallRadius_IofII} and \eqref{BackgroundPiSolution_SmallRadius_IIofII}, keeping only the most dominant terms in $e_{1,2,3}$ (equations \eqref{e1} through \eqref{e3}), to reformulate eq. \eqref{GreensFunctionEquation_HomogeneousRadialModeEquation} as
\begin{align}
\left( \partial_\xi^2 + \frac{1}{2 \xi} \partial_\xi + \frac{3}{4} - \frac{\ell(\ell+1)}{4 \xi^2} \right) R^{(1,2)}_\ell[\xi] = 0 \ .
\end{align}
One may rescale the solutions $R^{(1,2)}_\ell[\xi] \equiv \xi^{1/4} \mathcal{R}^{(1,2)}_\ell[\sqrt{3}\xi/2]$ and find that $\mathcal{R}^{(1,2)}_\ell[\sqrt{3}\xi/2]$ satisfies Bessel's equation
\begin{align}
\left( \partial_\zeta^2 + \frac{1}{\zeta} \partial_\zeta + \left( 1 - \frac{\left( \frac{2\ell+1}{4} \right)^2}{\zeta^2} \right) \right) \mathcal{R}^{(1,2)}_\ell[\zeta] = 0 \ ,
\end{align}
where $\zeta \equiv \sqrt{3}\xi/2$. Noting that the Wronskian between $J_\nu$ and $H^{(1)}_\nu$ is
\begin{align}
\label{Wronskian_BesseJ_HankeH1}
{\sf Wr}_{(z)}[ J_\nu, H^{(1)}_\nu ] = J_\nu[z] (H^{(1)}_\nu)'[z] - (J_\nu)'[z] H^{(1)}_\nu[z] = \frac{2i}{\pi z},
\end{align}
and that
\begin{align}
{\sf Wr}_{(z)}\left[ z^\frac{1}{4} J_\nu[z], z^\frac{1}{4} H^{(1)}_\nu[z] \right] = \sqrt{z} {\sf Wr}_{(z)}\left[ J_\nu[z], H^{(1)}_\nu[z] \right],
\end{align}
we conclude that the two linearly independent solutions normalized to obey eq. \eqref{WronskianCondition_Dynamic} are
\begin{align}
\label{HomogeneousSolutions_SmallRadii}
R_\ell^{(1)}[\xi] &\equiv \sqrt{\frac{i\pi}{2 \xi_v^{3/2}}} \xi^{\frac{1}{4}} H^{(1)}_{\frac{1}{4}\left( 2\ell+1 \right)}\left[ \sqrt{3}\xi/2 \right], \\
R_\ell^{(2)}[\xi] &\equiv \sqrt{\frac{i\pi}{2 \xi_v^{3/2}}} \xi^{\frac{1}{4}} J_{\frac{1}{4}\left( 2\ell+1 \right)}\left[ \sqrt{3} \xi/2 \right].
\end{align}
With these homogeneous solutions, the general solution for the radial Green's function, deep within the Vainshtein radius, $r,r'\ll r_v$, is
\begin{align}
\label{RadialGreensFunctionGeneralSolution_DeepInsideVainshtein}
\widetilde{g}_\ell[\xi,\xi']
&= \frac{i\pi}{2 \xi_v^{3/2}} \sqrt[4]{\xi \xi'} \Bigg( \mathcal{C}_\ell H^{(1)}_{\frac{1}{4}\left( 2\ell+1 \right)}\left[ \sqrt{3}\xi_>/2 \right] J_{\frac{1}{4}\left( 2\ell+1 \right)}\left[ \sqrt{3}\xi_</2 \right] - \left(1-\mathcal{C}_\ell\right) H^{(1)}_{\frac{1}{4}\left( 2\ell+1 \right)}\left[ \sqrt{3}\xi_</2 \right] J_{\frac{1}{4}\left( 2\ell+1 \right)}\left[ \sqrt{3}\xi_>/2 \right] \nonumber\\
&\qquad + C_\ell^\text{(JJ)} J_{\frac{1}{4}\left( 2\ell+1 \right)}\left[ \sqrt{3}\xi/2 \right] J_{\frac{1}{4}\left( 2\ell+1 \right)}\left[ \sqrt{3}\xi'/2 \right]
+ C_\ell^\text{(HH)} H^{(1)}_{\frac{1}{4}\left( 2\ell+1 \right)}\left[ \sqrt{3}\xi/2 \right] H^{(1)}_{\frac{1}{4}\left( 2\ell+1 \right)}\left[ \sqrt{3}\xi'/2 \right]
\Bigg) . 
\end{align}
Next, we recall the small argument limits ($|z|\ll 1$) of the Bessel and Hankel functions
\begin{align}
\label{BesselJ_SmallArgument}
J_{\frac{1}{4}\left( 2\ell+1 \right)}[z] &\to \frac{(z/2)^{\frac{1}{4}\left( 2\ell+1 \right)}}{\Gamma\left[\frac{1}{4}\left( 2\ell+5 \right)\right]} \left( 1 + \mathcal{O}\left[ z^2 \right] \right), \\
\label{HankelH1_SmallArgument}
H^{(1)}_{\frac{1}{4}\left( 2\ell+1 \right)}[z] &\to -\frac{i}{\pi} \Gamma\left[\frac{1}{4}\left( 2\ell+1 \right)\right] \left(\frac{2}{z}\right)^{\frac{1}{4}\left( 2\ell+1 \right)} \left( 1 + \mathcal{O}\left[ z^2 \right] \right) \nonumber\\
&\qquad
+ \left(\frac{z}{2}\right)^{\frac{1}{4}(2\ell+1)} \frac{(1+i\cot\left[ \frac{\pi}{4}(2\ell+1) \right])}{\Gamma[\frac{1}{4}(2\ell+5)]} \left( 1 + \mathcal{O}\left[ z^2 \right] \right) .
\end{align}
The $(2/z)^\nu$ piece of the small argument behavior of $H_\nu^{(1)}[z] \equiv J_\nu[z] + i N_\nu[z]$ can be traced to $J_{-\nu}[z]$. This implies that if we want a nonsingular solution as $r_</r_v \to 0$, we must set $\mathcal{C}_\ell = 1$ and $C^\text{(HH)}_\ell = 0$ for $\ell \geq 1$. For $\ell=0$, however, both $\xi^{1/4} J_{1/4}[\sqrt{3}\xi/2]$ and $\xi^{1/4} H_{1/4}^{(1)}[\sqrt{3}\xi/2]$ are nonsingular in the small radius limit; the former goes to zero and the latter to a constant. We must therefore write $R_0^{(2)}[r_< \ll r_v]$ as a linear combination of these two functions.

{\it The Radiative Limit} \qquad We may now fix the form of the radiative limit, $r_> \gg r_v \gg r_<$, of $\widetilde{g}_\ell$. This is crucial for studying the Galileon radiation seen by an asymptotic observer at $r \gg r_v$ generated by a source moving deep within the Vainshtein radius of the central mass ($r' \ll r_v$). Our previous discussion leads us to the forms
\begin{eqnarray}
\label{RadialGreensFunction_RadiativeLimit_GeneralForm}
\widetilde{g}_\ell[\xi,\xi'] &=&
\left\{
\begin{array}{lr}
h_0^{(1)}[\xi_>] 
\left( C_0^\text{(J)}\cdot \sqrt[4]{\xi_<} J_{\frac{1}{4}}\left[ \sqrt{3}\xi_</2 \right] 
+ C_0^\text{(H)}\cdot \sqrt[4]{\xi_<} H^{(1)}_{\frac{1}{4}}\left[ \sqrt{3}\xi_</2 \right] 
\right) & \ell=0 \\
& \\
h_\ell^{(1)}[\xi_>] 
C_\ell^\text{(J)}\cdot \sqrt[4]{\xi_<} J_{\frac{1}{4}(2\ell+1)}\left[ \sqrt{3}\xi_</2 \right] & \ell>0 \\
\end{array}
\right. \ .
\end{eqnarray}

\subsubsection{Static Green's Function}
\label{RadialGreensFunction_Section_Static}

The static limit of the Green's function, as defined in eq. \eqref{GreensFunctionEquation_StaticDef}, requires a slightly different treatment from the time dependent case because it amounts to setting $\omega = 0$ in frequency space, making the variables $\xi$, $\xi'$ and $\xi_v$ in eq. \eqref{GreensFunctionEquation_HomogeneousRadialModeEquation} ill defined. We therefore work instead with the original radial variables $r,r'$ and $r_v$, in terms of which \eqref{GreensFunctionEquation_RadialModeEquation} becomes, in the $\omega\rightarrow 0$ limit
\begin{align}
\label{GreensFunctionEquation_HomogeneousRadialModeEquation_Static}
\left( e_2 \partial_r^2 + \frac{2}{r} e_3 \partial_r - \frac{\ell(\ell+1)}{r^2} e_3 \right) R_\ell^{(1,2|s)}[r] = 0.
\end{align}
(Here $e_{2,3} = e_{2,3}[r,r_v]$ depend on $r,r_v$, not $\xi,\xi_v$.) We also express eq. \eqref{GreensFunction_RadialModeAnsatz} as
\begin{align}
\label{GreensFunction_RadialModeAnsatz_Static}
\widetilde{g}^\text{(s)}_\ell[r,r'] \equiv \lim_{\omega\to 0} \omega \widetilde{g}_\ell[\xi,\xi']
&= \mathcal{C}_\ell R_\ell^{(1|s)}[r_>] R_\ell^{(2|s)}[r_<] - (1-\mathcal{C}_\ell) R_\ell^{(1|s)}[r_<] R_\ell^{(2|s)}[r_>] \nonumber\\
&\qquad \qquad + C_\ell^{11} R_\ell^{(1|s)}[r] R_\ell^{(1|s)}[r'] + C_\ell^{22} R_\ell^{(2|s)}[r] R_\ell^{(2|s)}[r'], 
\end{align}
and eq. \eqref{WronskianCondition_Dynamic} as
\begin{align}
\label{WronskianCondition_Static}
e_2[r,r_v] \left( R_\ell^{(1|s)}[r] (R_\ell^{(2|s)})'[r] - (R_\ell^{(1|s)})'[r] R_\ell^{(2|s)}[r] \right) = \frac{1}{r^2}.
\end{align}
As we shall see, for $\ell \geq 1$, the radial static Green's function will be fixed once we demand that the solutions are regular for all radii $r,r'$. For $\ell=0$, regularity is irrelevant; instead, $\widetilde{g}^\text{(s)}_0$ will be determined by ensuring that eq. \eqref{ell=0_BoundaryCondition} is obtained and, for $r_> \to \infty$, that the Green's function goes to zero.

Multiplying both sides of eq. \eqref{GreensFunctionEquation_HomogeneousRadialModeEquation_Static} by $\sqrt{r^3(r^3+r_v^3)}$ (and dropping the labels) yields
\begin{align}
(r^3+r_v^3) R''[r] + \frac{4r^3+r_v^3}{4} \left( \frac{2}{r} R'[r] -\frac{\ell(\ell+1)}{r^2} R[r] \right) = 0
\end{align} 
and this equation may be readily solved in {\sf Mathematica} \cite{MMA}. The general homogeneous solutions to the static radial mode equation eq. \eqref{GreensFunctionEquation_HomogeneousRadialModeEquation_Static}, normalized to satisfy the condition in eq. \eqref{WronskianCondition_Static} are
\begin{align*}
R_\ell^{(1|s)}[r] &\equiv \sqrt{\frac{2}{(2\ell+1)r_v}} \left( \frac{r_v}{r} \right)^{\frac{\ell}{2}} \ _2F_1\left[ \frac{\ell}{6}+\frac{1}{3}, -\frac{\ell}{2}; \frac{5}{6}-\frac{\ell}{3}; -\frac{r^3}{r_v^3} \right], \\
R_\ell^{(2|s)}[r] &\equiv \sqrt{\frac{2}{(2\ell+1)r_v}} \left( \frac{r}{r_v} \right)^{\frac{\ell+1}{2}} \ _2F_1\left[ \frac{1}{6}-\frac{\ell}{6}, \frac{1}{2}+\frac{\ell}{2}; \frac{7}{6}+\frac{\ell}{3}; -\frac{r^3}{r_v^3} \right] .
\end{align*}
Since $\,_2F_1[\alpha,\beta;\gamma;z=0] = 1$, we see that the $(1-\mathcal{C}_\ell) R_\ell^{(1|s)}[r_<] R_\ell^{(2|s)}[r_>]$ term tends to $\sqrt{r_>/r_v} (r_>/r_<)^{\ell/2}$ and the $C_\ell^{11}$ term in eq. \eqref{GreensFunction_RadialModeAnsatz_Static} tends to $(r_v^2/(rr'))^{\ell/2}$, as $r/r_v,r'/r_v \to 0$. These two terms grow without bound, and therefore we must choose $\mathcal{C}_\ell=1$ and $C_\ell^{11} = 0$. 

Next, we use the identity
\begin{align}
\label{2F1_zTo1/z}
\,_2F_1[\alpha,\beta;\gamma;z] 
&= \frac{\Gamma[\gamma]\Gamma[\beta-\alpha]}{\Gamma[\beta]\Gamma[\gamma-\alpha]} (-z)^{-\alpha} \ _2F_1\left[\alpha,\alpha+1-\gamma;\alpha+1-\beta;\frac{1}{z}\right] \\
&\qquad \qquad
+ \frac{\Gamma[\gamma]\Gamma[\alpha-\beta]}{\Gamma[\alpha]\Gamma[\gamma-\beta]} (-z)^{-\beta} \ _2F_1\left[\beta,\beta+1-\gamma;\beta+1-\alpha;\frac{1}{z}\right] \nonumber
\end{align}
to recast the product $R_\ell^{(1|s)}[r_>] R_\ell^{(2|s)}[r_<]$ in eq. \eqref{GreensFunction_RadialModeAnsatz_Static} in terms of $_2F_1 [\alpha,\beta;\gamma;-(r_v/r_\lessgtr)^3]$. The two potentially divergent terms for $\ell\geq 1$ are the ones proportional to
\begin{align}
\label{StaticDivergences_I}
&\left(\frac{rr'}{r_v^2}\right)^\ell \left( 2 C_\ell^{22} \Gamma\left[\frac{\ell}{3}+\frac{1}{6}\right] \Gamma\left[ \frac{2 (\ell+2)}{3}\right] + \frac{\sqrt{\pi}\ell!}{\sin\left[\frac{1}{6}\pi (2 \ell+1) \right]} \right) \left( 1 + \mathcal{O}\left[ r_v^2/r_\lessgtr^3 \right] \right) \\
\label{StaticDivergences_II}
&\left(\frac{r_>}{r_<}\right)^\ell \frac{r_v}{r_<} \left( 2 C_\ell^{22} \Gamma\left[\frac{\ell}{3}+\frac{1}{6}\right] \Gamma\left[ \frac{2 (\ell+2)}{3}\right] + \frac{\sqrt{\pi}\ell!}{\sin\left[\frac{1}{6}\pi (2 \ell+1) \right]} \right) \left( 1 + \mathcal{O}\left[ r_v^2/r_\lessgtr^3 \right] \right) .
\end{align}
We may therefore collect the results
\begin{align}
\mathcal{C}_\ell = 1, \qquad C_\ell^{11} = 0, \qquad C_\ell^{22} = \frac{\ell! \Gamma[-\frac{1}{6}(2\ell+1)]}{2\sqrt{\pi} \Gamma[\frac{1}{3}(2\ell+1)]} ,
\end{align}
where we have used the $\Gamma$-function identity $\Gamma[z] \Gamma[1-z] = \pi \mathrm{cosec}[\pi z]$.

Notice that none of the regularity constraints apply for $\ell=0$. In fact, $R_0^{(1|s)}[r]$ is a constant. $R_0^{(2|s)}[r]$  varies as $\sqrt{2/r_v} \sqrt{r/r_v}$ for small $r/r_v$, and using eq. \eqref{2F1_zTo1/z}, we obtain the equivalent expression
\begin{align}
R_0^{(2|s)}[r] = \sqrt{\frac{2}{r_v}} \left(
\frac{\Gamma[\frac{1}{3}] \Gamma[\frac{7}{6}]}{\sqrt{\pi}} - \frac{r_v}{2r} \,_2F_1\left[ \frac{1}{3}, \frac{1}{2}; \frac{4}{3}; -\frac{r_v^3}{r^3}\right] 
\right),
\end{align}
which varies as $\sqrt{2/r_v}(\text{const.} - r_v/(2r))$ for large $r/r_v$. This implies that $R_0^{(2|s)}[r]$ is regular for both large and small $r/r_v$. To determine $\mathcal{C}_0$, $C_0^{11}$ and $C_0^{22}$ here, we recall the discussion towards the end of section \eqref{Setup}, that $\widetilde{g}_0^\text{(s)}[r,r'=0]/(4\pi)$ must correspond to the coefficient of the $\delta M/M$ piece of $\delta \Pb[r]$ in eq. \eqref{deltaPiBackground}; by spherical symmetry, the $\ell \geq 1$ do not contribute to the solution generated by a point mass at the origin. This implies $\mathcal{C}_0=0$ and $C_0^{11} = \Gamma[1/3]\Gamma[1/6]/(6 \sqrt{\pi})$. When $r_> \gg r_< \gg r_v$, by using the identity in eq. \eqref{2F1_zTo1/z} on $\widetilde{g}_0^\text{(s)}[r,r']/(4\pi)$, and setting the resulting $\,_2F_1$s to unity, we find that the only constant term (independent of both $r$ and $r'$) reads $C_0^{22} \Gamma^3[1/6]/(9 \cdot 2^{2/3} \sqrt{3\pi} r_v)$. Because we have already chosen the asymptotic boundary condition (see equations \eqref{BackgroundPiSolution_2F1} and \eqref{BackgroundPiSolution_2F1_LargeRadius}) that $\varphi[r \to \infty]$ generated by a point mass located at some finite $r'$ should approach zero, this implies $C_0^{22} = 0$. We see, at this point, that $\widetilde{g}^{(s)}_0[r,r']$ only depends on $r_>$ and not on $r_<$.

We may also arrive at the same result for $\widetilde{g}_0^\text{(s)}[r,r']/(4\pi)$, without invoking the background solution $\Pb$, if we refer to the curved spacetime picture described in section \eqref{CurvedSpacetime}. In particular, the static limit of eq. \eqref{GreensFunctionEquation_CurvedSpacePicture_NormalizeAtZero}, gotten by setting $\omega \to 0$, translates to
\begin{align}
-\lim_{r\to 0} \sqrt{r r_v^3} \partial_r \widetilde{g}_0^\text{(s)}[r,r' = 0] = 1 - \mathcal{C}_0 = 1 .
\end{align}
This immediately implies $\mathcal{C}_0 = 0$. Taking the $r_> \gg r_< \gg r_v$ limit tells us that, as $r_> \to \infty$, we are left with 
\begin{align}
\frac{2}{r_v} \left( C_0^{11} - \frac{\Gamma[\frac{1}{3}] \Gamma[\frac{7}{6}]}{\sqrt{\pi}} \right) 
+ C_0^{22}\left( \frac{2 \Gamma^2[\frac{1}{3}] \Gamma^2[\frac{7}{6}]}{\pi r_v} - \frac{\Gamma[\frac{1}{3}] \Gamma[\frac{7}{6}]}{\sqrt{\pi} r_<} \right).
\end{align}
Since we require $\varphi[r \to \infty] = 0$ for any finite radial location of the point mass, we must have $C_0^{11} = \Gamma[\frac{1}{3}] \Gamma[\frac{7}{6}]/\sqrt{\pi}$ and $C_0^{22} = 0$.

This completes the derivation of eq. \eqref{GreensFunctionResult_Static_Exact}.

With the exact solution to the static Green's function in hand, we may now take the limits $r,r' \ll r_v$, $r,r' \gg r_v$, and $r_> \gg r_v \gg r_<$. In the small radii limit, $r,r' \ll r_v$, we set the $\,_2F_1$s in eq. \eqref{GreensFunctionResult_Static_Exact} to unity, and drop the subleading $(rr'/r_v^2)^{(\ell+1)/2}$ term relative to the dominant $(r_</r_>)^{\ell/2} \sqrt{r_</r_v}$ term, to obtain
\begin{align}
\label{GreensFunctionResult_Static_Exact_v2}
G^\text{(static)}[\vec{x},\vec{x}']
&= -\frac{1}{2\pi r_v} \left( \sqrt{\frac{r}{r_v}} + \sqrt{\frac{r'}{r_v}} \right) 
+ \frac{2\sqrt{rr'}}{r_v^{3/2}} \sum_{\ell = 0}^\infty \sum_{m = -\ell}^\ell \frac{ Y_\ell^m[\theta,\phi] \overline{Y_\ell^m}[\theta',\phi'] }{ 2\ell+1 }
\frac{1}{\sqrt{r_>}} \left( \frac{r_<}{r_>} \right)^{\frac{\ell}{2}} ,
\end{align}
The infinite mode sum in eq. \eqref{GreensFunctionResult_Static_Exact_v2} may be collapsed into a closed expression by first summing over the azimuthal modes and then invoking the generating function of the Legendre polynomials. The result of the mode sum in eq. \eqref{GreensFunctionResult_Static_Exact_v2} is eq. \eqref{GreensFunctionResult_Static_rrp<<rv}.

In the same vein, the large radii $r,r' \gg r_v$ static Green's function may be summed into a closed form by similar means. In fact, our solution is consistent with our earlier observation that Galileon dynamics reduce to that of a minimally coupled massless scalar field in Minkowski; for since we know its static limit is eq. \eqref{GreensFunctionResult_Static_rrp>>rv}, our Galileon static Green's function ought to reduce to the same in this asymptotic limit to lowest order in $r_v/r_\lessgtr$. In detail, if one begins from eq. \eqref{GreensFunctionResult_Static_Exact}, applies the identity in eq. \eqref{2F1_zTo1/z} to the $_2F_1$s and then proceeds to set the transformed $_2F_1$s to unity -- because their arguments will go as $-(r_v/r)^3$, which is very small at large radii -- one finds a subleading term proportional to $r_v^{2\ell+2}/(rr')^{\ell+1}$ and a dominant term proportional to $(1/r_>)(r_</r_>)^\ell$. Keeping only the dominant term and again converting the sum over spherical harmonics into one over Legendre polynomials, followed by applying the latter's generating function, we reach eq. \eqref{GreensFunctionResult_Static_rrp>>rv}.

As for the case $r_> \gg r_v \gg r_<$, the result in eq. \eqref{GreensFunctionResult_Static_rp<<rv<<r} can be obtained by starting with the exact solution in eq. \eqref{GreensFunctionResult_Static_Exact}, but only applying the identity in eq. \eqref{2F1_zTo1/z} to $R^{(2|s)}_\ell[r_>]$, followed by setting the $_2F_1$s to unity.

\subsubsection{The WKB Green's Function}
\label{RadialGreensFunction_Section_WKB}

Next we consider the high frequency limit, $|\xi|,|\xi_v| \gg \max[1,\ell]$. We first rescale the mode functions via
\begin{align}
\label{WKB_RescaledSchrodingerForm}
R_\ell^{(1,2)}[\xi] \equiv \frac{\mathcal{R}_\ell^{(1,2)}[\xi]}{\left( \xi (\xi^3+\xi_v^3) \right)^{1/4}},
\end{align}
so that the Wronskian condition in eq. \eqref{WronskianCondition_Dynamic} becomes
$\mathcal{R}_\ell^{(1)}[\xi] (\mathcal{R}_\ell^{(2)})'[\xi] - (\mathcal{R}_\ell^{(1)})'[\xi] \mathcal{R}_\ell^{(2)}[\xi] = 1$,
and \eqref{GreensFunctionEquation_HomogeneousRadialModeEquation} reads
\begin{align}
0 = -\left(\mathcal{R}_\ell^{(1,2)}\right)''[\xi] 
+ \left( K[\xi,\xi_v] + U[\xi/\xi_v] \right) \mathcal{R}_\ell^{(1,2)}[\xi] \ .
\end{align}
Here $U$ has already been defined in eq. \eqref{WKBMomenta}, and we define
\begin{align}
\label{WKB_SubleadingK}
K[\xi,\xi_v] \equiv \frac{16 \xi^6 \ell (\ell+1)+4 \xi^3 \xi_v^3 (5 \ell(\ell+1)+6)+\xi_v^6 (4 \ell(\ell+1)-3)}{16 \xi^2 (\xi^3+\xi_v^3)^2}. 
\end{align}
For large $|\xi|,|\xi_v| \gg \max[1,\ell]$, we see that the denominator of $K$ scales as 8 powers of the large quantity $1/\delta \sim |\xi|,|\xi_v|$; while its numerator contains six powers of $1/\delta$ times terms of order unity, order $\ell$ and order $\ell^2$. This means that the largest possible scaling of $K$ is that it goes as $(\ell \delta)^2 \ll 1$. For $\ell=0$, $K$ scales as $\delta^2$. Since $U$ is of order unity, in the high frequency limit we may therefore discard $K$ relative to $U$.

Suppressing the irrelevant indices, we now seek to solve
\begin{align}
\label{WKB_Equation}
0 = -\epsilon^2 \mathcal{R}''[\xi] + U[\xi/\xi_v] \mathcal{R}[\xi] \ .
\end{align}
(Here and below, we are introducing a fictitious parameter $\epsilon$ that will be set to unity once the solutions to $\mathcal{R}$ are obtained.) Observe that, viewed as a function of $\xi$, the $U$ is very flat by assumption, because $\partial_\xi U[\xi/\xi_v] = U'[\xi/\xi_v]/\xi_v \ll 1$. This calls for the WKB method of solution, in which one uses the derivatives of $U$ with respect to $\xi$ as an expansion parameter. We therefore pose the ansatz
\begin{align}
\label{WKB_Ansatz}
\mathcal{R}[\xi] = \frac{e^{(i/\epsilon)\mathcal{S}[\xi]}}{\sqrt[4]{-U[\xi/\xi_v]}} \sum_{\ell=0}^\infty \epsilon^\ell \tau_{(\ell)}[\xi].
\end{align}
It is important to note from eq. \eqref{WKBMomenta} that $\sqrt{-U[\xi/\xi_v]}$ has no real zeros, though it has a singularity at $\xi/\xi_v = -1$ and a global minimum at $\sqrt{-U[\xi/\xi_v = 1/2]} = \sqrt{2/3}$. Both in eq. \eqref{WKB_Equation} and the ansatz of eq. \eqref{WKB_Ansatz},  $\epsilon$ will turn out to count derivatives, so that $1/\epsilon$ implies an integral. Inserting eq. \eqref{WKB_Ansatz} into eq. \eqref{WKB_Equation} and setting the coefficient of each distinct power $\epsilon^\ell$ to zero, the $\ell = 0$ term yields a relationship between $U$ and $\mathcal{S}'$, which we may integrate to obtain two solutions
\begin{align}
\label{WKB_PhaseSolution}
\mathcal{S}[\xi] = \pm \int^\xi \dd \xi'' \sqrt{-U[\xi''/\xi_v]} \ .
\end{align}
The $\ell=1$ term gives a differential relationship between $\tau_{(0)}/\sqrt[4]{-U[\xi/\xi_v]}$, its first derivative with respect to $\xi$, and $\mathcal{S}'$ and $\mathcal{S}''$. Through eq. \eqref{WKB_PhaseSolution}, this gives 
\begin{align}
\tau_{(0)} = \text{constant}.
\end{align}
By setting to zero the coefficients of $\epsilon^\ell$, for $\ell \geq 2$, we find a recursion relation obeyed by $\tau_{(\ell)}$,
\begin{align}
\label{WKB_RecursionRelation}
\tau_{(\ell)} = \mp \frac{1}{2} \int^\xi \frac{\dd \xi''}{\sqrt[4]{-U[\xi''/\xi_v]}} 
\frac{\dd^2}{\dd \xi''^2} \left( \frac{\tau_{(\ell-1)}[\xi'']}{\sqrt[4]{-U[\xi''/\xi_v]}} \right),
\end{align}
where $-$ (or $+$) is chosen if we chose the $+$ (or $-$) sign in eq. \eqref{WKB_PhaseSolution}. As advertised earlier, we see that every higher order in $\epsilon$ contains an additional derivative with respect to $\xi$; and the $1/\epsilon$ in the phase of eq. \eqref{WKB_Ansatz} is the integral in eq. \eqref{WKB_PhaseSolution}.

For our purposes, we shall work only to lowest order in the WKB approximation, just involving $\mathcal{S}$ and $\tau_{(0)}$; the solutions $R_\ell^{(1,2)}$ normalized to obey the Wronskian condition in eq. \eqref{WronskianCondition_Dynamic} are
\begin{align}
\label{HomogeneousSolutions_WKB}
R_\ell^{\binom{1}{2}}[\xi] 
= \frac{\exp\left[\pm i \xi_v \int_0^{r/r_v} \dd \vartheta \sqrt{-U[\vartheta]}\right]}{\sqrt[4]{-U[r/r_v]} \left( \xi (\xi^3+\xi_v^3) \right)^{1/4}}.
\end{align}
Let us pause to understand the large ($r/r_v \gg 1$) and small ($r/r_v \ll 1$) radius limits. Examining Fig.\eqref{WKBPotential} reminds us that $\sqrt{-U[r/r_v]}$ is basically flat for large $r/r_v \gg 1$. Together with the limit $\sqrt{-U[+\infty]} = 1$, we may infer, for some $\xi_0$ and $\xi$ such that $\xi/\xi_v > \xi_0/\xi_v \gg 1$,
\begin{align}
\label{FlatUIntegral_StepI}
\xi_v\int_0^{\xi/\xi_v} \sqrt{-U[\vartheta]} \dd \vartheta 
\approx \xi_v\int_0^{\xi_0/\xi_v} \sqrt{-U[\vartheta]} \dd \vartheta + \xi-\xi_0 
= \xi - \xi_v\int_0^{\xi_0/\xi_v} \left(1-\sqrt{-U[\vartheta]}\right) \dd \vartheta .
\end{align}
We will justify below that we may now further approximate this integral by extending the upper limit of integration $\xi_0/\xi_v$ to infinity,
\begin{align}
\label{FlatUIntegral}
\xi_v\int_0^{\xi/\xi_v} \sqrt{-U[\vartheta]} \dd \vartheta 
\approx \xi - \xi_v \mathcal{I}_\infty, \quad \xi/\xi_v \gg \ell
\end{align}
where $\mathcal{I}_\infty$ was defined in eq. \eqref{UInfinity}. Similarly, for $r/r_v \ll 1$, by the flatness of the potential $\sqrt{-U[\vartheta]}$ near $\vartheta=0$, we have
\begin{align}
\xi_v\int_0^{\xi/\xi_v} \sqrt{-U[\vartheta]} \dd \vartheta \approx \frac{\sqrt{3}}{2}\xi, \qquad \xi/\xi_v \ll 1 .
\end{align}
We thus have
\begin{eqnarray}
\label{HomogeneousSolutions_WKB_SmallAndLargeRadii}
R_\ell^{\binom{1}{2}}[\xi]  &\approx &
\left\{
\begin{array}{lr}
\frac{\exp\left[\pm i \frac{\sqrt{3}}{2}\xi \right]}{\sqrt{\sqrt{3}/2} \left( \xi \xi_v^3 \right)^{1/4}}, & r/r_v \ll 1 \\
& \\
\frac{\exp\left[\pm i \left( \xi - \xi_v \mathcal{I}_\infty \right) \right]}{\xi} & r/r_v \gg 1 \\
\end{array}
\right. \ .
\end{eqnarray}
The $r,r' \ll r_v$, $r,r' \ll r_v$ and $r_> \gg r_v \gg r_<$ limits reported in equations \eqref{GreensFunctionResult_WKB_LargeRadii}, \eqref{GreensFunctionResult_WKB_SmallRadii}, and \eqref{GreensFunctionResult_WKB_rp<<rv<<r} follow from equations \eqref{GreensFunctionResult_WKB} and \eqref{HomogeneousSolutions_WKB_SmallAndLargeRadii} once $C_\ell^{++}$ is computed.

It is important to observe that, in this high frequency limit we are working in, $R_\ell^{(1)}[\xi]$ (the $+$ sign solution) in eq. \eqref{HomogeneousSolutions_WKB} is proportional to $\xi^{1/4} H^{(1)}_{(2\ell+1)/4}[\sqrt{3}\xi/2]$ in the $r \ll r_v$ regime and to $h_\ell^{(1)}[\xi]$ in the $r \gg r_v$ regime. To validate this assertion, we merely need to compare the expressions in eq. \eqref{HomogeneousSolutions_WKB_SmallAndLargeRadii} against the high frequency limit of the Hankel functions. (Likewise, $R_\ell^{(2)}[\xi]$ is proportional to $h_\ell^{(2)}[\xi]$ and $\xi^{1/4} H^{(2)}_{(2\ell+1)/4}[\sqrt{3}\xi/2]$ in the limits $r \gg r_v$ and $r \ll r_v$ respectively.) Retarded boundary conditions mean, therefore, that in eq. \eqref{GreensFunction_RadialModeAnsatz} we need to set  $\mathcal{C}_\ell = 1$ and $C_\ell^{22} = 0$ for all $\ell\geq 0$. At this point, our WKB radial Green's function solution takes the form in eq. \eqref{GreensFunctionResult_WKB}. 

For $\ell \geq 1$,  $C_\ell^{++}$ may be fixed by regularity, demanding that the limit $r_</r_v \to 0$ yields a radial Green's function that is proportional to the high frequency behavior of $\xi_<^{1/4} J_{(2\ell+1)/4}[\sqrt{3}\xi_</2]$. From the form in eq \eqref{GreensFunctionResult_WKB} and the large argument limit of the Bessel function this translates to the consistency condition
\begin{align}
i e^{-i\sqrt{3}\xi_</2} + C_\ell^{++} e^{i\sqrt{3}\xi_</2} = \chi \sqrt{\frac{4}{\pi \sqrt{3}}} \cos\left[ \frac{\sqrt{3}\xi_<}{2} - \frac{\pi}{2} \frac{2\ell+3}{4} \right] ,
\end{align}
where $\chi$ is a constant. By converting the cosine into exponentials and equating the coefficients of $\exp[\pm i \sqrt{3}\xi_</2]$ on both sides, this lets us solve for both $C_\ell^{++}$ and $\chi$.

For $\ell=0$, the radial Green's function can now be proportional to a linear combination of the high frequency limits of $\xi_<^{1/4} J_{1/4}[ \sqrt{3} \xi_</2 ]$ and $\xi_<^{1/4} H^{(1)}_{1/4}[ \sqrt{3} \xi_</2 ]$, because as already discussed, both are non-singular in the zero radius limit. Let us consider setting $r_</r_v = 0$, eliminating the $\xi^{1/4} J_{1/4}$ term. If we now also take $r_> \ll r_v$, and if we remind ourselves of eq. \eqref{HomogeneousSolutions_WKB_SmallAndLargeRadii} and the large argument limit of $H^{(1)}_\nu$, we see that the WKB $C_0^{++}$ term must match onto the high frequency limit of the $\sqrt[4]{\xi\xi'} H^{(1)}_{1/4}[\sqrt{3}\xi/2] H^{(1)}_{1/4}[\sqrt{3}\xi'/2]$ term in eq. \eqref{RadialGreensFunctionGeneralSolution_DeepInsideVainshtein}, i.e. $C_0^{++} \propto C_0^\text{(HH)}$ when $|\xi_v| \gg \max[1,\ell]$. As we will discuss in section \eqref{CurvedSpacetime} below, the $\ell=0$ radial Green's function, which obeys an inhomogeneous Helmholtz equation in curved spacetime, must obey the Gauss' law in eq. \eqref{GreensFunctionEquation_CurvedSpacePicture_NormalizeAtZero}:
\begin{align}
-\lim_{\xi\to 0} \sqrt{\xi \xi_v^3} \partial_\xi \widetilde{g}_0[\xi,\xi' = 0] = 1,
\end{align}
where we have taken the small radius limit of the effective metric in eq. \eqref{EffectiveGeometry}. Because the $\xi^{1/4} J_{1/4}[\sqrt{3}\xi/2]$ vanishes when evaluated at $\xi=0$, this condition applied to eq. \eqref{RadialGreensFunctionGeneralSolution_DeepInsideVainshtein} allows us to solve for 
\begin{align}
\frac{i\pi}{2 \xi_v^{3/2}} C_0^\text{(HH)} = - \frac{(1+i)\pi}{4\xi_v^{3/2}},
\end{align}
which in turn yields the $C_0^{++}$ in eq. \eqref{GreensFunctionResult_WKB_C++}.

In the preceding discussion, we invoked the flatness of $\sqrt{-U[\vartheta]}$ at both large and small $\vartheta$. Let us pause to quantify this flatness by examining the following expressions
\begin{align}
\mathcal{F}_0[\vartheta] \equiv \xi_v\int_0^\vartheta \dd \vartheta' \left( \frac{\sqrt{3}}{2} - \sqrt{-U[\vartheta']} \right)
\end{align}
and
\begin{align}
\mathcal{F}_\infty[\vartheta] \equiv \xi_v\int_\vartheta^\infty \dd \vartheta' \left( 1 - \sqrt{-U[\vartheta']} \right).
\end{align} 
The reason for the $\xi_v$ in front of the integrals is that, because these expressions occur in phases (e.g. $\exp[i\xi_v \Phi_\lessgtr]$), it is not sufficient for the integrals themselves to be much less than unity. We need the entire expression to be small, so that very little oscillation occurs.

Computing the power series of $\mathcal{F}_0[\delta]$ and $\mathcal{F}_\infty[1/\delta]$ about $\delta = 0$ allows us to understand precisely how good an approximation the expressions in eq. \eqref{HomogeneousSolutions_WKB_SmallAndLargeRadii} are. For $r/r_v \ll 1$ we obtain
\begin{align}
\mathcal{F}_0[r/r_v] = \xi_v \frac{(\xi/\xi_v)^{5/2}}{5\sqrt{3}} + \mathcal{O}\left[ \xi_v(\xi/\xi_v)^4 \right]
\end{align}
and, for $r/r_v \gg 1$,
\begin{align}
\mathcal{F}_\infty[r/r_v] = \xi_v\frac{(\xi_v/\xi)^2}{8} + \mathcal{O}\left[ \xi_v(\xi_v/\xi)^5 \right].
\end{align}
Therefore, replacing $r_0/r_v$ with $+\infty$ in eq. \eqref{FlatUIntegral_StepI} results in an error in the phase that is of order $\xi_v (\xi_v/(\omega r_0))^2 \ll 1$, as long as we assume $|\xi_>| \gg \omega r_0 \gg |\xi_v|^{3/2}$ (i.e. the observer is far enough outside Vainshtein). Likewise, the approximation in $\xi_v \Phi_<$ for $r_</r_v \ll 1$, replacing $\sqrt{-U}$ with $\sqrt{3}/2$, makes an error in the phase that is of order $\xi_v (\xi_</\xi_v)^{5/2} \ll 1$ as long as $\ell \ll |\xi_v|^{3/5}$.

\subsubsection{Low and High Frequency Limits}

Because we were able to solve for the relevant mode functions (up to their overall $\xi_v$-dependent normalization constants) in the regions very close to and very far away from the the central mass, we were able to determine the form of the radial Green's function in the limits $r,r' \gg r_v$ (eq. \eqref{GreensFunctionResult_rrp>>rv}), $r,r' \ll r_v$ (eq.  \eqref{RadialGreensFunctionGeneralSolution_DeepInsideVainshtein}), and $r_> \gg r_v \gg r_<$ (eq. \eqref{RadialGreensFunction_RadiativeLimit_GeneralForm}). We may calculate the $\xi_v$-dependent constants appearing in these expressions at least in the low frequency $|\xi_v| \ll 1$ and high frequency $|\xi_v| \gg \max[1,\ell]$ limits by matching them onto the static and WKB results we have obtained in the previous two sections.

In the low frequency limit, we may simultaneously take the limit $|\xi_v| \ll 1$ and replace the Bessel and Hankel functions, and their spherical versions, with their small argument limits. The resulting expressions in equations \eqref{RadialGreensFunction_RadiativeLimit_GeneralForm}, \eqref{RadialGreensFunctionGeneralSolution_DeepInsideVainshtein}, and \eqref{GreensFunctionResult_rrp>>rv} can then be compared against the respective expressions in equations \eqref{GreensFunctionResult_Static_rp<<rv<<r}, (the small $r/r_v$, $r'/r_v$ limit of) \eqref{GreensFunctionResult_Static_Exact}, and \eqref{GreensFunctionResult_Static_rrp>>rv}.

Similarly, in the high frequency limit, we may simultaneously take the limit $|\xi_v| \gg 1$ and the resulting expressions in \eqref{RadialGreensFunction_RadiativeLimit_GeneralForm}, \eqref{RadialGreensFunctionGeneralSolution_DeepInsideVainshtein}, and \eqref{GreensFunctionResult_rrp>>rv} can then be compared with the respective expressions in equations \eqref{GreensFunctionResult_WKB_rp<<rv<<r}, \eqref{GreensFunctionResult_WKB_SmallRadii} and \eqref{GreensFunctionResult_WKB_LargeRadii}.

Notice that, upon these comparisons, for the radial Green's function evaluated deep inside Vainshtein, eq. \eqref{RadialGreensFunctionGeneralSolution_DeepInsideVainshtein}, $\mathcal{C}_0 = 1$ in the high frequency limit, while it goes to zero in the low frequency limit.

Finally, to arrive at equations \eqref{GreensFunctionResult_RadiativeLimit_IofIII}, \eqref{GreensFunctionResult_rrp<<rv_Static_ell=0}, and \eqref{GreensFunctionResult_rrp<<rv_WKB}, it is useful to write $H_{1/4}^{(1)} = J_{1/4} + i N_{1/4}$, and to use
\begin{align}
N_\nu[z] = \frac{J_\nu[z] \cos[\pi\nu] - J_{-\nu}[z]}{\sin[\pi\nu]}.
\end{align}

One may wonder why the result in eq. \eqref{GreensFunctionResult_rrp>>rv} is not simply the usual answer in flat Minkowski spacetime, i.e. $C_\ell^{(hh)}=0$, since if both observer and emitter are very far from the central mass $M$ the propagation of signals would not be expected to feel the presence of the central body $M$. The physical reason is that signals with wavelengths much longer than that of Vainshtein radius $|\xi_v| \ll 1$ indeed cannot resolve $r_v$ very well -- $C^{(hh)}_\ell$ is proportional to some integer power of the small quantity $\xi_v$ -- but once the wavelengths become much shorter than $r_v$ Galileon signals can resolve the Vainshtein scale and $C^{(hh)}_\ell$ is not small but becomes a mere phase.\footnote{A simpler toy example is to consider the theory of a minimally coupled massless scalar, with a spherical perfect absorber with radius $R_0$ centered at the origin of the spatial coordinate system. The radial retarded Green's function takes the form $ih_\ell^{(1)}[\xi_>] (j_\ell[\xi_<] - \chi h_\ell^{(1)}[\xi_<])$. Perfect absorber here means the scalar field observed on the surface of the sphere is identically zero, thereby imposing $\chi = j_\ell[\omega R_0]/h_\ell^{(1)}[\omega R_0]$. At low frequencies, $|\omega R_0| \ll 1$, $\chi \to i (\omega R_0)^{2\ell+1}/((2\ell-1)!!(2\ell+1)!!)$. At high frequencies, $|\omega R_0| \gg 1$, $\chi \to (1/2) (1 - \exp[i(\pi\ell-2\omega R_0)])$.}

We have now completed the derivation of the results in section \eqref{GreensFunctionResults}.

\subsection{The Curved Spacetime Picture}
\label{CurvedSpacetime}

Before we move on to investigate the radiation seen by an asymptotic observer generated by the motion of matter close to the central source $M$, we would like to discuss an alternative perspective for the Galileon propagating on the background $\Pb[r]$. We shall also discuss the existence of a Gauss' law for the Helmholtz equation obeyed by the $\ell=0$ mode of the radial Green's function.

By a direct calculation, it is possible to view the Green's function equation \eqref{GreensFunctionEquation} as that for a massless scalar wave equation in a curved spacetime. Specifically -- recalling equations \eqref{e1} through \eqref{e3} -- dividing both sides of \eqref{GreensFunctionEquation} by $\sqrt{e_1 e_2} e_3$, we arrive at
\begin{align}
\label{GreensFunctionEquation_CurvedSpacePicture}
\Box_x G[x,x'] = \Box_{x'} G[x,x'] = \frac{\delta^{(4)}[x-x']}{|\mathfrak{g}\mathfrak{g}'|^{1/4}} 
= \sqrt{\mathfrak{H}[r/r_v] \mathfrak{H}[r'/r_v]} \delta[t-t'] \frac{\delta[r-r']}{rr'} \delta^{(2)}[\widehat{x}-\widehat{x}'] 
\end{align}
with
\begin{align}
\label{mathfrakH}
\mathfrak{H}[\vartheta] &\equiv \frac{8 \vartheta^3}{1+4 \vartheta ^3} \sqrt{\frac{1+\vartheta^3}{3(1+2\vartheta^3) - 2 \sqrt{\vartheta^3(1+\vartheta^3)}}},
\end{align}
and
\begin{align}
\delta^{(2)}[\widehat{x}-\widehat{x}'] &\equiv \delta[\cos\theta-\cos\theta']\delta[\phi-\phi'], \\
\mathfrak{g} &\equiv \det \mathfrak{g}_{\alpha\beta}[x], \ \mathfrak{g}' \equiv \det \mathfrak{g}_{\alpha\beta}[x'].
\end{align}
We have denoted $\Box_x \equiv \mathfrak{g}^{\alpha\beta}[x] \nabla_{x^\alpha} \nabla_{x^\beta}$ and $\Box_{x'} \equiv \mathfrak{g}^{\alpha\beta}[x'] \nabla_{x'^\alpha} \nabla_{x'^\beta}$ to be the minimally coupled massless scalar wave operator in a curved spacetime geometry given by the metric
\begin{align}
\label{EffectiveGeometry}
\mathfrak{g}_{\alpha\beta} \dd x^\alpha \dd x^\beta 
&\equiv
e_1^{-\frac{1}{2}} e_2^{\frac{1}{2}} e_3 \dd t^2 
- e_1^{\frac{1}{2}} e_2^{-\frac{1}{2}} e_3 \dd r^2 
- e_1^{\frac{1}{2}} e_2^{\frac{1}{2}} r^2 \Omega_\text{AB} \dd x^\text{A} \dd x^\text{B}, \qquad x^\text{A} \equiv (\theta,\phi)
\end{align}
Observe that the $\mathfrak{H}[r'/r_v]$ (eq. \eqref{mathfrakH}) occurring in eq. \eqref{GreensFunctionEquation_CurvedSpacePicture} is $(8/\sqrt{3}) (r/r_v)^3$ in the small radii limit and unity at large radii. This is the Vainshtein effect at work: the magnitude of the point mass sourcing the Green's function grows weaker the closer it gets to the central mass $M$, but goes to unity far away from it. On the other hand, it is somewhat puzzling, in this curved spacetime picture, that a source located nearer and nearer to the spatial origin tends to zero strength; for instance, we have already remarked, towards the end of section \eqref{Setup}, that a static point source located at the origin must contribute to the background $\Pb[r]$ solution via a shifting of the mass, $M \to M + \delta M$.

To understand this we first re-express eq. \eqref{GreensFunctionEquation_CurvedSpacePicture} in accordance with our decomposition in eq. \eqref{GreensFunctionResult_ModeExpansion}; this means that we drop the integration symbol $\int \dd\omega/(2\pi)$, and replace $\delta[t-t']$ and $\delta^{(2)}[\widehat{x}-\widehat{x}']$ with, respectively, $e^{-i\omega(t-t')}$ and the spherical harmonic completeness relation. Because placing the point source at the origin means we have a spherically symmetric problem, this means only the monopole $\ell=0$ term in eq. \eqref{GreensFunctionEquation_CurvedSpacePicture} is relevant. Keeping the discussion general for now, let us merely assume the metric reads
\begin{align}
\mathfrak{g}_{\mu\nu} \dd x^\mu \dd x^\nu = g_{00} \dd t^2 + g_{rr} \dd r^2 + g_\text{AB} \dd x^\text{A} \dd x^\text{B},
\end{align}
and is time independent. Our frequency space curved spacetime equation is then
\begin{align}
\omega \left( -\omega^2 \sqrt{|g|} g^{00} \widetilde{g}_0[\xi,0] + \partial_r \left( \sqrt{|g|} g^{rr} \partial_r \widetilde{g}_0[\xi,0] \right) \right) = \delta[r],
\end{align}
which allows us to integrate over an infinitesimally small neighborhood about $r = 0$ to obtain the normalization condition:
\begin{align}
\label{GreensFunctionEquation_CurvedSpacePicture_NormalizeAtZero}
\lim_{r\to 0} \omega \sqrt{|g|} g^{rr} \partial_r \widetilde{g}_0[\xi,0] = 1 .
\end{align}
This indicates that, even though the measure $\mathfrak{H}[r/r_v]/r^2$ (see equations \eqref{GreensFunctionEquation_CurvedSpacePicture} and \eqref{mathfrakH}) tends to zero as $r\to 0$, a point mass sitting at the spatial origin must nonethless produce a non-zero (spherically symmetric) Galileon field, for otherwise eq. \eqref{GreensFunctionEquation_CurvedSpacePicture_NormalizeAtZero} cannot be satisfied.

That we are able to obtain an equivalent curved spacetime wave equation to eq. \eqref{GreensFunctionEquation}, is in fact one way to justify the ``measure'' $1/(rr')$ multiplying the $\delta$-functions on its right hand side. From general arguments due to Hadamard \cite{HadamardBook} -- see \cite{PoissonReview} for a review on Green's functions in curved spacetime -- we know that solutions exist for the massless scalar Green's function equation \eqref{GreensFunctionEquation_CurvedSpacePicture}. Since equations \eqref{GreensFunctionEquation} and \eqref{GreensFunctionEquation_CurvedSpacePicture} are equivalent, this means the $(rr')^{-1}$ is the correct measure. Moreover in this curved spacetime picture of the Galileon Green's function, we know that, when the observer at $x$ and the source at $x'$ can be connected by a unique geodesic, the retarded Green's function consists of the sum of two terms, namely
\begin{align}
\label{GreensFunction_HadamardForm}
G[x,x'] = \frac{\Theta[t-t']}{4\pi} \left( \delta[\sigma_{x,x'}] \sqrt{\Delta_{x,x'}} + \Theta[\sigma_{x,x'}] V_{x,x'} \right).
\end{align}
Here, $\sigma_{x,x'}$ is half the square of the geodesic distance from $x'$ to $x$. The first term after the equality describes propagation of Galileon signals on the null cone of the geometry in eq. \eqref{EffectiveGeometry}; $\Delta_{x,x'}$ is related to the evolution of the cross sectional area of null rays emanating from $x'$ to $x$. $V[x,x']$, known as the tail term, describes Galileons traveling inside the future null cone of $x'$. It is the solution to the homogeneous wave equation $\Box_x V = \Box_{x'} V = 0$, obeying non-trivial boundary conditions on the null cone of $x'$. We see that solving the retarded Galileon Green's function provides us information not only about the causal structure of Galileon signals but also about the effective spacetime in eq. \eqref{EffectiveGeometry}.

Via equations \eqref{BackgroundPiSolution_SmallRadius_IofII} through \eqref{BackgroundPiSolution_LargeRadius_IIofII}: we recover flat spacetime for the region well outside the Vainshtein radius ($r \gg r_v$),
\begin{align}
\label{geff_LargeRadius}
\mathfrak{g}_{\alpha\beta} \dd x^\alpha \dd x^\beta \approx \eta_{\mu\nu} \dd x^\mu \dd x^\nu
\end{align}
and for the region well within Vainshtein ($r \ll r_v$) we have instead
\begin{align}
\label{geff_SmallRadius}
\mathfrak{g}_{\alpha\beta} \dd x^\alpha \dd x^\beta 
\approx \left(\frac{r_v}{r}\right)^{\frac{3}{2}} \frac{1}{2\sqrt{3}} \left( \dd t^2 - \frac{3}{4} \dd r^2 - 3 r^2 \Omega_\text{AB} \dd x^\text{A} \dd x^\text{B} \right).
\end{align}
The $3/4$ in front of $\dd r^2$ indicates that, deep within the Vainshtein radius of the central object, Galileon waves propagating in the radial direction are superluminal, since $\dd r/\dd t = 2/\sqrt{3} > 1$. (We have already noted this, within the context of the WKB results, right after eq. \eqref{GreensFunctionResult_WKB_SmallRadii}.) In the same vein, it is worth mentioning that, if some method can be found to evaluate the infinite mode sum of the Green's function result for $r,r' \ll r_v$ described by equations \eqref{GreensFunctionResult_ModeExpansion} and \eqref{GreensFunctionResult_rrp<<rv_Static_ell=0} through \eqref{GreensFunctionResult_rrp<<rv_WKB}, we should obtain the Hadamard form in eq. \eqref{GreensFunction_HadamardForm} and this would give us a deeper insight into the superluminal properties of Galileon signals near the matter source.

Note that defining 
\begin{align}
\rho \equiv \sqrt{\sqrt{3}r/2}, \qquad \rho_v \equiv \sqrt{\sqrt{3}r_v/2}
\end{align}
transforms eq. \eqref{geff_SmallRadius} into
\begin{align}
\label{geff_SmallRadius_SpatiallyConformallyFlat}
\mathfrak{g}_{\alpha\beta} \dd x^\alpha \dd x^\beta 
				&= \frac{1}{2\sqrt{3}} \left(\frac{\rho_v}{\rho}\right)^3  \left( \dd t^2 - (2 \rho)^2 \delta_{ij} \dd \rho^i \dd \rho^j \right),
\end{align}
where the Cartesian components of the spatial coordinates are
\begin{align}
\rho^i &\equiv \rho (\sin\theta \cos\phi, \sin\theta \sin\phi, \cos\theta).
\end{align}
This small radius curved spacetime metric provides an alternate means of deriving the inhomogeneous portion of the static Green's function in eq. \eqref{GreensFunctionResult_Static_rrp<<rv}. To see this, first re-scale the static Green's function as 
\begin{align}
\label{GreensFunction_CurvedSpacePicture_SmallRadius}
G^\text{(static)}[\vec{x},\vec{x}'] \equiv \frac{\sqrt{3}}{\rho_v^3} \rho \rho' g^\text{(static)}[\rho,\theta,\phi;\rho',\theta',\phi'] \ .
\end{align}
Next, compute the static analog of \eqref{GreensFunctionEquation_CurvedSpacePicture}, namely
\begin{align}
\Box_x G^\text{(static)}[\vec{x},\vec{x}'] = |\mathfrak{g}[x] \mathfrak{g}[x']|^{-1/4} \delta^{(3)}[\vec{x}-\vec{x}'],
\end{align}
using the small radii metric in \eqref{geff_SmallRadius_SpatiallyConformallyFlat}. Thus, $g^\text{(static)}$ is the Green's function to the Laplacian in Euclidean 3-space
\begin{align}
\Delta_{\vec{\rho}} \equiv \delta^{ij} \frac{\partial}{\partial \rho^i} \frac{\partial}{\partial \rho^j},
\end{align}
namely
\begin{align}
-\Delta_{\vec{\rho}} g^\text{(static)}[\vec{\rho}, \vec{\rho}'] &= -\Delta_{\vec{\rho}'} g^\text{(static)}[\vec{\rho}, \vec{\rho}'] 
= \delta^{(3)}[\vec{\rho}-\vec{\rho}'].
\end{align}
Remembering the rescaling performed in \eqref{GreensFunction_CurvedSpacePicture_SmallRadius} and requiring that $G^\text{(static)}[\vec{x},\vec{x}'] = G^\text{(static)}[\vec{x}',\vec{x}]$ then fixes the general solution to take the form
\begin{align}
G^\text{(static)}[\vec{\rho},\vec{\rho}'] = \frac{\sqrt{3}}{4\pi \rho_v^3} \left( \frac{\rho \rho'}{|\vec{\rho}-\vec{\rho}'|} - \chi_0 - \chi_1 \left( \rho + \rho' \right) - \chi_2 \rho \rho' \right),
\end{align}
where $\chi_{0,1,2}$ are spacetime constants; these $\chi_{0,1,2}$ terms are homogeneous solutions, i.e. $\Box_x$ and $\Box_{x'}$ applied on them give zero. The $\chi_{0,1}$ may be fixed by placing the source at the spatial origin, $\rho' = 0$ or $\rho = 0$, and making sure that eq. \eqref{StaticGreensFunction_rp=0_SmallRadius} is recovered. (The $\chi_1$ may also be determined using the condition derived in eq. \eqref{GreensFunctionEquation_CurvedSpacePicture_NormalizeAtZero}. First replace $r$ with $\rho$; remember $G^\text{(static)}[\vec{x},\vec{0}] = \lim_{\omega\to 0} \omega \widetilde{g}_0[\xi,0]/(4\pi)$; and a short calculation yields $\sqrt{|g|} g^{\rho\rho} = -\rho_v^3/\sqrt{3}$. Altogether, $\chi_1 = 1$.) Finally, $\chi_2$ can be fixed by demanding that, for $r_> \to \infty$, the Green's function goes to zero. (We do not compute $\chi_{0,1,2}$ in detail, since we have already obtained the exact result in the previous sections.)

In equations \eqref{Xi_CurvedSpacetime} and \eqref{GradientsOfXi_CurvedSpacetime} we derive the minimally coupled scalar field generated by $n$ point masses in a generic curved spacetime in terms of the Hadamard form in eq. \eqref{GreensFunction_HadamardForm}. The reason for doing so is that the Galileon field $\varphi$ generated by the $n$ body dynamics is, in fact, related to eq. \eqref{GradientsOfXi_CurvedSpacetime} -- the curved geometry in question is eq. \eqref{EffectiveGeometry}.\footnote{It is not exactly the same expression because, in curved spacetime, the element of proper time is $\dd s = \sqrt{g_{\mu\nu}\dot{x}^\mu \dot{x}^\nu} \dd t$, whereas in this paper we are working in flat spacetime and the element of proper time is $\dd s = \sqrt{\eta_{\mu\nu}\dot{x}^\mu \dot{x}^\nu} \dd t$.} One cannot help but wonder if our mode expansion results in section \eqref{GreensFunctionResults} can be utilized, at least in some limits, to extract the various portions ($\sqrt{\Delta_{x,x'}}, \sigma_{x,x'}$, etc.) of the Hadamard form in eq. \eqref{GreensFunction_HadamardForm}. We leave these questions for possible future work.

\section{Radiative Processes}
\label{RadiationSection}

In the following two sections we wish to examine the Galileon waves produced by matter in motion, within two concrete scenarios. The first is the radiation made by acoustic waves propagating on the surface of the large central body of mass $M$ with radius $R_0$. The second is the radiation created by the movement of compact bodies orbiting this central body; for instance, this could describe our solar system, with planets orbiting around the Sun or a highly asymmetric mass ratio binary star system capable of also generating gravitational waves that could be observed by upcoming gravitational wave detectors.

We initiate the discussion by defining radiation to be the piece of the Galileon field that carries a non-trivial energy-momentum flux to infinity. Quantitatively, we only wish to consider the portion of $\varphi$ that contributes a non-zero power (per unit solid angle) at infinity,
\begin{align}
\label{AsymptoticPower}
\frac{\dd E}{\dd t \dd \Omega} \equiv \lim_{r \to \infty} r^2 \mathfrak{T}^{0r} = -\lim_{r \to \infty} r^2 \partial_r \varphi \partial_t \varphi.
\end{align}
(We have used eq. \eqref{GalileonAsymptoticStressTensor}.) As we will see more explicitly below, both $\partial_t \varphi$ and $\partial_r \varphi$ contain a power series in $1/r$, beginning at $1/r$. Even though the full Galileon field is $\Pi \equiv \Pb + \varphi$, according to eq. \eqref{BackgroundPiSolution_LargeRadius_IofII}, the derivative of the static background Galileon field goes as $\Pb'[r \to \infty] \sim 1/r^2$, it does not contribute to the asymptotic power we are currently after and hence would be neglected in the following discussion.

To highlight the importance of the non-linear self interactions of the Galileon field, we will also compute the radiation signature in the same setups arising from the minimally coupled massless scalar, so that we can compare the prediction of the two theories. That is, we will also consider the theory described by the following action
\begin{align}
\label{MinimallyCoupledMasslessScalar}
S_\Xi \equiv \int \dd^4 x \left( \frac{1}{2} (\partial\Xi)^2 + \Xi \frac{\delta T}{\mpl} \right) + \frac{M}{\mpl} \int \Pi \dd t'.
\end{align}
Compare this with the Galileon action $S_\Pi + S_M + \delta S$ in equations \eqref{Action_Pi}, \eqref{Action_CentralMass}, and \eqref{Action_Perturbations}. There, we had to first solve $\Pb$ sourced by $M$ and then proceed to perturb around it. Here, because the $\Xi$-theory is linear, the complete field $\Xi$ is gotten by superposing the field spawned by each individual source.

When we compute the radiation generated by the surface waves on $M$, we will assume that these waves are driven by external (non-gravitational) forces, so that $M$ itself can be considered to be a static source of the scalar field(s) (falling off as $1/r^2$), and do not produce any radiation due to backreaction. The only source of radiation there is $\delta T$ describing the waves themselves. When we compute the radiation of the $n$-body system, however, we need to remember that, even though $M \gg m_a$, and hence the location of $M$ never deviates far from the center of energy of the system -- the entire system is held together by (largely) conservative gravitational forces. In this case, we shall see it is important to include $M$ in the radiation calculation so as to enforce the conservation of linear momentum.

The radiative limits of the Green's function for $\varphi$ (equations \eqref{GreensFunctionResult_RadiativeLimit_IofIII} through \eqref{GreensFunctionResult_RadiativeLimit_IIIofIII}) and for $\Xi$ (equations \eqref{MinimallyCoupledMasslessScalarG_Fourier} and \eqref{AdditionTheorem}) both take the form
\begin{align}
G[x,x'] &= \int \frac{\dd \omega}{2\pi} e^{-i\omega(t-t')} \sum_{\ell,m} \widehat{\Omega}_\ell^m[\theta,\phi] (\widehat{\Omega}_\ell^m)^*[\theta',\phi'] \mathcal{R}^<_\ell[\xi_<] h_\ell^{(1)}[\xi_>] \ ,
\end{align}
where we have chosen to write the orthonormal angular mode functions as a linear combination of the usual spherical harmonics
\begin{align}
\widehat{\Omega}_\ell^m = \ _{(\ell)}L^m_{\phantom{m}n'} Y_\ell^{n'}.
\end{align}
(The $^*$ in $(\widehat{\Omega}_\ell^m)^*$ represents complex conjugation.) Here $_{(\ell)}L^m_{\phantom{m}n'}$ is a unitary $(2\ell+1)\times(2\ell+1)$ matrix, so $\{\widehat{\Omega}_\ell^m\}$ is as good an orthonormal basis as the spherical harmonics. For the minimally coupled massless scalar $\Xi$, equations \eqref{MinimallyCoupledMasslessScalarG_Fourier} and \eqref{AdditionTheorem} tell us
\begin{align}
\label{R<Xi}
\mathcal{R}^{(<|\Xi)}_\ell[\xi_<] = i\omega j_\ell[\xi_<] \ ,
\end{align}
while eq. \eqref{GreensFunctionResult_RadiativeLimit_IofIII} gives, for the Galileon $\varphi$,
\begin{align}
\label{R<Galileon_HighFreq}
\mathcal{R}^{(<|\varphi)}_\ell[\xi_<] 
= \omega C_\ell^\text{(rad)} \cdot \sqrt[4]{\xi_<} J_{\frac{\sigma_\ell}{4}(2\ell+1)}\left[ \sqrt{3}\xi_</2 \right],
\end{align}
where $\sigma_0 = -1$ and $\sigma_\ell = 1$ when $\ell \geq 1$. If we decompose the matter source in the same way that we decomposed the Green's function,
\begin{align}
\label{MatterDecomposition}
\delta T[x] = \int \frac{\dd \omega}{2\pi} e^{-i\omega t} \sum_{\ell,m} \widehat{\Omega}_\ell^m[\theta,\phi] \rho_\ell^m[\omega,r],
\end{align}
then by the orthonormality of both the exponentials and the angular mode functions, we may translate eq. \eqref{PiGeneralSolution} into the following solution for the $\varphi$ or $\Xi$ field evaluated at ($t,r,\theta,\phi$), sourced by $\delta T/\mpl$
\begin{align}
\label{Galileon_Xi_Solution}
\varphi[x],\Xi[x] 
&= \frac{1}{\mpl} \int \frac{\dd \omega}{2\pi} e^{-i\omega t} \sum_{\ell,m} \widehat{\Omega}_\ell^m[\theta,\phi] h_\ell^{(1)}[\xi] 
\int_0^\infty \dd r' r'^2 \mathcal{R}^<_\ell[\xi']  \rho_\ell^m[\omega,r'].
\end{align}
The asymptotic behavior of the spherical Hankel function then implies that the scalar field, for a fixed angular frequency $\omega$, is indeed proportional to a finite power series in $1/r$, and the series begins at $1/r$. This leading order $1/r$ piece of the time and radial derivatives is precisely what we call radiation, because when inserted into eq. \eqref{AsymptoticPower}, the factors of $r$ cancel and what remains is a finite amount of energy transported to infinity. Any part of the fields containing a higher power than $1/r$ would make a contribution to $r^2 \mathfrak{T}^{0r}$ that decays to zero at infinity. 

Moreover, since taking a time derivative brings down a $-i\omega$ and taking a $r$-derivative -- to lowest order in $1/r$ -- brings down a $+i\omega$, we deduce that the radiative part of the fields obey the relationship,
\begin{align}
(\partial_t\varphi)_\text{radiation} = -(\partial_r\varphi)_\text{radiation}
\end{align}
with $\Xi$ respecting the same equation. Hence it suffices to display only the time derivatives,
\begin{align}
\label{TimeDerivativesOfRadiation}
\partial_t \varphi, \partial_t \Xi &=
-\frac{1}{\mpl} \int \frac{\dd \omega}{2\pi} e^{-i\omega t} \sum_{\ell,m} \widehat{\Omega}_\ell^m[\theta,\phi]
(-i)^\ell\frac{e^{i\xi}}{r} \left(1+\mathcal{O}\left[\frac{1}{r}\right] \right) 
\int_0^\infty \dd r' r'^2 \mathcal{R}^<_\ell[\xi']  \rho_\ell^m[\omega,r'].
\end{align}
Because it contains an $r$-dependent exponential $e^{i\xi}$, one may worry that the Fourier integral in eq. \eqref{TimeDerivativesOfRadiation} would somehow introduce additional terms that go as $1/r$ (after the $\omega$-integral has been performed), and therefore that dropping the higher powers of $1/r$ in eq. \eqref{TimeDerivativesOfRadiation} at this stage is premature. However, we will see from the integral in eq. \eqref{TotalEnergyRadiated_MasterIntegral} below that the $e^{i\xi}$ drops out of the expression for total energy, and therefore $r$ does not take part in the resulting $\omega$-integral; in particular, the higher powers of $1/r$ that have been discarded in eq. \eqref{TimeDerivativesOfRadiation} would indeed decay away once we take the $r \to \infty$ limit.

In general, we expect the Fourier integral in eq. \eqref{TimeDerivativesOfRadiation} to be extremely difficult to evaluate. We will instead consider what the spectrum of radiation emitted from a particular system is. The spectrum is indeed a physical observable, since observations usually take place over enough cycles of the radiation field for a Fourier analysis to be done.

The total energy per solid angle radiated to infinity is the integral
\begin{align}
\frac{\dd E}{\dd \Omega} = -\lim_{r\to\infty} \int_{-\infty}^\infty r^2 \partial_r \varphi \partial_t \varphi \dd t.
\end{align}
(The same expression holds for $\Xi$.) A few standard Fourier identities allow us to express the total scalar energy radiated per unit solid angle, from eq. \eqref{TimeDerivativesOfRadiation}, as the following integral over all angular frequencies
\begin{align}
\label{TotalEnergyRadiated_MasterIntegral}
\frac{\dd E[\varphi \text{ or } \Xi]}{\dd \Omega} 
&= \frac{1}{\mpl^2} \int \frac{\dd \omega}{2\pi} \Bigg\vert \sum_{\ell,m} \widehat{\Omega}_\ell^m[\theta,\phi] (-i)^\ell 
\int_0^\infty \dd r' r'^2 \mathcal{R}^<_\ell[\xi']  \rho_\ell^m[\omega,r'] \Bigg\vert^2. 
\end{align}
In the subsequent two sections, we perform the decomposition in eq. \eqref{MatterDecomposition} for surface waves on the spherical mass $M$ as well as that of $n$ compact bodies orbiting it, and proceed to apply them to eq. \eqref{TotalEnergyRadiated_MasterIntegral}.

\subsection{Surface Waves On Spherical Body}

In this section we will describe surface (acoustic) waves propagating on the large central mass $M$ of radius $R_0$ by
\begin{align}
\label{Source_SurfaceWaves}
\delta T_\odot[x] \equiv \frac{M}{V_\odot} \delta R \ \delta[r - R_0].
\end{align}
where $V_\odot$ is the volume of $M$,
\begin{align}
V_\odot \equiv \frac{4}{3}\pi R_0^3.
\end{align}
Denote by $R[t,\widehat{x}]$ the radius of the mass $M$ at a given time $t$ and direction $(\theta,\phi)$ from the spatial center of the coordinate system. For a non-relativistic system, which we shall assume is the case in this section, the trace of the stress-energy tensor $\delta T$ primarily describes its mass density $\delta T_{00}$. Then eq. \eqref{Source_SurfaceWaves} may be interpreted as describing surface waves of very small fluctuations $\delta R[t,\widehat{x}] \equiv R[t,\widehat{x}]-R_0$ around the mean radius $R_0$, on an otherwise perfectly spherical body. We will decompose these undulations $\delta R$ as
\begin{align}
\label{Source_SurfaceWaves_deltaR}
&\delta R[t,\widehat{x}] = R_0
\sum_{\ell=1}^\infty \sum_{m=0}^\ell \left(
\widehat{A}_\ell^m[\widehat{x}] a_\ell^m \cos\left[ \Omega_\ell^m t + \Phi_\ell^m \right] + \widehat{B}_\ell^m[\widehat{x}] b_\ell^m \cos\left[ \Omega'^m_\ell t + \Psi_\ell^m \right]
\right),
\end{align}
with $|a_\ell^m|, |b_\ell^m| \ll 1$. The ``rotated'' spherical harmonics $\{\widehat{A}_\ell^m,\widehat{B}_\ell^m\}$ are defined as, whenever $m\neq 0$,
\begin{align}
\label{RotatedSphericalHarmonics_IofII}
\widehat{A}_\ell^m &\equiv i^m (Y_\ell^m + Y_\ell^{-m})/\sqrt{2}, \nonumber\\
\widehat{B}_\ell^m &\equiv i^{m+1} (Y_\ell^m - Y_\ell^{-m})/\sqrt{2},
\end{align}
whereas for $m=0$,
\begin{align}
\label{RotatedSphericalHarmonics_IIofII}
\widehat{A}_\ell^0 &\equiv Y_\ell^0 = (\widehat{A}_\ell^0)^*, \qquad B_\ell^0 \equiv 0.
\end{align}
Because $\overline{Y_\ell^m} = (-)^m Y_\ell^{-m}$, the $\widehat{A}_\ell^m$ and $\widehat{B}_\ell^m$ are real. This is the primary reason for using them instead of the usual $Y_\ell^m$, because now $a_\ell^m$ and $b_\ell^m$ can directly be read off as dimensionless amplitudes of a particular mode of vibration, with respective oscillation frequencies $\Omega_\ell^m$, $\Omega'^m_\ell$, and phases $\Phi_\ell^m$ and $\Psi_\ell^m$. For technical convenience, we will assume all oscillation frequencies $\{\Omega_\ell^m,\Omega'^m_\ell\}$ are distinct and positive.

It is also worthwhile to observe we have not allowed a monopole $\ell=0$ term in the infinite sum eq. \eqref{Source_SurfaceWaves_deltaR}: in particular, $\int \dd^3 x \delta T[t,\vec{x}] = 0$, and the total mass is a constant. While our scalar field theories do not enjoy any sort of symmetry giving rise to a conservation law for the associated scalar charge, the coupling to the scalar fields here is of (sub-)gravitational strength, and thus very weak. Therefore the requirement from the known laws of physics that mass is a conserved quantity, in the Minkowski spacetime we are working in, therefore takes precedence.

By a direct calculation, one can check that our surface waves have the following decomposition
\begin{align}
&\delta T_\odot[x] = \int \frac{\dd \omega}{2\pi} e^{-i\omega t} \frac{M R_0}{V_\odot} \delta[R_0-r] \pi \\
&\times \sum_{\ell=1}^\infty \sum_{m=0}^\ell
\left( \widehat{A}_\ell^m[\widehat{x}] a_\ell^m \left( e^{-i\Phi_\ell^m}\delta[\omega-\Omega_\ell^m] + e^{i\Phi_\ell^m}\delta[\omega+\Omega_\ell^m] \right) 
+ \widehat{B}_\ell^m[\widehat{x}] b_\ell^m \left( e^{-i\Psi_\ell^m}\delta[\omega-\Omega'^m_\ell] + e^{i\Psi_\ell^m}\delta[\omega+\Omega'^m_\ell] \right) \right). \nonumber
\end{align}
Here, the analog of the $\rho_\ell^m$s appearing within the formula in eq. \eqref{TotalEnergyRadiated_MasterIntegral}, are the coefficients of $\int (\dd\omega/(2\pi))e^{-i\omega t} \widehat{A}_\ell^m$ and $\int (\dd\omega/(2\pi))e^{-i\omega t} \widehat{B}_\ell^m$. When taking the square $|\dots|^2$ in eq. \eqref{TotalEnergyRadiated_MasterIntegral}, we  encounter cross terms involving the $\delta$-functions. However, since we have assumed that all frequencies are distinct and positive, the arguments of the $\delta$-functions cannot be simultaneously zero unless the frequencies are in fact the same. This collapses the summations into a single $\ell$- and a single $m$-sum. It remains to deal with squares of the form $((2\pi)\delta[\omega \pm \omega'])^2$, with $\omega'$ being one of the $\Omega_\ell^m$s or $\Omega'^m_\ell$s. We will treat one of the $(2\pi)\delta[\omega \pm \omega']$ as the total duration of time, since
\begin{align}
(2\pi)\delta[\omega = 0] = \int \dd t \lim_{\omega\to 0} e^{-i\omega t} = \text{total time elapsed}.
\end{align}
Dividing both sides of eq. \eqref{TotalEnergyRadiated_MasterIntegral} by total time, i.e. $(2\pi)\delta[\omega = 0]$, then gives us back power radiated in scalar waves per unit solid angle.

{\it Minimally Coupled Massless Scalar} \qquad The result for $\Xi$, from eq. \eqref{R<Xi}, is then
\begin{align}
\label{RadiationFromSurfaceWaves_Xi_Full}
\frac{\dd E[\Xi]}{\dd t\dd \Omega} 
	&= \frac{9 G_\text{N} M^2}{\pi R_0^2} 
\sum_{\ell=1}^\infty \sum_{m=0}^\ell \Bigg\{ \left\vert \widehat{A}_\ell^m[\widehat{x}] a_\ell^m \Omega_\ell^m R_0 j_\ell[\Omega_\ell^m R_0] \right\vert^2 
		+ \left\vert \widehat{B}_\ell^m[\widehat{x}] b_\ell^m \Omega'^m_\ell R_0 j_\ell[\Omega'^m_\ell R_0] \right\vert^2 \Bigg\}.
\end{align}
(We have exploited the fact that, because $j_\ell[z]$ is $z^\ell$ times a power series in $z^2$, $|j_\ell[-z]| = |j_\ell[z]|$.)

For non-relativistic systems, a substantial subset of the oscillation periods $\sim 1/\Omega_\ell^m,1/\Omega'^m_\ell$ ought to be much longer than the light crossing time $\sim R_0$. Therefore we can treat the small dimensionless quantities $\Omega_\ell^m R_0, \Omega'^m_\ell R_0 \ll 1$ as expansion parameters. This leads us, in this non-relativistic limit, to
\begin{align}
\label{RadiationFromSurfaceWaves_Xi_NR}
&\frac{\dd E[\Xi]}{\dd t\dd \Omega} 
= \frac{9 G_\text{N} M^2}{\pi R_0^2} \sum_{\ell=1}^\infty \sum_{m=0}^\ell \frac{1}{((2\ell+1)!!)^2} 
\Bigg\{ \left\vert \widehat{A}_\ell^m[\widehat{x}] a_\ell^m \left( \Omega_\ell^m R_0 \right)^{\ell+1} \right\vert^2 + \left\vert \widehat{B}_\ell^m[\widehat{x}] b_\ell^m \left( \Omega'^m_\ell R_0 \right)^{\ell+1} \right\vert^2 \Bigg\}.
\end{align}

{\it Galileons} \qquad For the Galileon $\varphi$, because the Vainshtein radius depends on $\Lambda$ in eq. \eqref{Action_Pi} -- a free parameter in this paper -- it is not necessary that $\Omega_\ell^m r_v, \Omega'^m_\ell r_v \ll 1 \leq \ell$. This prompts us to write the energy output $\dd E_\ell^m/\dd t \dd\Omega$ due to Galileons as a function of the mode numbers $(\ell,m)$ instead. We will assume that the surface waves are non-relativistic, and thus the ratio of the radius to that time of oscillation is a small number (i.e. $\Omega_\ell^m R_0, \ \Omega'^m_\ell R_0 \ll 1$), so that we may replace the Bessel and Hankel functions with their small argument limits.

When $\Omega_\ell^m r_v, \Omega'^m_\ell r_v \ll 1$ and $\ell \geq 1$,
\begin{align}
\label{RadiationFromSurfaceWaves_Galileon_FullLowFreq_NR}
\frac{\dd E_\ell^m[\varphi]}{\dd t\dd \Omega} 
	&= \frac{288 G_\text{N} M^2}{R_0^2} \frac{R_0}{r_v} \left( \frac{\Gamma \left[-\frac{2 \ell}{3}-\frac{1}{3}\right] \Gamma \left[\frac{5}{6}-\frac{\ell}{3}\right]}{2^{3\ell+\frac{7}{2}} \Gamma\left[\frac{\ell }{2}+\frac{3}{4}\right] \Gamma\left[\frac{\ell}{2}+\frac{5}{4}\right] \Gamma[-\ell]} \left(\frac{\cos\left[\frac{1}{6} \pi (2 \ell +1)\right]}{\sin [\pi  \ell]} +1\right) \right)^2 \\
&\times \Bigg\{ \left(\Omega_\ell^m R_0\right)^{\ell+2} \left( \Omega_\ell^m r_v \right)^\ell \left\vert \widehat{A}_\ell^m[\widehat{x}] a_\ell^m \right\vert^2 + \left(\Omega'^m_\ell R_0 \right)^{\ell+2} \left( \Omega'^m_\ell r_v \right)^\ell \left\vert \widehat{B}_\ell^m[\widehat{x}] b_\ell^m  \right\vert^2 \Bigg\}. \nonumber
\end{align}
Notice that for the $\ell=1$ case, the Vainshtein radius $r_v$ drops out; low frequency oscillations are thus unscreened to the lowest order. At higher than dipole order, $\ell \geq 2$, by comparing the $(R_0/r_v) (\Omega_\ell^m R_0)^{\ell+2} (\Omega_\ell^m r_v)^\ell$ in the Galileon result in eq. \eqref{RadiationFromSurfaceWaves_Galileon_FullLowFreq_NR} to the analogous $(\Omega_\ell^m R_0)^{2\ell+2}$ in eq. \eqref{RadiationFromSurfaceWaves_Xi_NR} for the non-interacting massless scalar, we find that Galileon power is actually enhanced (up to $\ell$-dependent numerical factors) by the ratio $(r_v/R_0)^{\ell-1} \gg 1$. 

When $\Omega_\ell^m r_v, \Omega'^m_\ell r_v \gg \max[1,\ell]$, and $\ell \geq 1$,
\begin{align}
\label{RadiationFromSurfaceWaves_Galileon_FullHighFreq_NR}
\frac{\dd E_\ell^m[\varphi]}{\dd t\dd \Omega} 
&= \frac{G_\text{N} M^2}{R_0^2}
\frac{3^{\frac{\ell}{2}+\frac{9}{4}}}{4^{\ell+1} \Gamma\left[ \frac{\ell}{2} + \frac{5}{4} \right]^2}
\Bigg\{ \left\vert \widehat{A}_\ell^m[\widehat{x}] a_\ell^m \right\vert^2 \frac{\left( \Omega_\ell^m R_0 \right)^{\ell+3}}{(\Omega_\ell^m r_v)^\frac{3}{2}}
	+ \left\vert \widehat{B}_\ell^m[\widehat{x}] b_\ell^m \right\vert^2 \frac{\left( \Omega'^m_\ell R_0 \right)^{\ell+3}}{(\Omega'^m_\ell r_v)^{\frac{3}{2}}} \Bigg\} .
\end{align}
For a fixed $\ell$ mode, we may take the ratio of the $\Xi$ results in eq. \eqref{RadiationFromSurfaceWaves_Xi_NR} to the ones here for high frequency oscillations: up to numerical $\ell$-dependent factors, we obtain $(\Omega_\ell^m R_0)^{\ell-1} (\Omega_\ell^m r_v)^{3/2}$. For $\ell=1$, this ratio is $(\Omega_\ell^m r_v)^{3/2} \gg 1$, telling us the dipole term in eq. \eqref{RadiationFromSurfaceWaves_Galileon_FullHighFreq_NR} is Vainshtein screened. However, in this non-relativistic limit, $\Omega_\ell^m R_0 \ll 1$, once $\ell$ is large enough that $(\Omega_\ell^m R_0)^{\ell-1} (\Omega_\ell^m r_v)^{3/2} \ll 1$, we see that high multipole Galileon radiation becomes Vainshtein amplified relative to their non-interacting cousins. We also note that, while low frequency oscillations in eq. \eqref{RadiationFromSurfaceWaves_Galileon_FullLowFreq_NR} contain integer powers of frequency, here the radiation spectrum contain $(\Omega_\ell^m R_0)^{3/2}$ and $(\Omega'^m_\ell R_0)^{3/2}$.

The expressions in equations \eqref{RadiationFromSurfaceWaves_Galileon_FullLowFreq_NR} and \eqref{RadiationFromSurfaceWaves_Galileon_FullHighFreq_NR} came directly from the Bessel and Hankel mode functions satisfying the original wave equations of $\varphi$ and $\Xi$. Together with the presence of the combinations $\Omega_\ell^m R_0$ and $\Omega'^m_\ell R_0$, these facts teach us that radiation generated by the surface vibrations of the massive object $M$ directly probes not only the vibrations themselves but also the dynamics of our field theories -- in the Galileon case, it carries information about the theory operating deep within the Vainshtein radius of $M$, where the self interactions of the full $\Pi$ theory are dominant.

At small ratios of the Vainshtein radius to oscillation time scale (eq. \eqref{RadiationFromSurfaceWaves_Galileon_FullLowFreq_NR}), we see that the Galileon radiation spectrum, like its $\Xi$ cousin in eq. \eqref{RadiationFromSurfaceWaves_Xi_NR}, contains only integer powers of the oscillation frequencies. However, for very large Vainshtein radius to oscillation time scale ratios (eq. \eqref{RadiationFromSurfaceWaves_Galileon_FullHighFreq_NR}), the power emitted begins to contain fractional powers of angular frequencies; this indicates there must be a change in the spectral index if one is able to probe Galileon radiation over a broad bandwidth. This can be traced to the Galileon radial mode functions $\sqrt[4]{\xi} J_{(1/4)(2\ell+1)}[\sqrt{3}\xi/2]$ within the Vainshtein radius, $r\ll r_v$. We will witness this phenomenon again in the following section on the $n$ body radiative problem.

\subsection{$n$ Point Masses Orbiting Within The Vainshtein Radius Of Large Central Mass}
\label{NBodyProblem_Radiation}

In this section, we consider an arbitrary number of compact bodies of masses $\{m_a|a=1,2,\dots,n\}$ orbiting around the central object $M$.\footnote{The gravitational waves analog to this section can be found in~\cite{Wagoner:1976am}. The conservative aspect of the $n$ body (weak field) gravitational problem has a long history; see, for instance, \cite{Chu:2008xm} and the references within.} We will assume this system is held together by only gravitational and scalar forces, and we will further approximate these compact bodies as point masses, with spatial position vectors $\{\vec{y}_a | a=1,2,\dots,n\}$. If we work within the non-relativistic (i.e., slow motion) approximation, valid for a wide range of astrophysical dynamics, including that of our solar system -- the Galileon and gravitational interactions are described by 
\begin{align}
\label{nBodyStressEnergy}
\sum_{a=1}^n m_a \int \dd t \left( 
1 - \frac{1}{2} \left( \dot{\vec{y}}_a^2 - \frac{h_{00}}{\mpl} \right) + \dots 
\right) \left( \frac{\varphi \text{ or } \Xi}{\mpl} - 1 \right) .
\end{align}
The effects of gravitation are encoded in the proper time element $\dd s_a = \dd t\sqrt{g_{\mu\nu} \dot{y}_a^\mu \dot{y}_a^\nu}$, which we have expanded in powers of velocities and the graviton field $h_{\mu\nu}/\mpl$. We have assumed gravity is weak, $g_{\mu\nu} = \eta_{\mu\nu} + h_{\mu\nu}/\mpl$, where $|h_{\mu\nu}/\mpl| \ll 1$. The virial theorem tells us that the potential $h_{\mu\nu}/\mpl$ scales as the square of the typical velocities $v_a^2 \equiv (\dd \vec{y}_a/\dd t)^2$, so that the ``$+\dots$" in eq. \eqref{nBodyStressEnergy} can be understood to scale as $v_a^3$ and higher. (The scalar potential $\Xi/\mpl$ would scale similarly; but the static portion of $\varphi/\mpl$ would be considerably weaker because of Vainshtein screening.)

Because we are interested in Galileon and not in gravitational radiation, we may ``integrate out'' the gravitational field $h_{00}/\mpl$. To the lowest order in the non-relativistic expansion, this amounts to replacing the $h_{00}/\mpl$ evaluated on the $a$th point mass world line with the gravitational potential exerted on $m_a$ by $M$; the potentials due to the other compact bodies, as long as they are distant enough, scale as $m_b/M$, $b \neq a$, relative to that due to $M$, and hence are subdominant. (We ignore the possible self-force contribution.) The Newtonian energy per unit mass $E_a/m_a \equiv (1/2)(\dot{\vec{y}}_a^2 + h_{00}/\mpl)$ is a constant (up to corrections of $\mathcal{O}[v_a^4]$) -- this means we may replace $(1/2)(\dot{\vec{y}}_a^2 - h_{00}/\mpl)$ in eq. \eqref{nBodyStressEnergy} with $\dot{\vec{y}}_a^2 - E_a/m_a$.\footnote{The need for including the gravitational potentials, in order for the ensuing analysis to be consistent with energy conservation, has been emphasized in \cite{deRham:2012fw}, and our discussion here overlaps with that treatment.} At this point, what is relevant for the Galileon radiation problem is the interaction
\begin{align}
\label{deltaT_nBody}
\int \dd^4 x \frac{\varphi \text{ or } \Xi}{\mpl} \delta T_m \equiv 
\sum_{a=1}^n m_a \int \dd t F_a[t] \frac{\varphi \text{ or } \Xi}{\mpl},
\end{align}
with the non-relativistic expansion denoted by
\begin{align}
\label{F_a_Expansion}
F_a[t] \equiv 1 - \dot{\vec{y}}_a^2 + \frac{E_a}{m_a} + \mathcal{O}\left[ v_a^3, v_a^2 \frac{m}{M} \right] .
\end{align}

By a direct calculation, the trace of the associated stress-energy in eq. \eqref{deltaT_nBody} is
\begin{align}
\delta T_m[t,\vec{x}] &= \sum_a m_a \int \dd t' F_a[t'] \delta[t - x^0[t']] \delta^{(3)}[\vec{x}-\vec{y}_a[t']] .
\end{align}
We rewrite this as
\begin{align}
\delta T_m[t,\vec{x}] = \int \frac{\dd \omega}{2\pi} e^{-i\omega t} \sum_{\ell = 0}^\infty \sum_{m=-\ell}^\ell Y_\ell^m[\widehat{x}] \rho_\ell^m[\omega,r],
\end{align}
where, by a change of variables from proper time to coordinate time,
\begin{align}
\label{rho_PointMass}
\rho_\ell^m[\omega,r]
&= \sum_a m_a \int \dd t' F_a[t'] e^{i\omega t'} \frac{\delta\left[ r - |\vec{y}_a[t']| \right]}{r |\vec{y}_a[t']|} \overline{Y_\ell^m}[\widehat{y}_a[t']].
\end{align}
With this $\rho_\ell^m$, the total scalar energy emitted by $n$ bodies orbiting around the parent body $M$ is therefore given by \eqref{TotalEnergyRadiated_MasterIntegral}.
\begin{align}
\label{RadiationFromNBodyProblem_PhiOrXi}
&\frac{\dd E[\varphi \text{ or } \Xi]}{\dd \Omega} 
= 32\pi G_\text{N} \int \frac{\dd \omega}{2\pi} \Bigg\vert \sum_{a=1}^n m_a \int \dd t' F_a[t']
e^{i\omega t'} \sum_{\ell=0}^\infty \sum_{m=-\ell}^\ell Y_\ell^m[\widehat{x}] \overline{Y_\ell^m}[\widehat{y}_a[t']] (-i)^\ell \mathcal{R}^<_\ell\big[ \omega |\vec{y}_a[t']| \big] \Bigg\vert^2.
\end{align}

{\it Minimally Coupled Massless Scalar} \qquad Using equations \eqref{R<Xi}, \eqref{F_a_Expansion} and 
\begin{align}
\label{SphericalWaveExpansionOfPlaneWave}
e^{-i\vec{k}\cdot\vec{x}} = 4\pi \sum_{\ell=0}^\infty \sum_{m=-\ell}^\ell (-i)^\ell j_\ell[|\vec{k}|r] Y_\ell^m[\widehat{k}] \overline{Y_\ell^m}[\widehat{x}] 
\end{align}
eq. \eqref{RadiationFromNBodyProblem_PhiOrXi} becomes
\begin{align}
\label{RadiationFromNBodyProblem_Xi}
&\frac{\dd E[\Xi]}{\dd \Omega} 
= \frac{2 G_\text{N}}{\pi} \int \frac{\dd \omega}{2\pi} \Bigg\vert \sum_{a=1}^n m_a 
 \int \dd t' e^{i\omega t'} \partial_{t'}\left\{ F_a[t'] \exp\left[-i\omega \widehat{x}\cdot\vec{y}_a[t'] \right] \right\} \Bigg\vert^2 \ , 
\end{align}
where we have also used the fact that every integer power of $\omega$ occurring within our integrand may be replaced with a time derivative (namely, $i\partial_{t'}$) acting on $t'$ dependent factors. Let us now Taylor expand $\exp\left[-i\omega \widehat{x}\cdot\vec{y}_a[t'] \right]$, and convert each additional power of $\omega$ into an additional $i \partial_{t'}$. Roughly speaking, each time derivative should scale as
\begin{align}
\label{TimeD_Omega_v/r_Scaling}
\partial_t \sim \omega \sim v_a/r_a ,
\end{align}
where $r_a$ is the typical orbital radii of the compact bodies in motion. 

We have previously highlighted that the $n$-body system under consideration is held together by (largely) conservative forces, and it is therefore important to include the central body $M$ in our radiative calculations to respect the conservation of linear momentum. This is an appropriate place to consider how the backreaction on the central mass $M$ by the compact bodies orbiting it affects our analysis here and below. What we shall do is to extend the sum $\sum_{1 \leq a \leq n} \to \sum_{0 \leq a \leq n}$ to include that of $M$ itself, via the definitions
\begin{align}
\vec{y}_0 \equiv \text{spatial location of $M$}, \qquad m_0 \equiv M.
\end{align}
For the minimally coupled scalar case at hand, this is nothing but the principle of superposition, since the theory is linear. For the Galileon case, this amounts to including in the matter perturbations $\delta T_m$ in eq. \eqref{deltaT_nBody} the time dependent multipole moments of the central mass $M$,\footnote{Observe the hierachy: $|\vec{y}_0| \ll |\vec{y}_a| \ll r_v$, where $a=1,2,3,\dots,n$.} induced by the gravitational and scalar forces of its $n$ planetary companions pushing it away from the center of the spatial coordinate system. (That all our results in this section depend on at least two time derivatives of the spatial coordinates of $M$ and the $n$ compact bodies corroborates this interepretation.) This is subtly different from the non-self-interacting $\Xi$ case because, to capture the Vainshtein mechanism, we first had to assume that $M$ was motionless so that we could solve for the background $\Pb$ it generated, before proceeding to compute the Galileon Green's function. It is for this reason we have phrased our discussion of including $M$ in the sum over $a$ in terms of a backreaction on the motion of $M$. More quantitatively, the stress energy of the central mass is given by $M \delta^{(3)}[\vec{x}-\vec{y}_0]$, and we may Taylor expand it about $\vec{y}_0 = \vec{0}$. The lowest order term is $M \delta^{(3)}[\vec{x}]$, which is the source of the background field $\Pb$; we are thus treating every higher term in the expansion (which is necessarily proportional to powers of $\vec{y}_a[t]$) as part of the matter perturbation $\delta T_m$.

By Taylor expanding $\exp\left[-i\omega \widehat{x}\cdot\vec{y}_a[t'] \right]$, followed by converting all $\omega$'s into time derivatives, we find that the non-relativistic scalar energy loss per unit solid angle from our $n$-body system is given by the expansion:
\begin{align}
\label{RadiationFromNBodyProblem_Xi_NR}
&\frac{\dd E[\Xi]}{\dd \Omega} 
= \frac{2 G_\text{N}}{\pi} \int \frac{\dd \omega}{2\pi} \left\vert \sum_{a=0}^n m_a \int \dd t' e^{i\omega t'} 
\left( \widehat{x}\cdot\frac{\dd^2 \vec{y}_a}{\dd t'^2} + \frac{1}{2}\frac{\dd^3 \left( \widehat{x}\cdot\vec{y}_a \right)^2}{\dd t'^3} - \frac{\dd \dot{\vec{y}}^2_a}{\dd t'} + \mathcal{O}\left[ v_a^4, v_a^3 \frac{m}{M} \right] \right)
\right\vert^2,
\end{align}
The importance of including $M$ in the sum over $a$ is now manifest, for the first term involving the sum of all forces must yield zero, $\sum_a m_a (\dd^2 \vec{y}_a/\dd t'^2) \cdot \widehat{x} = 0$; in a spacetime translation symmetric background, linear momentum is conserved and Newton's third law must be obeyed. Therefore $\Xi$ radiation really begins at $\mathcal{O}\left[ v_a^3 \right]$.

As a check of the formalism here, in eq. \eqref{Xi_RadiationFormula_NR} below we shall re-derive the analog of eq. \eqref{RadiationFromNBodyProblem_Xi_NR} directly from the position space Green's function in eq. \eqref{MinimallyCoupledMasslessScalarG_Fourier}, but (for simplicity) without including the gravitational potential in the proper time element.\footnote{Be aware that the frequency space analysis here does not capture the dependence on the approximate retarded time $t'=t-r$.} We see that, at this order in the non-relativistic expansion, including the gravitational potential merely changes the $-(1/2)\dd \dot{\vec{y}}_a^2/\dd t'$ in eq. \eqref{Xi_RadiationFormula_NR} to $-\dd \dot{\vec{y}}_a^2/\dd t'$ in eq. \eqref{RadiationFromNBodyProblem_Xi_NR}.

Before we proceed to the Galileon case, let us note that we could have obtained eq. \eqref{RadiationFromNBodyProblem_Xi_NR} directly from the infinite $\ell$-sum in eq. \eqref{RadiationFromNBodyProblem_PhiOrXi}, if we Taylor expand the spherical Bessel functions and use the explicit polynomial expressions for the $P_\ell$s. (The reason for collapsing the infinite $\ell$-sum into an exponential in eq. \eqref{RadiationFromNBodyProblem_Xi} is to emphasize that, because $\exp\left[-i\omega \widehat{x}\cdot\vec{y}_a[t'] \right]$ admits a Taylor expansion in integer powers of $\omega$, the radiation spectrum in \eqref{RadiationFromNBodyProblem_Xi} depends on frequency solely through time derivatives acting on the $\vec{y}_a$s; this will not be the case for Galileons.) Because we are seeking an answer accurate up to $\mathcal{O}[v_a^3]$, by counting powers of $\omega$, we may infer that up to the $\ell=2$ terms of the sum are needed. Specifically, the $m_a \dd^2 \vec{y}_a/\dd t'^2$ term comes from the leading order piece of the $\ell=1$ term; whereas the $\dd^3 \left( \widehat{x}\cdot\vec{y}_a \right)^2/\dd t'^3$ from the leading order piece of the $\ell=2$ term; and the $\dd \dot{\vec{y}}_a^2/\dd t'$ comes from the $\ell=0$ term with $F_a$ included. Furthermore, there is cancellation between the $\ell=0$ first order correction term involving $(\omega |\vec{y}_a|)^2$ and that from the $\ell=2$ term.

{\it Galileons} \qquad In parallel with the treatment for $\Xi$, we will assume that our astrophysical $n$-body system is non-relativistic, so that the orbital time scale $1/\omega$ is very small compared to the light crossing time $\sim |\vec{y}_a|$, allowing us to replace the Bessel and Hankel functions with their small argument limits. 

For low frequencies, $|\xi_v| \ll 1$, we have
\begin{align}
\label{RadiationFromNBodyProblem_Galileon_NR_LowFreq}
\frac{\dd E[\varphi]}{\dd \Omega \dd (\omega/2\pi)} 
= \frac{2 G_\text{N}}{\pi} \left\vert \sum_{a=1}^n m_a \int \dd t' e^{i\omega t'} \sum_{\ell=0}^\infty \partial_{t'}^{\ell+1} \left\{ F_a[t'] \mathcal{M}_\ell[t'] \right\}
\right\vert^2,
\end{align}
where
\begin{eqnarray}
\mathcal{M}_\ell[t']  & = &
\left\{
\begin{array}{lr}
1 -\frac{(\omega |\vec{y}_a[t']|)^2}{4} + \dots, & \ell=0 \\
& \\
\sqrt{\frac{|\vec{y}_a[t']|}{r_v}} \left( r_v |\vec{y}_a[t']| \right)^{\frac{\ell}{2}} P_\ell\left[ \widehat{x}\cdot\widehat{y}_a[t'] \right]
\frac{\sqrt{\pi} (2\ell+1) \Gamma \left[-\frac{2 \ell}{3}-\frac{1}{3}\right] \Gamma \left[\frac{5}{6}-\frac{\ell}{3}\right]}{2^{3 \ell +\frac{3}{2}} \Gamma\left[\frac{\ell }{2}+\frac{3}{4}\right] \Gamma\left[\frac{\ell}{2}+\frac{5}{4}\right] \Gamma[-\ell]} \left(\frac{\cos\left[\frac{1}{6} \pi (2 \ell +1)\right]}{\sin [\pi  \ell]} +1\right), & \ell >0 \\
\end{array}
\right. \ ,
\end{eqnarray}
and we have converted the spherical harmonics sum to one over Legendre polynomials. The presence of $\partial_{t'}^\ell$ implies the magnitude of the $\ell$th channel is suppressed by $\sqrt{r/r_v} (r/\tau)^{\ell/2}$ relative to the lowest $\ell=0$ mode, where $r$ is the typical orbital radius and $\tau$ is the typical orbital time scale; the factor of $(r_v/\tau)^{\ell/2}$ further suppresses the power loss because the motion is highly non-relativistic. (We have already noted that the leading monopole and dipole terms in the radial Galileon Green's function in the non-relativistic limit matches that of its non-interacting cousin.) 

Developing the non-relativistic expansion up to $\mathcal{O}[v_a^3]$ requires the monopole, dipole and quadrupole terms ($\ell = 0,1,2$): in the low frequency limit, we gather
\begin{align}
\label{RadiationFromNBodyProblem_Galileon_NR_LowFreq_ell=1}
&\frac{\dd E[\varphi]}{\dd \Omega \dd (\omega/(2\pi))} 
= \frac{2 G_\text{N}}{\pi} \Bigg\vert \sum_{a=0}^n m_a \int \dd t' e^{i\omega t'} \Bigg( \widehat{x}\cdot\frac{\dd^2 \vec{y}_a}{\dd t'^2} \\
&\qquad
+\frac{1}{4} \frac{\dd^3 \vec{y}_a^2}{\dd t'^3} - \frac{\dd \dot{\vec{y}}_a^2}{\dd t'}
+ \frac{\sqrt{3}\Gamma[-\frac{2}{3}] \Gamma[\frac{1}{6}]}{40\pi^{3/2}} \frac{\dd^3}{\dd t'^3}\left( \sqrt{\frac{r_v}{|\vec{y}_a[t']|}}\left( \vec{y}_a^2 - 3 \left(\widehat{x}\cdot\vec{y}_a\right)^2 \right) \right) + \mathcal{O}\left[ v_a^4, v_a^3 \frac{m}{M} \right]
\Bigg) \Bigg\vert^2, \qquad |\xi_v| \ll 1. \nonumber
\end{align}
Comparing equations \eqref{RadiationFromNBodyProblem_Xi_NR} and \eqref{RadiationFromNBodyProblem_Galileon_NR_LowFreq_ell=1} informs us that very low frequency Galileon waves travel essentially unscreened, even though they are generated deep within the Vainshtein radius of the system. The $\sqrt{r_v/|\vec{y}_a|}$ factor in eq. \eqref{RadiationFromNBodyProblem_Galileon_NR_LowFreq_ell=1} may even yield Vainshtein enhancement relative to its cousin $\Xi$. Moreover, we shall now argue that unlike eq. \eqref{RadiationFromNBodyProblem_Xi_NR}, the first term $\sum_a m_a (\dd^2 \vec{y}_a/\dd t'^2) \cdot \widehat{x}$ here in eq. \eqref{RadiationFromNBodyProblem_Galileon_NR_LowFreq_ell=1} and in eq. \eqref{RadiationFromNBodyProblem_Galileon_NR_HighFreq_ell=1} below, is small -- it scales as $m/M$ -- but is no longer exactly zero. The primary reason is that, while gravitational forces between any two objects obey Newton's third law, Galileon forces between the compact bodies (in the non-relativistic limit) do not. This is because the background $\Pb$ does not respect spatial translation symmetry; this statement can even be checked explicitly by taking the gradients of, say, the static Green's function in eq. \eqref{GreensFunctionResult_Static_rrp<<rv} with respect to both $\vec{x}$ and $\vec{x}'$, and noting they do not give equal and opposite spatial vectors. It does turn out, however, that the force between each compact body and the central mass is equal and opposite: without loss of generality we may consider some small mass $m$ lying on the positive $z$-axis. The problem is now cylindrically symmetric, which tells us the force on $M$ due to $m$; and the force on $m$ due to $M$, must both point along the $z$-axis. The $\int \dd t (M \text{ or } m)\Pi/\mpl$ coupling tells us the $z$-component of the force on $m$ due to $M$, to leading order, is simply $(m/\mpl) \Pb' = -M m/(2 \pi \mpl^2 r_v^{3/2} \sqrt{r'})$ (refer to $\Pb'$ in eq. \eqref{BackgroundPiSolution_SmallRadius_IofII}), where $r'$ is the radial location of $m$. The force on $M$ due to $m$ is, in the non-relativistic limit, given by first invoking the cylindrical symmetry to replace $|\sqrt{r/r_v}\widehat{x}-\sqrt{r'/r_v}\widehat{x}'|$ in eq. \eqref{GreensFunctionResult_Static_rrp<<rv} with $|\sqrt{r/r_v}-\sqrt{r'/r_v}|$, and then computing $\lim_{r \to 0} (Mm/\mpl^2) \partial_r G^\text{(static)}[\vec{x},\vec{x}'] = M m/(2 \pi \mpl^2 r_v^{3/2} \sqrt{r'})$.\footnote{The reader concerned about the domain of validity of the $\varphi$ solution generated by $m$, whose gradient is responsible for the force acting on $M$, can perform the following order-of-magnitude check. Replace the $\Pi$ in the Lagrangian density of the action in eq. \eqref{Action_Pi} with the total field of $M$ and $m$, i.e. $\Pi \to \Pb + (m/\mpl) G^\text{(static)}[\vec{x},\vec{x}']$, with $G^\text{(static)}[\vec{x},\vec{x}']$ given by eq. \eqref{GreensFunctionResult_Static_rrp<<rv}. After expanding about $r=0$, divide the dominant piece of the quadratic-in-$G^\text{(static)}$ portion of the resulting Lagrangian density by the dominant piece of the cubic-in-$G^\text{(static)}$ term. For fixed $r'$, the radial location of $m$, one should find the ratio to go as $M r'/(m r) \gg 1$ -- i.e. the linear solution offered by $G^\text{(static)}$ should be an excellent description of the force of $m$ on $M$. In the same spirit, one may also expand this same quadratic-to-cubic ratio about $r=r'$ and find that, in the $r,r' \ll r_v$ limit, nonlinearities begin to render the solution offered by $G^\text{(static)}$ invalid at distances closer to $m$ than $\sqrt{m/M} r'$.} Note that even though Newton's third law is not obeyed between the compact bodies, because the theory we started with in eq. \eqref{FullAction} was defined in flat spacetime, total linear and angular momentum must still be conserved. What must happen is that the radiation generated at $\mathcal{O}[v_a^2]$ carries away some of the linear momentum.

At high frequencies but non-relativistic orbital speeds, i.e. $r_v \gg 1/|\omega| \gg |\vec{y}_a|$,
\begin{align}
\label{RadiationFromNBodyProblem_Galileon_NR_HighFreq}
\frac{\dd E[\varphi]}{\dd \Omega \dd (\omega/2\pi)} 
= \frac{32 G_\text{N}}{|\omega r_v|^{3/2}} \left\vert \sum_{a=1}^n m_a \int \dd t' e^{i\omega t'} 
\sum_{\ell=0}^\infty (-i)^\ell \partial_{t'} \left\{ F_a[t'] \mathcal{M}_\ell[t'] \right\}
\right\vert^2,
\end{align}
where
\begin{eqnarray}
\mathcal{M}_\ell[t'] & = &
\left\{
\begin{array}{lr}
\frac{(-3)^{7/8}}{16\Gamma[\frac{7}{4}]}\left( 1 - \frac{(\omega |\vec{y}_a[t']|)^2}{4} + \dots \right), & \ell=0 \\
& \\
\frac{3^{\frac{\ell}{4}+\frac{1}{8}}}{2^{\ell+1} \Gamma[\frac{\ell}{2}+\frac{1}{4}]} e^{i\pi\frac{2\ell+5}{8}} \left( \omega |\vec{y}_a[t']| \right)^{\frac{\ell+1}{2}} P_\ell\left[\widehat{x}\cdot\widehat{y}_a[t']\right] , & \ell >0 \\
\end{array}
\right. \ .
\end{eqnarray}
Here, the prefactor $1/|\xi_v|^{3/2} \ll 1$ exhibits Vainshtein screening of high frequency Galileon radiation. Counting powers of $\omega$ tells us that an $\mathcal{O}[v_a^3]$-accurate answer receives contributions from the $\ell=0,1,2$ terms:
\begin{align}
\label{RadiationFromNBodyProblem_Galileon_NR_HighFreq_ell=1}
&\frac{\dd E[\varphi]}{\dd \Omega \dd (\omega/(2\pi))} 
= \frac{2 \cdot 3^{\frac{3}{4}} G_\text{N}}{|\omega r_v|^{3/2} \Gamma[\frac{3}{4}]^2} \Bigg\vert \sum_{a=0}^n m_a \int \dd t' e^{i\omega t'} \Bigg( 
\widehat{x}\cdot\frac{\dd^2 \vec{y}_a}{\dd t'^2} \\
&\qquad
- \frac{\sqrt[4]{-3} \Gamma[\frac{3}{4}]}{\Gamma[\frac{1}{4}]} \frac{\dd^3}{\dd t'^3} \left( \frac{\vec{y}_a^2 - 3 \left( \widehat{x}\cdot\vec{y}_a\right)^2}{\sqrt{\omega|\vec{y}_a|}} \right)
+ \frac{1}{2\sqrt{3}} \left( \frac{1}{2} \frac{\dd^3 \vec{y}_a^2}{\dd t'^3} - 2 \frac{\dd \dot{\vec{y}}_a^2}{\dd t'} \right)
+ \mathcal{O}\left[ v_a^{7/2}, v_a^3 \frac{m}{M} \right]
\Bigg) \Bigg\vert^2, \qquad |\xi_v| \gg 1. \nonumber
\end{align}
(The $m_a \dd^2 \vec{y}_a/\dd t'^2$ came from $\ell=1$; the second group involving $1/\sqrt{\omega|\vec{y}_a|}$ from $\ell=2$; and the final group from $\ell=0$ with $F_a$ included.) Comparing equations \eqref{RadiationFromNBodyProblem_Galileon_NR_LowFreq_ell=1} and \eqref{RadiationFromNBodyProblem_Galileon_NR_HighFreq_ell=1}, we find, as we did in the previous section, that the slope of the Galileon power spectrum depends on the size of $|\xi_v|$. At low frequencies the radiation is very similar to that of the non-interacting case, while at high frequencies, the spectrum acquires an overall suppression factor of $1/|\xi_v|^{3/2}$. The phenomenology of Galileon radiation appears to be richer than its minimally coupled massless counterpart due to the existence of the additional length scale $r_v$ in the problem. Also note the presence of fractional powers of $\omega$ that cannot be associated with time derivatives -- this aspect of the Galileon radiation has no analog in its non-interacting cousin nor in gravitational waves propagating on flat spacetime.

{\it Gravitational Waves} \qquad Let us also record the power spectrum of GW emission. Arguments based on conservation of energy and the validity of Newton's third law leads us to infer that the lowest order answer arises from the third time derivative of the quadrupole moment of the system. Misner, Thorne and Wheeler \cite{MTW} equations 36.1 and 36.2 tell us
\begin{align}
\label{GWQuadrupoleFormula}
\frac{\dd E}{\dd (\omega/(2\pi))} = \frac{G_\text{N}}{5} \left\vert \widetilde{\dddot{Q}_{ij}}[\omega] \right\vert^2,
\end{align}
where $\widetilde{\dddot{Q}_{ij}}$ is the Fourier transform of the triple time derivative of the quadrupole moment,
\begin{align}
\widetilde{\dddot{Q}_{ij}}[\omega] &\equiv \int \dd t' e^{i\omega t'} 
\sum_{a=1}^n m_a \frac{\dd^3}{\dd t'^3} \left( y_{ai}[t'] y_{aj}[t'] - \frac{1}{3} \delta_{ij} \vec{y}_a^2[t'] \right) .
\end{align}
As we have seen, the forces acting between compact bodies orbiting around a central mass $M$ no longer obey Newton's third law if Galileons exist ($\sum_a m_a \dd^2 \vec{y}_a/\dd t'^2 \neq 0$) -- this in turn means, eq. \eqref{GWQuadrupoleFormula} may no longer be the leading order answer to the GW spectrum. We hope to return to this issue in the future.

\section{Summary and discussion}

Within the next decade or so, gravitational wave detectors are expected to begin hearing signals from astrophysical systems such as inspiralling compact binaries. It is therefore an appropriate time to explore the possibility that there could be emission of radiation due to additional degrees of freedom coming from the various modifications of General Relativity that has been proposed in the literature to date. Our work initiates such an investigation for Galileons, a class of scalar field theories that exhibit what is known as the Vainshtein mechanism, within the context where there is a large central mass $M$.

We have constructed the Galileon retarded Green's function satisfying the linearized equations of motion about the background field of this central mass. The main results are described in section \eqref{GreensFunctionResults}. For the radiation problem, where the observer is situated at a very large distance from $M$ and the source generating Galileon waves is located well within the Vainshtein radius of $M$, the primary results can be found in equations  \eqref{GreensFunctionResult_RadiativeLimit_IofIII} through \eqref{GreensFunctionResult_RadiativeLimit_IIIofIII}. We have also obtained the exact static Green's function, the WKB limit of the retarded Green's function, and the Green's function evaluated both deep inside and far outside the Vainshtein radius of $M$.

We have used this radiative Green's function to obtain the frequency spectrum of Galileon radiation emitted from (acoustic) waves on the surface of the spherical mass $M$, described by equations \eqref{Source_SurfaceWaves}, \eqref{Source_SurfaceWaves_deltaR}, \eqref{RotatedSphericalHarmonics_IofII}, and \eqref{RotatedSphericalHarmonics_IIofII}. The power dissipated per unit solid angle per mode $(\ell,m)$, in the non-relativistic limit, can be found in eq. \eqref{RadiationFromSurfaceWaves_Galileon_FullLowFreq_NR} for the case where the $r_v$-to-oscillation-time-scale ratio was much smaller than unity; and in \eqref{RadiationFromSurfaceWaves_Galileon_FullHighFreq_NR} for that ratio very large. To illustrate the importance of the nonlinearities of the Galileon interaction for the problem at hand, we have also calculated for comparison the power emitted if we replace the Galileon with a non-interacting massless scalar, with the relevant results given in equations \eqref{RadiationFromSurfaceWaves_Xi_Full} and \eqref{RadiationFromSurfaceWaves_Xi_NR}.

A particularly interesting application of our results is to the radiation spectrum due to the motion of $n$ point masses gravitationally bound to the central mass $M$. This is the dissipative aspect of the Galileon modified gravitational dynamics whose conservative portion we investigate in a separate paper \cite{ACHT}. We have focused on the non-relativistic limit, and have found the energy in Galileon radiation lost to infinity in two regimes. For small $r_v$-to-orbital-time-scale ratios, this is given by eq. \eqref{RadiationFromNBodyProblem_Galileon_NR_LowFreq}, and the lowest order answer is \eqref{RadiationFromNBodyProblem_Galileon_NR_LowFreq_ell=1}. For large $r_v$-to-orbital-time-scale ratios, the answer is eq. \eqref{RadiationFromNBodyProblem_Galileon_NR_HighFreq}, with the lowest order result in eq. \eqref{RadiationFromNBodyProblem_Galileon_NR_HighFreq_ell=1}. For comparison, the non-interacting massless scalar result is eq. \eqref{RadiationFromNBodyProblem_Xi_NR} and that of the quadrupole radiation formula for gravitational waves is eq. \eqref{GWQuadrupoleFormula}.

In these radiative processes, we find that the non-interacting massless scalar and Galileon radiation are comparable in the non-relativistic, low frequency, and low multipole regime. Moreover, in this limit Galileon radiation is actually amplified relative to its non-interacting counterpart for higher multipoles -- these findings are a direct consequence of the structure of the Galileon radial Green's function, which we have already highlighted in the discussion surrounding equations \eqref{VainshteinAmplification_IofII} and \eqref{VainshteinAmplification_IIofII}. In the high frequency limit, we confirm the anticipated Vainshtein screening of the Galileo radiation at low multipole orders; demonstrating it to be of $\mathcal{O}\left[ \xi_v^{-3/2} \right]$ relative to its non-interacting counterpart. At high enough multipoles, however, high frequency Galileon radiation becomes enhanced relative to its non-interacting counterpart. Moreover, for the astrophysical $n$ body system, where Newton's third law is obeyed between each compact body and $M$ but not between the compact bodies themselves, the leading $\mathcal{O}[v_a^2]$ terms in the radiation formulas, equations \eqref{RadiationFromNBodyProblem_Galileon_NR_LowFreq_ell=1} and \eqref{RadiationFromNBodyProblem_Galileon_NR_HighFreq_ell=1}, are small (scaling as $\mathcal{O}[m/M]$) but non-zero, in contrast with the non-self-interacting scalar case, where $\sum_a m_a \dd^2 \vec{y}_a/\dd t'^2 = 0$.

Having developed some quantitative understanding of the production of Galileon radiation in this paper, let us remark that, if Galileon waves exist, they are in principle detectable by GW detectors. About flat Minkowski spacetime, the sum of the graviton-matter and Galileon-matter coupling (eq. \eqref{Action_Perturbations}) is
\begin{align}
S_\text{I} \equiv -\frac{1}{2} \int \frac{T^{\mu\nu}}{\mpl} h_{\mu\nu}^\text{(eff)} \dd^4 x , \qquad
h_{\mu\nu}^\text{(eff)} \equiv h_{\mu\nu} - 2 \Pi \eta_{\mu\nu}.
\end{align}
This implies that, if Galileons are present, ordinary matter experiences an effective weakly curved metric of the form 
\begin{align}
g_{\mu\nu} = \eta_{\mu\nu} + \frac{h_{\mu\nu}^\text{(eff)}}{\mpl} ,
\end{align}
and the tidal forces experienced by the arms of the interferometers of GW detectors would now be due to both the transverse-traceless graviton $h^\text{(TT)}_{\mu\nu}$ and the Galileon waves $\varphi$.

Some of the calculations we have carried out have overlap with other recent work on the same topic \cite{deRham:2012fw}. In this paper, we have sought to understand the Galileon radiation spectrum in both the high and low frequency limits, and have found that the slope of the power spectrum of Galileon radiation should be a non-trivial function of $\xi_v$, the ratio of the Vainshtein radius $r_v$ to the typical wavelength of the emitted waves. (For instance, the radiative limit of our retarded Green's function, namely, eq. \eqref{GreensFunctionResult_RadiativeLimit_IofIII} and the coefficients $C_\ell^\text{(rad)}$s described in \eqref{GreensFunctionResult_RadiativeLimit_IIofIII} and \eqref{GreensFunctionResult_RadiativeLimit_IIIofIII}, have very different $\xi_v$-dependence for $|\xi_v| \ll 1$ compared to $|\xi_v| \gg 1$.) In \cite{deRham:2012fw}, the focus was on the power loss from binary pulsar systems such as PSR B1913+16, and thus the authors carried out an analysis valid in the high frequency limit~\cite{Private}.\footnote{As explained in \cite{deRham:2012fw}, for Galileons to be relevant for cosmology, it is often assumed that $\Lambda \sim \mpl H_0^2 \sim 1/(10^3 \text{km})$, where $H_0 \sim 10^{-33}$ eV is the current Hubble parameter. For binary pulars like PSR B1913+16, with masses on the order of a few solar masses, the corresponding Vainshtein radius is $r_v \sim \mathcal{O}[ 10^3 ]$ light years. Because the period of typical binaries ($\sim 1/\omega$) are of the order of a few hours, therefore $|\xi_v| \gg 1$.} In the limit $|\xi_v| \gg 1$ we find that the Galileon power scales as $|\xi_v|^{-3/2}$, in agreement with the results of~\cite{deRham:2012fw}. 

We note that, a priori, caution is required in interpreting the results both in this paper and in \cite{deRham:2012fw} as describing a comparable mass binary pulsar system, because it is unclear if such a binary can be treated within the perturbative framework, given that the nonlinearities of Galileons are very important when the motion is taking place deep within the collective Vainshtein radius of the system itself.\footnote{In more detail, in \cite{deRham:2012fw} the binary system, with masses $M_1$ and $M_2$, is modeled by adding and subtracting to the binaries' stress energies a monopole term with stress energy given by $T_0[\vec{x}] = (M_1+M_2) \delta^{(3)}[\vec{x}]$. The field generated by $T_0$ is then used as a background, and the stress-energies of the pair of point masses themselves minus the stress energy of the central monopole (see their eq. (2.6)) are treated as perturbations. However, since, as in this paper, the linearized equation of motion about the background of the monopole are solved, the subtracted monopole really plays no role, and this scheme is really equivalent to the setup where there is one central mass $M \equiv M_1+M_2$ and two masses in orbit around it, one of mass $M_1$ and the other $M_2$. But since $M_1$ and $M_2$ are comparable in magnitude to $M_1+M_2$, and there is no small dimensionless ratio one may use as an expansion parameter, it is not evident for the general binary problem that the $M_{1,2}$ are mere perturbations on top of the $M$. More quantitatively, recall the mode functions evaluated deep within the Vainshtein radius can be expressed in a separated form, reflecting the spherical symmetry of the background. In a comparable mass relativistic binary system, this spherical symmetry is completely absent. Thus, the separation of variables technique may not be useful in solving the general binary problem.} However, the authors of \cite{deRham:2012fw} have explicitly verified the validity of their perturbative scheme, and will present it in an upcoming publication \cite{Private,dRMT}.

{\it Future Work} \qquad It would be worthwhile to convert the frequency space calculation in this paper to a real time one, in order to better understanding the physical meaning behind the fractional powers of angular frequency found in the Galileon emission spectrum. To this end, the identification of the correct contour prescriptions in the Fourier integral of eq. \eqref{GreensFunctionResult_ModeExpansion} would be necessary. We also have left unexplored a large range of $|\xi_v|$, as we have only examined the extreme limits $|\xi_v| \ll 1$ and $|\xi_v| \gg 1$. Furthermore, since the nonlinearities of the Galileon theory play such a crucial role in its dynamics, we hope in the near future to go beyond the linear analysis about the background $\Pb$ due to $M$. We could compute, say within the Born approximation, the first correction arising from the cubic self-interactions in eq. \eqref{Action_Pi} to the wave solutions we have obtained here, so as to better understand the domain of validity of the results in this paper.

There are a number of interesting further directions for future work. In 4 dimensional flat spacetime, one should introduce the quartic and quintic Galileon terms and carry out an analogous analysis to that performed here. One may also wish to develop an understanding of the backreaction of the power loss on the motion of the $n$ point masses. In this context, there are other processes one could consider -- for example, one could carry out a calculation analogous to the one found in \cite{Peters:1970mx} and \cite{Kovacs:1978eu} for the gravitational case, in which one small mass scatters off $M$, producing Galileon bremsstrahlung radiation in the process.

\section{Acknowledgments}

We would like to thank Melinda Andrews, Lasha Berezhiani, Claudia de Rham, Kurt Hinterbichler, Justin Khoury, Denis Klevers, Andrew Matas, and Andrew Tolley for helpful discussions. This work was supported in part by the US Department of Energy, NSF grant PHY-0930521, and NASA ATP grant NNX11AI95G. We have made extensive use of {\sf Mathematica} \cite{MMA} to carry out the calculations in this paper. 

\appendix

\section{Minimally Coupled Massless Scalar Radiation: Spacetime Calculation}
\label{GradientsOfMinimallyCoupledMasslessScalarField}

The central goal of this section is an alternate derivation of eq. \eqref{RadiationFromNBodyProblem_Xi_NR}, but (for technical simplicity) without the inclusion of gravitational interactions. We wish to compute the non-relativistic power emitted in non-interacting massless scalar radiation by the motion of $n$ compact bodies. We will do so by finding the scalar field and its first derivatives generated by these point masses in flat spacetime; and from these gradients construct $r^2 \mathfrak{T}^{0r}$. However, we will first derive a expression in curved spacetime and proceed to specialize to Minkowski spacetime. The reason for doing so is that the Galileon $\varphi$ field generated by the $n$ body system we examined in section \eqref{NBodyProblem_Radiation} is related to the problem of a non-interacting scalar in the geometry given in eq. \eqref{EffectiveGeometry}.

Let the $n$ masses be $\{m_a|a=1,2,\dots,n\}$ and their spacetime locations $\{y^\mu_a\}$. The massless scalar theory in question is the curved spacetime generalization of the $\Xi$-theory in eq. \eqref{MinimallyCoupledMasslessScalar}, namely
\begin{align}
S'_\Xi &\equiv \frac{1}{2} \int \dd^4 x|g|^{1/2} \nabla^\mu \Xi \nabla_\mu \Xi + \sum_{a=1}^n \frac{m_a}{\mpl} \int \dd s_a \Xi,
\end{align}
with proper times
\begin{align}
\dd s_a \equiv \dd t\sqrt{g_{\mu\nu} \frac{\dd y^\mu_a}{\dd t} \frac{\dd y^\nu_a}{\dd t}}.
\end{align}
From the Hadamard form of the scalar Green's function in eq. \eqref{GreensFunction_HadamardForm}, 
\begin{align}
G_{x,x'} = \frac{\Theta[t-t']}{4\pi}\left( \delta[\sigma] U_{x,x'} + \Theta[\sigma] V_{x,x'}\right),
\end{align}
with
\begin{align}
U_{x,x'} \equiv \sqrt{\Delta_{x,x'}}
\end{align}
(for an explanation of these symbols, see \cite{PoissonReview}), the solution to $\Xi$ is given by the integral
\begin{align}
\label{XiCurvedSpacetimeIntegralRep}
\Xi[x]
&= \sum_{a=1}^n \frac{m_a}{4\pi \mpl} \int \dd s_a \Theta[t-y_a^0[s_a]] 
\left( \delta[\sigma_{x,y_a}] U_{x,y_a} + \Theta[\sigma_{x,y_a}] V_{x,y_a} \right)\nonumber,
\end{align}
For a fixed $x$, $\sigma_{x,y_a} = $ constant defines the geodesic joining $x$ (the observation point) to $y_a$ (the source point), provided this geodesic is unique (an assumption used when deriving the Hadamard form in eq. \eqref{GreensFunction_HadamardForm}). Thus it must be possible to invert $\sigma_{x,y_a}$ for $s_a$ and vice versa. In particular, in the following, we will need
\begin{align}
\frac{\dd \sigma_{x,y_a}}{\dd s_a} = \frac{\dd y_a^{\alpha'}}{\dd s_a} \nabla_{\alpha'} \sigma[x,x' = y_a] \ .
\end{align}
and hence
\begin{align}
\frac{\dd s_a}{\dd \sigma_{x,y_a}} = \left(\frac{\dd \sigma_{x,y_a}}{\dd s_a}\right)^{-1}.
\end{align}
The null cone piece of $\Xi$ involves $\delta[\sigma_{x,y_a}]$ which we may then write
\begin{align}
\delta[\sigma_{x,y_a}] = \left. \frac{\delta[s_a-\tau_a]}{|\dd \sigma_{x,y_a}/\dd s_a|} \right\vert_{s_a = \tau_a},
\end{align}
where $\tau_a$, the retarded time, is defined to be the proper time of the $a$th mass when it lies on the backward null cone of the observer at $x$,
\begin{align}
\sigma[x,y_a[\tau_a]] \equiv 0.
\end{align}
The ``backward'' part of the requirement is reinforced by the step function $\Theta[t-y_a^0]$ in eq. \eqref{XiCurvedSpacetimeIntegralRep}.

The curved spacetime scalar field produced by the $n$ point masses is therefore
\begin{align}
\label{Xi_CurvedSpacetime}
\Xi[x] &= \sum_{a=1}^n \frac{m_a}{4\pi\mpl} \Bigg\{ \left( \frac{\Delta^{1/2}[x,y_a]}{|\dd \sigma_{x,y_a}/\dd s_a|} \right)_{s_a = \tau_a}
	+ \int_{-\infty}^{\tau_a^-} \dd s' V[x,y_a[s']] \Bigg\}.
\end{align}
We may obtain the gradients of $\Xi$ by differentiating the integral representation in eq. \eqref{XiCurvedSpacetimeIntegralRep}. 
\begin{align}
\nabla_\alpha \Xi[x]
&= \sum_a \frac{m_a}{4\pi\mpl} \int \dd s_a \Bigg( \nabla_\alpha \sigma \delta'[\sigma] U 	
+ \delta[\sigma] \nabla_\alpha U + \delta[\sigma] \nabla_\alpha \sigma V + \Theta[\sigma] \nabla_\alpha V \Bigg).
\end{align}
The $\delta'[\sigma]$ term may be re-written, holding $x$ fixed,
\begin{align}
\delta'[\sigma_{x,y_a}] = \frac{\dd s_a}{\dd \sigma_{x,y_a}} \frac{\dd}{\dd s_a} \frac{\delta[s_a-\tau_a]}{|\dd \sigma[s_a=\tau_a]/\dd s_a|},
\end{align}
which allows us to integrate by parts to obtain
\begin{widetext}
\begin{align}
\label{GradientsOfXi_CurvedSpacetime}
&\nabla_{\alpha} \Xi[x] = \sum_{a=1}^n \frac{m_a}{4\pi\mpl} \Bigg( \Bigg\{
\frac{\nabla_{\alpha} U_{x,y_a}}{|\dd \sigma_{x,y_a}/\dd s_a|}
+ \frac{\nabla_{\rho'} \nabla_{\alpha} \sigma_{x,y_a} U_{x,y_a}
+ \nabla_{\rho'} U_{x,y_a} \nabla_{\alpha} \sigma_{x,y_a}}{(\dd \sigma_{x,y_a}/\dd s_a)^2} \frac{\dd y_a^{\rho'}}{\dd s_a} \\
&\qquad + \frac{\nabla_{\alpha} \sigma_{x,y_a} U_{x,y_a}}{|\dd \sigma_{x,y_a}/\dd s_a|^3} \left(
\frac{\dd^2 y_a^{\lambda'}}{\dd s_a^2} \nabla_{\lambda'} \sigma_{x,y_a} + \frac{\dd y_a^{\lambda'}}{\dd s_a} \frac{\dd y_a^{\kappa'}}{\dd s_a} \nabla_{\kappa'} \nabla_{\lambda'} \sigma_{x,y_a}
\right) + \frac{\nabla_{\alpha} \sigma_{x,y_a} V_{x,y_a}}{|\dd \sigma_{x,y_a}/\dd s_a|} \bigg\}\bigg\vert_{s_a=\tau_a} 
+ \int_{-\infty}^{\tau_a^-} \dd s_a \nabla_\alpha V_{x,y_a} \Bigg), \nonumber
\end{align}
\end{widetext}
with $U_{x,x'} \equiv \sqrt{\Delta_{x,x'}}$. The primed derivatives are with respect to $y_a$ (the $a$th point mass location) and the unprimed ones with respect to $x$ (the observer location).

{\it Minkowski spacetime} \qquad In Minkowski spacetime, $U=1$ and $\nabla U = V = \nabla V = 0$, while the world function reads
\begin{align}
\sigma_{x,x'} = \frac{1}{2} \eta_{\mu\nu} (x-x')^\mu (x-x')^\nu
\end{align}
so that its derivatives are
\begin{align}
\nabla_\mu \sigma_{x,y_a} = (x-y_a)_\mu, \quad \nabla_{\mu'} \sigma_{x,y_a} = (y_a-x)_\mu, \quad
\nabla_{\mu} \nabla_{\nu'} \sigma_{x,y_a} = -\eta_{\mu\nu},
\end{align}
and
\begin{align}
\frac{\dd \sigma_{x,y_a}}{\dd s_a} = -\frac{\dd y_a^{\nu'}}{\dd s_a} (x-y_a)_\nu.
\end{align}

For convenience we shall use a dot to represent a derivative with respect to $s_a$. The gradient of $\Xi$ in Minkowski spacetime generated by $n$ point masses is given by
\begin{align}
\label{GradientsOfXi_Minkowski}
\nabla^\mu \Xi[x]
&= -\sum_{a=1}^n \frac{m_a}{4\pi\mpl} \Bigg\{
\frac{\dot{y}_a^\mu}{(\dot{y}_a^\kappa (y_a-x)_\kappa)^2} 
-\frac{(x-y_a)^\mu}{|\dot{y}_a^\kappa (y_a-x)_\kappa|^3} \left(
\ddot{y}_a^\lambda (y_a-x)_\lambda + \dot{y}_a^2 \right) \Bigg\} \Bigg\vert_{s_a=\tau_a} \ ,
\end{align}
where the retarded proper time $\tau_a$ is the solution of the equation
\begin{align}
\label{RetardedTimeConditionInMinkowski}
t-y_a^0[\tau_a] = |\vec{x}-\vec{y}_a[\tau_a]|.
\end{align}
Comparing equations \eqref{GradientsOfXi_CurvedSpacetime} and \eqref{GradientsOfXi_Minkowski}, we see that in flat spacetime, the field detected by an observer (or felt by some test mass) at $x$ is a function of the positions of the $n$ point masses evaluated at the appropriate retarded times -- that is how long it takes the signal to reach the observer from the source. Whereas in curved spacetime, where Huygens' principle no longer holds, the problem of motion is a richer one, because the proper time integral involving the tail term $V_{x,y_a}$ in eq. \eqref{GradientsOfXi_CurvedSpacetime} receives contributions from the point masses' entire past histories. (The electromagnetic counterpart to this statement can be found in \cite{DeWittBrehme}, in which the $A_\mu$ and $F_{\mu\nu}$ analogues of equations \eqref{Xi_CurvedSpacetime} and \eqref{GradientsOfXi_CurvedSpacetime} are derived.)

We are now ready to determine the power radiated by the $n$ point masses moving in flat spacetime. Let us assume the motion of these $n$ bodies is confined within some finite spatial volume containing the center of the spatial coordinate system; this will certainly be true when these $n$ bodies are bound together by their mutual gravity, which is the central theme of this paper. We wish to extract the piece of $\Xi$ that represents radiation. As already explained earlier, in Minkowski spacetime, the radiative piece of $\Xi$ is the $1/r$ piece. From the result in eq. \eqref{GradientsOfXi_Minkowski}, we see that this can be identified by counting powers of $(x-y_a)$; this isolates the acceleration $\ddot{y}_a^\lambda$ term. If we let the observer lie at some very large radius, we deduce $\ddot{y}_a^\lambda (y_a-x)_\lambda = -|\vec{x}-\vec{y}_a| \ddot{y}_a^0 + \ddot{\vec{y}}_a\cdot(\vec{x}-\vec{y}_a) \to r(-\ddot{y}_a^0 + \ddot{\vec{y}}_a\cdot\widehat{x})$ and $\ddot{y}_a^\lambda (y_a-x)_\lambda = -|\vec{x}-\vec{y}_a| \dot{y}_a^0 + \dot{\vec{y}}_a\cdot(\vec{x}-\vec{y}_a) \to r(-\dot{y}_a^0 + \dot{\vec{y}}_a\cdot\widehat{x})$. Recalling the notation $\dd y_a^\mu/\dd s_a \equiv \dot{y}_a^\mu$ and $\dd^2 y_a^\mu/\dd s_a^2 \equiv \ddot{y}_a^\mu$,
\begin{align}
\label{GradientsOfXi_Minkowski}
\nabla^\mu \Xi[r\to\infty]
&= -\frac{\left( 1,\widehat{x} \right)}{4\pi r} \left.\sum_{a=1}^n \frac{m_a}{\mpl}
\frac{\ddot{y}_a^\alpha \eta_{\alpha\beta} \left( 1, \widehat{x} \right)^\beta}{|\dot{y}_a^0 - \dot{\vec{y}}_a\cdot\widehat{x}|^3} \right\vert_{s_a=\tau_a} \ ,
\end{align}
and the power radiated to infinity per unit solid angle $r^2 \mathfrak{T}^{0r} = r^2 \nabla^r \Xi \nabla^t \Xi$ is
\begin{align}
\label{Xi_RadiationFormula_Exact}
\frac{\dd E}{\dd t\dd \Omega} &= \frac{2 G_\text{N}}{\pi} \left( \left.\sum_{a=1}^n m_a
\frac{\ddot{y}_a^\alpha \eta_{\alpha\beta} \left( 1, \widehat{x} \right)^\beta}{\left\vert \dot{y}_a^\mu \eta_{\mu\nu} \left( 1, \widehat{x} \right)^\nu \right\vert^3} \right\vert_{s_a=\tau_a} \right)^2.
\end{align}
To take the non-relativistic limit, we first recall that $\dd/\dd s_a = (1-(\dd \vec{y}_a/\dd t')^2)^{-1/2} \dd/\dd t'$, where the retarded time $t'$ satisfies eq. \eqref{RetardedTimeConditionInMinkowski}, i.e. $t' = t - |\vec{x}-\vec{y}_a[t']|$. That retarded time depends on the trajectory $\vec{y}_a$ means, upon carrying out the retarded time derivatives with respect to $t'$ in eq. \eqref{Xi_RadiationFormula_Exact}, we still need to Taylor expand every time dependent expression in powers of $|\vec{y}_a[t']|/r$, because the latter expansion introduces further time derivatives. It is at this point that terms containing the same number of time derivatives (now evaluated at the approximate retarded time $t' = t-r$) are considered to be of the same order in the non-relativistic expansion. Up to $\mathcal{O}[v_a^3]$ we have
\begin{align}
\label{Xi_RadiationFormula_NR}
\frac{\dd E}{\dd t\dd \Omega} &= \frac{2 G_\text{N}}{\pi} \left( \left.\sum_{a=1}^n m_a
\left( \widehat{x}\cdot\frac{\dd^2 \vec{y}_a}{\dd t'^2} 
+ \frac{1}{2} \frac{\dd^3 \left(\widehat{x}\cdot\vec{y}_a\right)^2}{\dd t'^3} 
- \frac{1}{2}\frac{\dd}{\dd t'} \left(\frac{\dd\vec{y}_a}{\dd t'}\right)^2 + \mathcal{O}[v_a^4] \right)
\right\vert_{t'=t-r} \right)^2.
\end{align}

\section{Green's Function of Linear Second Order Ordinary Differential Equation}

\label{ODEGreensFunction_Appendix}

In this section we review the construction of the Green's function for linear second order ordinary differential equations (ODEs) in terms of their homogeneous solutions. Consider the differential operator
\begin{align}
\label{ODEDifferentialOperator}
\mathfrak{D}_z \equiv p_2[z] \frac{\dd^2}{\dd z^2} + p_1[z] \frac{\dd}{\dd z} + p_0[z] \ ,
\end{align}
where $p_{0,1,2}[z]$ are smooth functions of $z$. The homogeneous solution $f$ to the corresponding linear second order ODE satisfies
$\mathfrak{D}_z f[z] = 0$, and the associated symmetric Green's function obeys
\begin{equation}
\label{ODEGreensFunctionEquation}
\mathfrak{D}_z G[z,z'] = \mathfrak{D}_{z'} G[z,z'] = \lambda[z] \delta[z-z'] = \lambda[z'] \delta[z-z'] \ .
\end{equation}

To solve this, first assume $z > z'$, so that $\delta[z-z'] = 0$. The problem is thus reduced to solving the homogeneous equation $\mathfrak{D}_z G[z,z'] = \mathfrak{D}_{z'} G[z,z'] = 0$. Let $f_{1,2}[z]$ be a pair of linearly independent homogeneous solutions, i.e.
\begin{align}
\mathfrak{D}_z f_1[z] = \mathfrak{D}_z f_2[z] = 0, \quad {\sf Wr}_{(z)}[f_1,f_2] \neq 0,
\end{align}
where the Wronskian {\sf Wr} is defined to be
\begin{align}
{\sf Wr}_{(z)}[f_1,f_2] \equiv f_1[z] (f_2)'[z] - (f_1)'[z] f_2[z].
\end{align}
Because $\mathfrak{D}_z G[z,z'] = 0$, we must have
\begin{align}
G[z > z'] = \alpha^\text{I} f_\text{I}[z], \qquad \text{I} = 1,2
\end{align}
where the $\alpha^\text{I}$s are $z$-independent. Similarly, $\mathfrak{D}_{z'} G[z,z'] = 0$ means
\begin{align}
\alpha^\text{I} = A_>^\text{IJ} f_\text{J}[z'], \qquad \text{J} = 1,2,
\end{align}
where $A_>^\text{IJ}$ is now a $2\times 2$ matrix of $z,z'$-independent constants. The same argument would hand us, for $z < z'$,
\begin{align}
G[z < z'] = A_<^\text{IJ} f_\text{I}[z] f_\text{J}[z'], \qquad \text{I},\text{J} = 1,2
\end{align}
If $G[z,z']$ were not continuous at $z=z'$ its second derivative with respect to $z$ or $z'$ there, and hence the second line of eq. \eqref{ODEGreensFunctionEquation}, would involve a derivative of $\delta[z-z']$. That implies we may assume $G[z,z']$ is continuous at $z=z'$. $A_<^\text{IJ} f_\text{I}[z] f_\text{J}[z] = A_>^\text{IJ} f_\text{I}[z] f_\text{J}[z]$ imposes the conditions
\begin{align}
\label{ODEGreensFunctionEquation_A_I}
A_>^\text{11} = A_<^\text{11}, \ A_>^\text{22} = A_<^\text{22} \\
\label{ODEGreensFunctionEquation_A_II}
A_>^\text{12} + A_>^\text{21} = A_<^\text{12} + A_<^\text{21}.
\end{align}
We now integrate $\mathfrak{D}_z G[z,z'] = \lambda[z] \delta[z-z']$ around the neighborhood of $z=z'$, applying integrating-by-parts as many times necessary to shift all the derivatives acting on $G[z,z']$ onto the $p_{2,1}$. By continuity, the $p_0$ term, $[p_1 G]_{z=z'^-}^{z=z'^+}$, $[(p_2)' G]_{z=z'^-}^{z=z'^+}$ and the remaining integral involving $G$ (with no derivatives acting it) drop out -- assuming $p_{1,2}$ are smooth -- leaving us with
\begin{align}
\label{ODEGreensFunctionEquation_JunctionCondition_I}
\lambda[z'] = p_2[z'] \left[ \frac{\partial G[z,z']}{\partial z} \right]_{z=z'^-}^{z=z'^+}.
\end{align}
Employing the continuity conditions in equations \eqref{ODEGreensFunctionEquation_A_I}, one would find the $A_>^{11}$ and $A_>^{22}$ do not contribute to $\lambda[z']$. We may use \eqref{ODEGreensFunctionEquation_A_II} to eliminate, say $A_>^{12}$, and find \eqref{ODEGreensFunctionEquation_JunctionCondition_I} becomes
\begin{align}
\label{ODEGreensFunctionEquation_JunctionCondition_II}
\lambda[z'] = -p_2[z'] (A_<^{21}-A_>^{21}) {\sf Wr}_{(z')}[f_1,f_2]
\end{align}
Because any ``rotation'' of the pair $f_{1,2}$, i.e. the pair $\{q_\text{I} \equiv Q_\text{I}^{\phantom{I}\text{J}} f_\text{J}| \text{I}=1,2\}$ for any $2\times 2$ invertible $Q$, is an equally valid pair of linearly independent homogeneous solutions, without loss of generality we may choose $A_>^{21} = A_<^{21}-1$, such that now
\begin{align}
\label{ODEGreensFunction_Measure}
\lambda[z] = -p_2[z] {\sf Wr}_{(z)}[f_1, f_2]
\end{align}
Therefore, the general solution to eq. \eqref{ODEGreensFunctionEquation} is 
\begin{align}
\label{ODEGreensFunctionSolution}
G[z,z'] &= \mathcal{C} f_1[z_>] f_2[z_<] - (1-\mathcal{C}) f_1[z_<] f_2[z_>] \nonumber\\
&\qquad + C^{11} f_1[z] f_1[z'] + C^{22} f_2[z] f_2[z'],
\end{align}
where $z_>$ (or $z_<$) is the greater (or smaller) of the pair $(z,z')$, and $\mathcal{C}$, $C^{11}$ and $C^{22}$ are arbitrary constants. These coefficients will be fixed by the boundary conditions of the given physical problem. 

By using the differential equation obeyed by the $f_{1,2}$, one may readily show that
\begin{align}
\frac{\dd}{\dd z} {\sf Wr}_{(z)}[f_1,f_2] = -\frac{p_1[z]}{p_2[z]} {\sf Wr}_{(z)}[f_1,f_2].
\end{align}
This in turn means the Wronskian of two linearly independent solutions, and hence the measure $\lambda[z]$, can be solved up to an overall constant, without first solving for the homogeneous solutions. Recalling eq. \eqref{ODEGreensFunction_Measure}, we gather
\begin{align}
\label{ODEWronskianSolution}
\lambda[z] = \chi \ p_2[z] \exp\left[ -\int^z \dd z'' \frac{p_1[z'']}{p_2[z'']} \right],
\end{align}
where $\chi$ is the constant. In solving for $G[z,z']$, we will thus first choose a value for $\chi$, and use this choice to fix the normalization of the product of the solutions $f_1[z_>] f_2[z_<]$ in eq. \eqref{ODEGreensFunctionSolution}.

We wish to emphasize, because we are constructing a symmetric Green's function, one that is a Green's function with respect to both variables $z$ and $z'$, we have just seen that one does not have a choice in picking the measure $\lambda[z]$, but rather $\lambda[z]$ is fixed up to an overall numerical constant by $p_2[z]$ in eq. \eqref{ODEDifferentialOperator} and the Wronskian of any two linearly independent homogeneous solutions.

To summarize, once a pair of homogeneous solutions $f_{1,2}$ are known, the symmetric Green's function has the general solution given in equation \eqref{ODEGreensFunctionSolution}. The measure $\lambda[z]$ multiplying the $\delta$-functions in eq. \eqref{ODEGreensFunctionEquation} is given by eq. \eqref{ODEWronskianSolution}, and the overall constant $\chi$ there needs to be chosen. In a given problem -- for this paper it is the solution of $\widetilde{g}_\ell[\xi, \xi']$ in eq. \eqref{GreensFunctionResult_ModeExpansion} -- the analog of $\mathcal{C}$, $C^\text{11}$, and $C^\text{22}$ (or the $A_{1,2}$ and $B_{1,2}$) will be fixed by appropriate regularity and boundary conditions. Because $\lambda[z]$ has been computed, the overall normalization of the products $f_1[z_>] f_2[z_<]$ is determined by the Wronskian condition in eq. \eqref{ODEGreensFunction_Measure}.

{\it $\delta$-Function Measure} \qquad Let us conclude this section by justifying the $(rr')^{-1}$ measure on the right hand side of eq. \eqref{GreensFunctionEquation}. Equation \eqref{ODEWronskianSolution} informs us that we can determine this measure up to a constant, by integrating the ratio of $-2e_3/\xi$ to $-e_2$. This may be achieved by using the explicit expressions in \eqref{e1} through \eqref{e3} yielding
\begin{align}
\label{RadialModeFunction_WronskianIntegral}
\exp\left[ -\int^\xi \frac{2 e_3[\xi'',\xi_v]}{\xi'' e_2[\xi'',\xi_v]} \dd \xi'' \right]
= \left( \xi(\xi^3+\xi_v^3)\right)^{-\frac{1}{2}}.
\end{align}
This immediately tells us that the measure multiplying the $\delta$-functions in eq. \eqref{GreensFunctionEquation} is
\begin{align}
\chi e_2[\xi,\xi_v] \exp\left[ -\int^\xi \frac{2 e_3[\xi'',\xi_v]}{\xi'' e_2[\xi'',\xi_v]} \dd \xi'' \right]
= \frac{\chi}{\xi^2},
\end{align}
where $\chi$ is a constant. Far away from the central mass $M$, using equations \eqref{BackgroundPiSolution_LargeRadius_IofII} and \eqref{BackgroundPiSolution_LargeRadius_IIofII} to keep only the most dominant terms in $e_{1,2,3}$ (equations \eqref{e1} through \eqref{e3}), the left hand side of eq. \eqref{GreensFunctionEquation} yields, as expected, the flat spacetime minimally coupled massless scalar wave equation
\begin{align}
\label{GreensFunctionEquation_Minkowski}
\eta^{\mu\nu} \partial_\mu \partial_\nu G[x,x'] &= \frac{\chi}{rr'} \delta[t-t']\delta[r-r'] \delta[\cos\theta-\cos\theta'] \delta[\phi-\phi'] \ .
\end{align}
Therefore choosing $\chi = 1$ amounts to adhering to the usual convention.


\begin{thebibliography}{99}

\bibitem{Nicolis:2008in} 
  A.~Nicolis, R.~Rattazzi and E.~Trincherini,
  %``The Galileon as a local modification of gravity,''
  Phys.\ Rev.\ D {\bf 79}, 064036 (2009)  [arXiv:0811.2197 [hep-th]].  %%CITATION = ARXIV:0811.2197;%%

\bibitem{Dvali:2000hr} 
  G.~R.~Dvali, G.~Gabadadze and M.~Porrati,
  %``4-D gravity on a brane in 5-D Minkowski space,''
  Phys.\ Lett.\ B {\bf 485}, 208 (2000)  [hep-th/0005016].  %%CITATION = HEP-TH/0005016;%%
  
  %\cite{Vainshtein:1972sx}
\bibitem{Vainshtein:1972sx} 
  A.~I.~Vainshtein,
  ``To the problem of nonvanishing gravitation mass,''
  Phys.\ Lett.\ B {\bf 39}, 393 (1972).
  %%CITATION = PHLTA,B39,393;%%

%\cite{Deffayet:2001uk}
\bibitem{Deffayet:2001uk} 
  C.~Deffayet, G.~R.~Dvali, G.~Gabadadze and A.~I.~Vainshtein,
  ``Nonperturbative continuity in graviton mass versus perturbative discontinuity,''
  Phys.\ Rev.\ D {\bf 65}, 044026 (2002)
  \href{http://arxiv.org/abs/hep-th/0106001}{[hep-th/0106001]}.
  %%CITATION = HEP-TH/0106001;%%

%\cite{Luty:2003vm}
\bibitem{Luty:2003vm} 
  M.~A.~Luty, M.~Porrati and R.~Rattazzi,
  ``Strong interactions and stability in the DGP model,''
  JHEP {\bf 0309}, 029 (2003)
  \href{http://arxiv.org/abs/hep-th/0303116}{[hep-th/0303116]}.
  %%CITATION = HEP-TH/0303116;%%
  
  %\cite{Nicolis:2004qq}
\bibitem{Nicolis:2004qq} 
  A.~Nicolis and R.~Rattazzi,
  ``Classical and quantum consistency of the DGP model,''
  JHEP {\bf 0406}, 059 (2004)
  \href{http://arxiv.org/abs/hep-th/0404159}{[hep-th/0404159]}.
  %%CITATION = HEP-TH/0404159;%%
 
%\cite{Deffayet:2009wt}
\bibitem{Deffayet:2009wt} 
  C.~Deffayet, G.~Esposito-Farese and A.~Vikman,
  ``Covariant Galileon,''
  Phys.\ Rev.\ D {\bf 79}, 084003 (2009)
  \href{http://arxiv.org/abs/0901.1314}{[arXiv:0901.1314 [hep-th]]}.
  %%CITATION = ARXIV:0901.1314;%%
  
   %\cite{Deffayet:2009mn}
\bibitem{Deffayet:2009mn} 
  C.~Deffayet, S.~Deser and G.~Esposito-Farese,
  ``Generalized Galileons: All scalar models whose curved background extensions maintain second-order field equations and stress-tensors,''
  Phys.\ Rev.\ D {\bf 80}, 064015 (2009)
  \href{http://arxiv.org/abs/0906.1967}{[arXiv:0906.1967 [gr-qc]]}.
  %%CITATION = ARXIV:0906.1967;%%
 
 %\cite{deRham:2010eu}
\bibitem{deRham:2010eu} 
  C.~de Rham and A.~J.~Tolley,
  ``DBI and the Galileon reunited,''
  JCAP {\bf 1005}, 015 (2010)
  \href{http://arxiv.org/abs/1003.5917}{[arXiv:1003.5917 [hep-th]]}.
  %%CITATION = ARXIV:1003.5917;%%
  
  %\cite{Deffayet:2010zh}
\bibitem{Deffayet:2010zh} 
  C.~Deffayet, S.~Deser and G.~Esposito-Farese,
  ``Arbitrary $p$-form Galileons,''
  Phys.\ Rev.\ D {\bf 82}, 061501 (2010)
  \href{http://arxiv.org/abs/1007.5278}{[arXiv:1007.5278 [gr-qc]]}.
  %%CITATION = ARXIV:1007.5278;%%
  
   %\cite{Padilla:2010de}
\bibitem{Padilla:2010de} 
  A.~Padilla, P.~M.~Saffin and S.~-Y.~Zhou,
  ``Bi-galileon theory I: Motivation and formulation,''
  JHEP {\bf 1012}, 031 (2010)
  \href{http://arxiv.org/abs/1007.5424}{[arXiv:1007.5424 [hep-th]]}.
  %%CITATION = ARXIV:1007.5424;%%
    
%\cite{Deffayet:2010qz}
\bibitem{Deffayet:2010qz} 
  C.~Deffayet, O.~Pujolas, I.~Sawicki and A.~Vikman,
  ``Imperfect Dark Energy from Kinetic Gravity Braiding,''
  JCAP {\bf 1010}, 026 (2010)
  \href{http://arxiv.org/abs/1008.0048}{[arXiv:1008.0048 [hep-th]]}.
  %%CITATION = ARXIV:1008.0048;%%

%\cite{Hinterbichler:2010xn}
\bibitem{Hinterbichler:2010xn} 
  K.~Hinterbichler, M.~Trodden and D.~Wesley,
  ``Multi-field galileons and higher co-dimension branes,''
  Phys.\ Rev.\ D {\bf 82}, 124018 (2010)
  \href{http://arxiv.org/abs/1008.1305}{[arXiv:1008.1305 [hep-th]]}.
  %%CITATION = ARXIV:1008.1305;%%
  
  %\cite{Goon:2010xh}
\bibitem{Goon:2010xh} 
  G.~L.~Goon, K.~Hinterbichler and M.~Trodden,
  ``Stability and superluminality of spherical DBI galileon solutions,''
  Phys.\ Rev.\ D {\bf 83}, 085015 (2011)
  \href{http://arxiv.org/abs/1008.4580}{[arXiv:1008.4580 [hep-th]]}.
  %%CITATION = ARXIV:1008.4580;%%
    
%\cite{Khoury:2011da}
\bibitem{Khoury:2011da} 
  J.~Khoury, J.~-L.~Lehners and B.~A.~Ovrut,
  ``Supersymmetric Galileons,''
  Phys.\ Rev.\ D {\bf 84}, 043521 (2011)
  \href{http://arxiv.org/abs/1103.0003}{[arXiv:1103.0003 [hep-th]]}.
  %%CITATION = ARXIV:1103.0003;%%

%\cite{Goon:2011qf}
\bibitem{Goon:2011qf} 
  G.~Goon, K.~Hinterbichler and M.~Trodden,
  ``Symmetries for Galileons and DBI scalars on curved space,''
  JCAP {\bf 1107}, 017 (2011)
  \href{http://arxiv.org/abs/1103.5745}{[arXiv:1103.5745 [hep-th]]}.
  %%CITATION = ARXIV:1103.5745;%%
  
%\cite{Trodden:2011xh}
\bibitem{Trodden:2011xh}
  M.~Trodden and K.~Hinterbichler,
  ``Generalizing Galileons,''
  Class.\ Quant.\ Grav.\  {\bf 28} (2011) 204003
  \href{http://arxiv.org/abs/1104.2088}{[arXiv:1104.2088 [hep-th]]}.
  %%CITATION = ARXIV:1104.2088;%%  

%\cite{Goon:2011uw}
\bibitem{Goon:2011uw} 
  G.~Goon, K.~Hinterbichler and M.~Trodden,
  ``A New Class of Effective Field Theories from Embedded Branes,''
  Phys.\ Rev.\ Lett.\  {\bf 106}, 231102 (2011)
  \href{http://arxiv.org/abs/1103.6029}{[arXiv:1103.6029 [hep-th]]}.
  %%CITATION = ARXIV:1103.6029;%%    

%\cite{Burrage:2011bt}
\bibitem{Burrage:2011bt} 
  C.~Burrage, C.~de Rham and L.~Heisenberg,
  ``de Sitter Galileon,''
  JCAP {\bf 1105}, 025 (2011)
  \href{http://arxiv.org/abs/1104.0155}{[arXiv:1104.0155 [hep-th]]}.
  %%CITATION = ARXIV:1104.0155;%%

%\cite{Goon:2011xf}
\bibitem{Goon:2011xf} 
  G.~Goon, K.~Hinterbichler and M.~Trodden,
  ``Galileons on Cosmological Backgrounds,''
  JCAP {\bf 1112}, 004 (2011)
  \href{http://arxiv.org/abs/1109.3450}{[arXiv:1109.3450 [hep-th]]}.
  %%CITATION = ARXIV:1109.3450;%%

%\cite{Zhou:2011ix}
\bibitem{Zhou:2011ix} 
  S.~-Y.~Zhou and E.~J.~Copeland,
  ``Galileons with Gauge Symmetries,''
 \href{http://arxiv.org/abs/1112.0968}{arXiv:1112.0968 [hep-th]}.
  %%CITATION = ARXIV:1112.0968;%%

%\cite{Goon:2012mu}
\bibitem{Goon:2012mu} 
  G.~Goon, K.~Hinterbichler, A.~Joyce and M.~Trodden,
  ``Gauged Galileons From Branes,''
  \href{http://arxiv.org/abs/1201.0015}{arXiv:1201.0015 [hep-th]}.
  %%CITATION = ARXIV:1201.0015;%%

%\cite{Goon:2012dy}
\bibitem{Goon:2012dy} 
  G.~Goon, K.~Hinterbichler, A.~Joyce and M.~Trodden,
  %``Galileons as Wess-Zumino Terms,''
  JHEP {\bf 1206}, 004 (2012)
  [arXiv:1203.3191 [hep-th]].
  %%CITATION = ARXIV:1203.3191;%%

%\cite{Gabadadze:2012tr}
\bibitem{Gabadadze:2012tr} 
  G.~Gabadadze, K.~Hinterbichler, J.~Khoury, D.~Pirtskhalava and M.~Trodden,
  %``A Covariant Master Theory for Novel Galilean Invariant Models and Massive Gravity,''
  arXiv:1208.5773 [hep-th].
  %%CITATION = ARXIV:1208.5773;%%
  
    %\cite{deRham:2010ik}
\bibitem{deRham:2010ik} 
  C.~de Rham and G.~Gabadadze,
  ``Generalization of the Fierz-Pauli Action,''
  Phys.\ Rev.\ D {\bf 82}, 044020 (2010)
  \href{http://arxiv.org/abs/1007.0443}{[arXiv:1007.0443 [hep-th]]}.
  %%CITATION = ARXIV:1007.0443;%%

%\cite{deRham:2010kj}
\bibitem{deRham:2010kj} 
  C.~de Rham, G.~Gabadadze and A.~J.~Tolley,
  ``Resummation of Massive Gravity,''
  Phys.\ Rev.\ Lett.\  {\bf 106}, 231101 (2011)
  \href{http://arxiv.org/abs/1011.1232}{[arXiv:1011.1232 [hep-th]]}.
  %%CITATION = ARXIV:1011.1232;%%

%\cite{Hinterbichler:2011tt}
\bibitem{Hinterbichler:2011tt} 
  K.~Hinterbichler,
  ``Theoretical Aspects of Massive Gravity,''
  \href{http://arxiv.org/abs/1105.3735}{arXiv:1105.3735 [hep-th]}.
  %%CITATION = ARXIV:1105.3735;%%

\bibitem{ACHT}
	M.~Andrews, Y.-Z.~Chu, K.~Hinterbichler, and M.~Trodden, (in progress).
  
\bibitem{deRham:2012fw} 
  C.~de Rham, A.~J.~Tolley and D.~H.~Wesley,
  %``Vainshtein Mechanism in Binary Pulsars,''
  arXiv:1208.0580 [gr-qc].
  %%CITATION = ARXIV:1208.0580;%%

\bibitem{MMA}
	Wolfram Research, Inc., Mathematica, Version 8.0, Champaign, IL (2011). 

\bibitem{Wagoner:1976am} 
  R.~V.~Wagoner and C.~M.~Will,
  %``PostNewtonian Gravitational Radiation from Orbiting Point Masses,''  
  Astrophys.\ J.\  {\bf 210}, 764 (1976)  [Erratum-ibid.\  {\bf 215}, 984 (1977)].  %%CITATION = ASJOA,210,764;%%

\bibitem{Chu:2008xm} 
  Y.~-Z.~Chu,
  %``The n-body problem in General Relativity up to the second post-Newtonian order from perturbative field theory,''
  Phys.\ Rev.\ D {\bf 79}, 044031 (2009)
  [arXiv:0812.0012 [gr-qc]].
  %%CITATION = ARXIV:0812.0012;%%

\bibitem{HadamardBook}
	J.~Hadamard, ``{\it Lectures on Cauchy's problem in linear partial differential equations}'' 
	(Yale University Press, New Haven, 1923).

\bibitem{PoissonReview} 
  E.~Poisson,
  %``The Motion of point particles in curved space-time,''  
  Living Rev.\ Rel.\  {\bf 7}, 6 (2004)  [gr-qc/0306052].  %%CITATION = GR-QC/0306052;%%

  E.~Poisson, A.~Pound and I.~Vega,
  %``The Motion of point particles in curved spacetime,''  
  Living Rev.\ Rel.\  {\bf 14}, 7 (2011)  [arXiv:1102.0529 [gr-qc]].
  %%CITATION = ARXIV:1102.0529;%%

\bibitem{MTW}
  C.~W.~Misner, K.~S.~Thorne and J.~A.~Wheeler,
  %``Gravitation,''  
  San Francisco 1973, 1279p

\bibitem{Peters:1970mx} 
  P.~C.~Peters,
  %``Relativistic gravitational bremsstrahlung,''  
  Phys.\ Rev.\ D {\bf 1}, 1559 (1970).  %%CITATION = PHRVA,D1,1559;%%

\bibitem{Kovacs:1978eu} 
  S.~J.~Kovacs and K.~S.~Thorne,
  %``The Generation of Gravitational Waves. 4. Bremsstrahlung,''
  Astrophys.\ J.\  {\bf 224}, 62 (1978).  %%CITATION = ASJOA,224,62;%%

	S.~J.~Kovacs and K.~S.~Thorne,
  %``The Generation of Gravitational Waves. 3. Derivation of Bremsstrahlung Formulas,'' 
  Astrophys.\ J.\  {\bf 217}, 252 (1977).  %%CITATION = ASJOA,217,252;%%

\bibitem{Private}
	Private communication with Claudia de Rham and Andrew Tolley.

\bibitem{dRMT}
	C.~de Rham, A.~Matas, and A.~Tolley, in progress (2012)

\bibitem{DeWittBrehme} 
  B.~S.~DeWitt and R.~W.~Brehme,
  %``Radiation damping in a gravitational field,''
  Annals Phys.\  {\bf 9}, 220 (1960).  %%CITATION = APNYA,9,220;%%


\end{thebibliography}
\end{document}